# TRENDS IN eBUSINESS AND eGOVERNMENT

## Edited by
## Ömer AYDIN

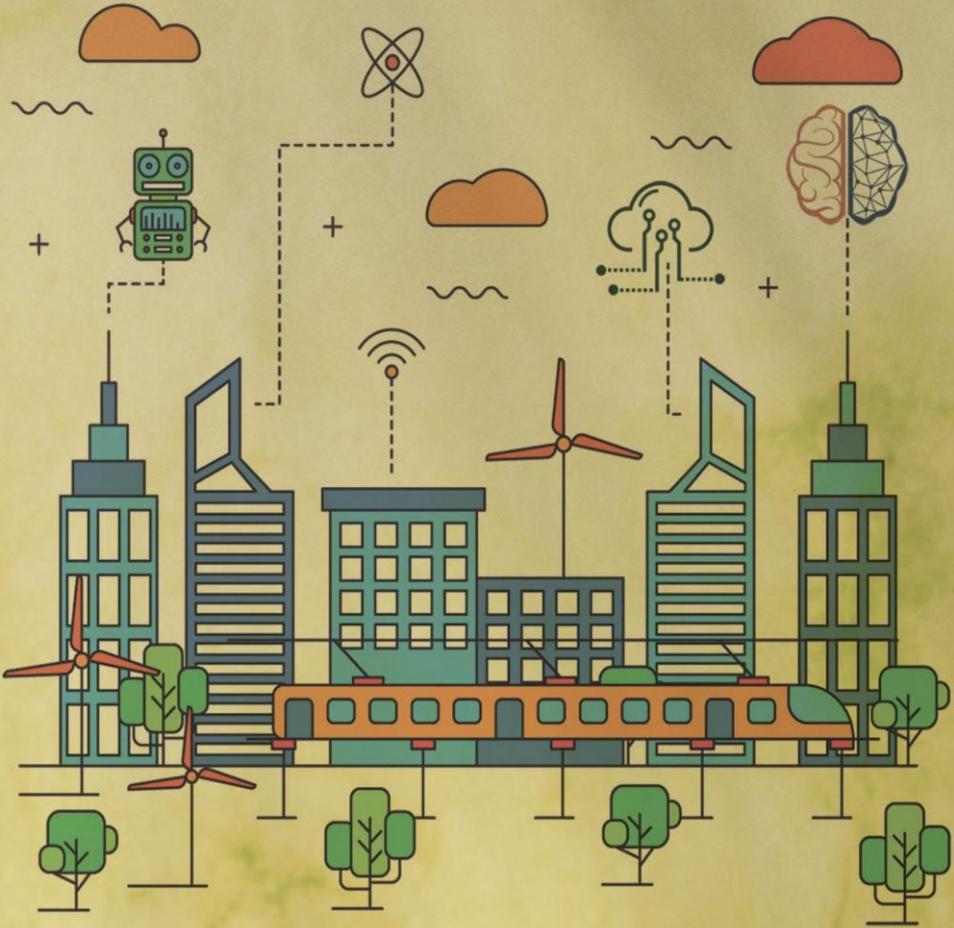

**EFEACADEMY**

# TRENDS IN eBUSINESS AND eGOVERNMENT

**Edited by**

Dr. Ömer AYDIN

İstanbul

December, 2020

# TRENDS IN eBUSINESS AND eGOVERNMENT

# TRENDS IN eBUSINESS AND eGOVERNMENT

# FOREWORD

Technology affects all areas. Business and government processes are changing with the use of the internet, mobile devices, internet of things, blockchain, machine learning, artificial intelligence and many other new technologies. In this book, we aimed to focus the use of technology, new trends in business life and government also we publish high-quality research studies in all sub-areas of Information Systems, Knowledge Management, eBusiness, eCommerce, eMarketing, mCommerce, eGovernment, ePublic Services, eGovernance etc. In the chapters of the book, original studies on electronic business and government are included. Studies give an idea about technologies that can be used in this field and new methods that can be applied. This book covers some of the topics listed in the "Call for book" web page that can be accessed on http://www.izmiracademy.com/1/trends-in-ebusiness-and-egovernment/. All other required information is listed on that page.

This book is an edited book that has been reviewed through a double-blind peer-review process. Each book chapter was reviewed by at least two different reviewers who are experts in their field. The chapters in the book have been edited and this publication has emerged as a result. The book consists of 7 chapters. Book chapter authors are reputable scientists from different countries of the world.

The first chapter is a critical review and a case study in e-Business, with special attention to the digital currencies resource and its possibilities. It is an example of technovation for the improvement of personnel income and motivation, as a good practice of CSR 3.0. The study explains how it works this win-win practice, with a real example of a Spanish company. The second chapter attempts to incorporate the Unified Theory of Acceptance and Use of Technology (UTAUT) model with perceived risk theory (security risk and privacy risk) to explore its impact towards the intention to use m-government services. Age, gender and education level were also adopted as moderator variables to provide an in-depth understanding of citizens' preference in m-government services. Partial Least Square (PLS) Structural Equation Modelling method was conducted. The third chapter aims to assess the level of gender inclusivity in the municipal e-procurement processes in the City of Johannesburg as a case study. It uses a Gender and Development (GAD) approach. Among the questions raised in the chapter are whether gender mainstreaming is considered in the municipal procurement processes; and if there are any initiatives in place to capacitate men and women to ensure their participation in the e-procurement processes. The fourth chapter examines the impediments that derail the intensive uptake of eLearning programmes in a particular higher education institution. The study adopted an inductive research paradigm that followed a qualitative research strategy. Data were collected by means of one-on-one in-depth interviews from selected faculty members at a nominated institution of higher learning. The fifth chapter investigated the role of Knowledge Management Systems (KMS) in enhancing the export performance of firms operating within the manufacturing sector in Zimbabwe. The study used a quantitative approach in which a survey questionnaire was distributed to 555 managers drawn from 185 manufacturing firms based in Harare. Data analyses involved the use of descriptive statistics, Spearman correlations and regression analysis. In the sixth chapter, a survey was undertaken on 131 small and medium-sized enterprises (SMEs) from Pelagonija region in order to determine the current level of SME digitalization within the region. It is aimed to compare with European Union (EU) average and to make conclusions on the impact of the SME digitalization to region gross domestic product (GDP) growth as well as revenues collection. The last chapter's purpose was to develop a measuring and modelling framework/instrument of Internet banking service quality (IBSQ) for the South African banking sector. Snowball and convenience sampling, both non-probability techniques were used to recruit participants for the study. A total of 310 Internet banking customer responses were utilised in the analysis.

Dr. Ömer AYDIN

December, 2020

İzmir, Turkey

# CONTENT

# Trends in eBusiness: digital currencies for a CSR 3.0 good practice[*]

Antonio Sánchez-Bayón[1] , Miguel Ángel García-Ramos Lucero[2]

## Abstract

This is a critical review and a case study in e-Business, with special attention to the digital currencies resource and its possibilities. It is an example of technovation for the improvement of personnel income and motivation, as a good practice of CSR *3.0*. The study explains how it works this win-win practice, with a real example of a Spanish company. In this case, there are benefits for the whole stakeholders, the environment, other companies and the next generations.
**Jel Codes:** D24, D31, I31, J3, K0, L2, M14, O15, O33.

## INTRODUCTION: Is there a resistance to the digital transition & business culture transformation?

This is a critical and practical study to support the digital transition and the eBusiness in the timeline of Horizon 2030 (H2030)[3]. The attention is focused on the digital currency, as an example of technovation for personnel income and a good practice of *CSR 3.0,* which distingues the enterprises oriented towards people wellness & happiness management. In other publications, there was a general contextualization:

- an initial balance of globalization and its changes (with its troubles and challenges to reach the *knowledge society*, Sánchez-Bayón, 2016);

- the attention to the main economic-business transformations (Andreu & Sánchez-Bayón, 2019);

- an introduction to the current stage of *post-globalization* (as a transitory period of global convergence: from the *value crisis of 2008* to *H2030*, Valero and Sánchez-Bayón, 2018);

---

[*] Paper written for the PhD dissertation in Economics; with the support of the research group GESCE-URJC (URL: https://gestion2.urjc.es/pdi/groups-investigacion/gesce).

[1] SJD (UCM), PhD in Theology (UM), PhD in Humanities (UVA), PhD in Philosophy (UCM) & PhDc. in Economics (UVA & UCM). Prof. Applied Economics at URJC (office J49, Vicalvaro campus, Paseo de Artilleros s/n, 28032 Madrid). Email: Antonio.sbayon@urjc.es; Orcid: 0000-0003-4855-8356

[2] PhDc. in Economy (UCJC). Prof. Finances at EAE Business School (Madrid campus, c/Joaquin Costa 42, Madrid). Email: magarcia@campus.eae.es; Orcid: 0000-0002-8671-0374

[3] It is the date established as the point of no-return: the nations, which accept the projects and alliances in international fora and institutions (e.g. United Nations: Sustainable Development Goals, Global Compact, Future of Work), in the way to get the convergence for knowledge society.

- an explanation about the *4th. industrial & technological revolution* and the digital economy (included, *gig phase, wellbeing economics*, etc., Sánchez-Bayón, 2019a & 2019b);

- even a range of innovation trends in business culture, occupational wellbeing and organizational health & wellness (González & Sánchez-Bayón, 2019).

In this occasion, it is offered a case of business practice, based in *social-business digital currencies* (SBDC –beyond the current social currencies-). They are useful as a refutation to the proposals of *mainstrain* academics (from welfare state economics-WSE) and the *new-luddite* militants (opposed to technological advances because they are considered a violation of working conditions, destroying jobs and increasing social disparity, Bailey, 1998. Sale, 1996). Current and previous luddites (see table 1), they are wrong, since the disappearance of jobs in one sector leads to the appearance of new jobs in emerging sectors; the same ones being more suitable for human inventiveness. For example, a tenant farmer with no limit of hours and with a subsistence production to become an industrial worker with shifts and an steady salary (2º industrial rev.), going through being an office clerk with fixed hours and income that allows savings (3º industrial rev.), even professionals with financial and schedule freedom (4º industrial rev.). Actually, the relationship between technological advances and labor wellbeing is not proportionally inverse, but exponentially convergent. The more technological advances take place, the more global wealth increases (both in terms of income and benefits to be enjoyed); and the greater convergence takes place in the planetary standard of living, thus increasing the wellbeing of humanity and its life expectancy. Those are two of the major components of the measurement of the global happiness index. In addition, both were announced by Bentham and Malthus in the 19th century, and they were the inspiration to measure the development, since the 1960s, by the *Organization of Economic Cooperation and Development-OECD*, and worldwide since 2012 (Rojas, 2014. VV.AA.a, 2012. VV.AA.b, 2020). Such a phenomenon, by which artificial intelligence has to overcome and replace the human being in tedious tasks - doing it even better - is called *singularity* (Kurzweil, 2005), and its point of no return is predicted for 2030 (coinciding with the rest of planetary convergence plans, such as *Global Compact-United Nations/UN, Future of work-International Labour Organization/ILO, Green Compact-Europena Union/EU*, etc.). As evidence we find the reports of specialized international organizations, such as the World Bank or the International Monetary Fund, as well as the indexes evolution such as the Gini-OECD coefficient (which is decreasing as the Lorenz curve flattens worldwide) or the human development index-UN (that is being improving yearly, see Table 1).

In this framework of global improvement, universalized since globalization, a risk exists as the *Easterlin paradox*, if only attention is paid to production and not to people's happiness (Easterlin, 1974. Easterlin et al 2010). This alarm was useful for the States part of the OECD, which after the great crisis of the 70s (and its stagflation), led to the reformulation of the Business Schools of the Anglo-Saxon and Nordic countries (Sánchez-Bayón et al., 2020). Currently, after globalization, the collection of all revolutions (see table 1), they have spread worldwide, taking place in a concentrated manner, allowing the accelerated growth of the vast World region of *the Trans-Pacific Area* (Pacific, Oceania and Southeast Asia), which by H2030 it will have surpassed the *Atlantic World* (origin of the aforementioned revolutions). Now, for this increase to become well-being, it requires a reinvestment of the wealth generated in infrastructure and education services, health, transport, housing, etc. What is more a profound change in business culture is needed, so that organizations stop being exclusively results-oriented and no longer treat their workers as mere human resources (interchangeable pieces of a mechanical system), thus beginning to pay attention to talent and motivation of employees and their organizational well-being (Sánchez-Bayón, 2019a).

**Table 1.** Comparison relating industrial and technological revolutions (Sánchez-Bayón, 2019a).

| Revolutions | Features | Macro and social indexes |
|---|---|---|
| 1st Rev. (1790-1870, Atlantic Europe) | coal and steam engine; it goes from the countryside to urban workshops (highlighting the textile sector); civil service leasing contracts (for agreed days and benefits); estates and unions slow their progress | Less than 1,200 million people, with a world GDP per capita of less than $ 1,000. |
| 2nd Rev. (1880-1950, in Europe and the Anglo-Saxon world) | oil, electricity and assembly line, it goes from workshops to factories (highlighting the automobile sector); proper employment contracts (under a protective legal regime); its advance slows down (with accelerations and recessions) wars and state interventions. | At the beginning of the c. XX the world population was of 2,000 million people approx., With a GDP per capita over 1,000 $ |
| 3rd Rev. (1960-2008, in the West – especially Asian tigers) | computing and robotization, plus nuclear and renewable energy; It goes from factories to centralized techno-bureaucratic headquarters and offshored production and sales modules, plus the emergence of *malls* or shopping centers, with a diversity of labor relations and employability (civil and commercial contracts, labor, civil servants, etc.). State interventions continue to alter their progress (this is WSE's golden age). | At the turn of the millennium, the worldwide population was over 6,000 million inhabitants and its GDP per capita was close to $ 10,000 |
| 4th Rev. (2008-2030, planetarium) | internet, programming (especially, *block-chain* since 2009) and mobiles (*smartphone* as an office), it is the era of social networks, *apps & everywhere commerce-enc* or virtual continuous marketing, giving the return of the professional (*knowmads v. freeriders*), who can be a commission agent, biller, affiliate, etc. (New formulas for the regulation of mixed labor relations emerge, eg. *click-pay, flexecurity, part-time jobs mix*). It is also the period of the emergence of *smart-contracts & DAO* (smart contracts, like codes in the cloud, whose parts are artificial intelligence, which operate from the Stock Market to driving with no driver). | We are currently more than 7,400 million population on the planet, with a per capita GDP of more than $ 13,500. |

This has been the deficit of the corporate production model of the *Asian tigers* (e.g. Japan, South Korea, Singapore, Taiwan, Hong-Kong), which carried out their revolutions after World War II, putting themselves at the level of the most developed countries (OECD), but running out of its model for not advancing to the stage of the digital economy relative to the economy of happiness (serve as evidence, from the alienation of the South Korean chaebol model, to the death of work due to *karoshi* - excess work - or *karojisatsu* - suicide by working conditions- of Japanese corporations, Frank, 2014. Amagasa et al, 2005).

The first stage of the digital economy has been the *gig phase* or bowling phase (Sánchez-Bayón, 2019b), which includes:

a) the collaborative and circular economy-CCE (it is based on social networks and platforms, recycling shared goods and services, e.g. AirBnB, Uber, Wallapop);

b) the autonomous economy-AE (it is based on *big-data, internet of things-IoT, artificial intelligence-IA, augmented reallyty/AR-virtual reallyty/VR-mixed reallyty/MR*, etc., articulated through 5G, *block-chain, smart-contracts* and DAOs, e.g. funds of investment in autonomous car fleet, *fintech*);

c) the orange economy-OE (it is based on talent and creativity applied to experience and entertainment, eg. gastronomy, tourism, video games, festivals).

With post-globalization, there is implementing around the World a collaborative intelligence network (e.g. *Global Compact-UN, Wellbeing Economics Alliance-Wold Economic Forum*) to share experiences and good practices that allow progress towards the next stage of the digital economy, such as the authentic welfare economy or *wellbeing economics*. This new stage includes expressions like talent & happiness management (Cubeiro, 2012. Frey, 2018). Other expression of the wellbeing economics, it is the *corporate social responsibility 3.0* (CSR 3.0). In this case, it pays attention to the social-business digital currencies (SBDC), as a resource to improve the remuneration of employees and, at the same time, to care the environmental and other social benefits. Therefore, previously, what is understood and how digital currencies operate will be clarified, as well as its contribution to the promotion of CSR 3.0 (which is typical of companies oriented towards people, their talent and their happiness), in addition to illuminating about the hypothetical initially raised paradox.

## CRYPTOCURRENCIES V. DIGITAL CURRENCIES: what they are, how they operate and what are their implications in digital economy

Cryptocurrencies, or better said (for this study) digital currencies, are autonomous, virtual and decentralized pecuniary units, which can be exchanged as a payment instrument for means of a system or *network* of electronic operations that does not require financial intermediaries (since all participants are notaries). The cited features are summarized in their digital condition, as they are carried out through an electronic procedure (such as a transfer or card payment). As for its origin and development, find the synopsis of the Table 2.

**Table 2.** Development of digital currencies (own elaboration for this research)

| Stages | Milestones and features | Relevant cases |
|---|---|---|
| Background (1988) | *Denationalization of money* (advertisement of non-national currencies, Hayek, 1976). Cover of *The Economist* (predicting the appearance of a currency that would displace national ones). | Reference baskets of currencies are introduced, which will give way to cases such as the ECU (antecedent of the euro and with which it could be operated via stock exchange interconnection systems). |
| 1998 | The word cryptocurrency is introduced and consolidated | Appearance of the Wei Dai B-Money system |
| 2008/2009 | The first paper on Bitcoin is published | Satoshi Nakamoto spreads Bitcoin and his first operation takes place on metzdowd.com |
| May 22, 2010 | First real transaction with Bitcoin | Some pizzas were paid with Bitcoins |
| December 2017 | Price of derivatives contracts on Bitcoin | Futures on Bitcoin are traded in CME and CBOE |

**Figure 1.** Evolution of the price of Bitcoin (own elaboration for this research)

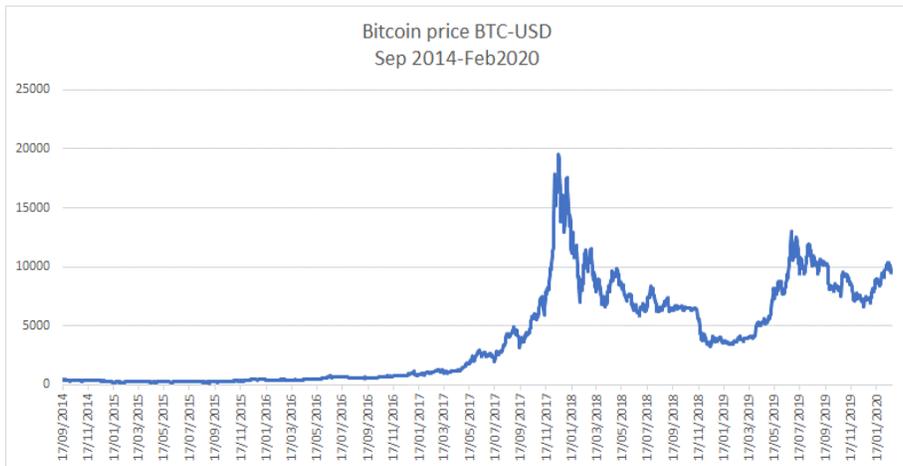

This kind of instruments were conceptualized and developed to be used as traditional currencies, as conventional currencies to acquire goods and services, with the difference that they avoid entering into commissions from financial intermediaries and under a novel technology called *blockchain*.

At present they can be related more to a commodity or financial asset than to a foreign currency, since many traders acquire cryptocurrencies seeking to generate a return (capital gain) derived from their price, they are more used for speculation than to be a means of payment in commercial transactions. There is a lot of financial literature on the intention of users when they exchange their domestic currencies for digital ones. There are empirical studies (Glaser et al, 2014) that, mainly users or investors with little information or academic preparation are not interested in an alternative transaction system, what they want is to participate in cryptocurrencies as an investment vehicle. The question of the usefulness of cryptocurrencies as an exchange practice (which has aroused the interest of regulators) is their immense volatility, which leads to think that it is used as a speculative investment. In 2012, the ECB said that Bitcoin should be considered a high risk system for its participants from a financial perspective. It was even hinted at its similarity to a Ponzi scheme (ECB, 2012). China, in 2013, announced a ban on Bitcoin as a currency for financial institutions (Ruwitch and Sweeney, 2013). If this was the case at the beginning of the boom in digital currencies (and other crypto assets), one can imagine the current protectionism of Central Banks before the implementation of payment systems such as *Google pay* and its *G currency*, or before Facebook's *Libra* (as it comes happening since 2018) - it will be understood why it is interesting to keep the suffix "crypto", as a connotation, when it is actually due to the encryption code-.

As for the so-called crypto assets, they are a set of crypto currencies together with other forms of goods and services which use cryptography (blockchain technology) for their operation. These crypto assets include cryptocurrencies and tokens. *Tokens* are a value unit of private entity for exchanges. William Mougayar defined the token as a unit of value, which helps an organization to govern the business model and give more power to its users to interact with its products, while facilitating the distribution of benefits among all its shareholders (Mougavar, 2016). Tokens have different uses and utilities in the blockchain (internal unit of account, intermediating transactions between buyers and sellers in the internal markets of the platform, or granting rights to token holders) but, regardless of their use, the tokens have been revealed as an effective method for technology start-ups to raise capital at the earliest stage of their business cycle. Instead of making capital increases or trying to convince Venture Capital funds (venture capital), blockchain companies are frequently financed through ICOs: *initial coin offering*. Tokens are offered at auction and used to fund the projects. In 2016, 250 million dollars were raised with this financing methodology for SMEs (Conley, 2017). The development of tokens, both as currencies and financial assets, respond to a compliance framework; however, there are issues such as capital increases through ICOs that are still not legally well defined (at least in Spain). If the tokens are currencies, ICOs must comply with know-your-customer legislation and anti-money laundering rules. If it is a financial asset, they will be subject to the legislation of the regulators (e.g. SEC, CNMV).

Whether *crypto-tokens* or crypto assets are currencies, financial assets, or a different and new asset, it also affects how they should be analyzed from an economic point of view. Although there is a theoretical body of knowledge with strong academic foundations in monetary and financial economics, how the theory should be applied to crypto assets and ICOs is still beginning to be explored. There is hardly any quantitative research on the estimation of cryptocurrencies/assets for potential investors, how start-ups should structure ICOs, etc. As has already been mentioned, the technology under which cryptocurrencies work is called blockchain, or chain of blocks. This blockchain removes all intermediaries through complete decentralization.

So that it can be understood in a better way, a chain of blocks is like a record book which is the blocks themselves that are connected and encrypted, like a distributed and secure database. In order for the blockchain to function properly, the information must be verified by several users, and in each block there are a large number of transactions. As more and more transactions are completed, the block reaches the point where it no longer supports more and there it must be validated and sealed, this is what users do when mining. What is mining?

Mining describes how blocks are generated within the blockchain. The chain contains blocks with information and transactions. In order for transactions to flow, we need confirmation from the miners. The so-called miners compete among themselves in order to have the right to create a new block in the chain (Zheng et al, 2017). It is a P2P (peer to peer) network, miners compete with each other. The first to create a valid block and seal it receives cryptocurrencies.

A very important aspect that we want to highlight is the high level of security of the system. The blockchain is unbreakable against a possible modification of the account book and the theft of Bitcoins turns unfeasible (Berentsen, 2018). The miners collect pending transactions (of Bitcoins for example), verify the legitimacy and chain it into what they call a "candidate for a block" in order to win new issues of the cryptocurrency if they convince the rest of the participants of the network or chain to add their candidate to the blockchain block. Access is usually free (in Bitcoin), you do not need authorization to become a miner, just download the software and the most recent copy of the chain. In practice, however, there are few and huge miners that produce most of the new blocks accepted in the chain due to the high competitiveness that allows profitability thanks to the economies of scale at the hardware and electricity level.

Regulation is extremely important, digital currencies can significantly reduce tax revenues for nations and is a serious danger to the banking sector (as has been recognized by the Bank of America), mainly in the current scenario of low interest rates and growth in developed countries. For instance, John Cryan, former Chairman of Deutsche Bank, warned about the possibility of traditional banknotes and coins disappearing in ten years due to their inefficiency. Cryptos and Blockchain increase the uncertainty in the financial autonomy of national economies, as recently recognized by Margarita Delgado, Deputy Governor of the Bank of Spain, speaking of Libra as a serious danger for monetary policy (see below). Proof of this are, for example, the recent reactions of the ECB, the IMF and the Fed regarding Facebook's Libra cryptocurrency, which shows concern regarding the financial system due to digital assets created by technology multinationals with strong penetration power in the monetary use of the population. The concerns relate to the protection of customer data, protection against money laundering or potential abuse of a leading position. Basically, there is a *trade-off* between cost reduction, speed of financial transactions and the menace of international financial volatility due to credit risk due to lack of support from a public institution.

The European Banking Authority defined 70 risks, divided into several categories based on who or what is threatened by them (Lansky, 2018). Today's payment systems are one of the groups. Obviously, the traditional banking system is fading, its business model is outdated due to new technologies and cryptocurrencies mean a radical change in transactions and business models. It is certain that digital assets have risks, but we cannot forget that currently around 2 billion people do not have access to the banking system. New technologies allow them to participate in the international economy and lift them out of poverty. Peer-to-peer currency networks are becoming increasingly common, making centralized control of funding difficult and posing a serious threat to the financial industry.

If they want to survive, large banks must digitize and provide real-time services such as those offered by cryptocurrencies and are already investing in research and development of Blockchain technology. Institutions that adapt will survive, the rest will die.

## USE OF DIGITAL CURRENCIES TO IMPROVE EMPLOYEES' COMPENSATION

The advantages of digital currencies as a medium of exchange in the financial and monetary system are, first of all, *in relation to the cost of transactions.* The technology provides high cost efficiency in international transactions to digital currencies compared to traditional instruments. According to Enciso (2018), cryptocurrencies are a contribution to the economic development of countries because it becomes an alternative stock exchange with costs that reach a 50% reduction in relation to the traditional stock market.

As the CEO of the Andreessen-Horowitz Venture Capital Manager, Marc Andreessen, said about Bitcoin: it introduces value into the system, transfers the value, the receiver obtains the value, no need for authorization and in many cases, free of commissions. The last advantage is of high importance. Bitcoin is the first internet payment system where transactions can be made with little or no fees. Traditional transaction systems charge commissions of 2/3% and this is in developed countries. In many other countries, modern payment systems do not exist or fees are much higher (Andreessen, 2014).

Second, one of the main advantages refers to their *decentralized nature,* that is, they are not controlled or administered by any government or public administration. Decentralization lies in the open nature of the code of its protocol, which means that the programming code is freely available to access and redistribute it. Nature of the system is based on the so-called collaborative economy, because any collaborator (with hardware provision) can process transactions on the Blockchain and obtain remuneration for it (what we have previously called mining). The reason for this simile is because, as in a mineral mine, the *commodity* decreases as it is exploited, the bitcoin algorithm is designed so that in 2,140 all Bitcoins are taken for granted.

That open technology provides the third advantage for collaborators, its infallibility (Lakomski-Daguerre and Desmedt, 2015): since any attempt to manipulate transactions results in a computer block incompatible with the previous and the next. For this reason, cryptography supporters call these systems "trust-less", because it means the substitution of a computer code in the trust of the public collateral of the traditional currency. Dozen authors portray Blockchain as the "reliable protocol" and Blockchain makes the network more than the internet of information, the internet of money. The birth of this technology stems from the loss of confidence in businesses and other institutions after the 2008 financial crisis (as indicated by the Edelman Confidence Barometer).

The fourth advantage for economic agents submits to transactions privacy, their anonymous nature. The right to privacy and anonymity arouses enormous interest in economic and commercial transactions and the World in general. There are plenty of examples of the monitoring of public government entities to prevent criminal and terrorist activities and of marketing companies to profile different users. We consider useful to clarify the distinction between privacy and anonymity in the context of financial transactions (Gallardo et al, 2019). Anonymity refers to the lack of knowledge towards the actor or actors who take part in it. Privacy submits to whether the product and quantity of the transaction are unknown, but not its actors. In relation to cryptocurrencies, transactions are anonymous, identities are not registered, but each transaction is registered in an electronic book of public nature. The anonymous nature alters the regulatory capacity in the financial field and therefore is used for the payment of criminal transactions. The fifth application with which the payment system is improved is because with cryptocurrencies all transactions are carried out from person to person, there are no intermediaries. It is a "peer to peer" (P2P) system. In addition, with which we enter the sixth advantage, the faster transactions in relation to fiat currencies. This project for the technological advancement of means of payment would be in line with technological development, economic globalization and the necessary agility of transactions today, and would be a rational evolution of the monetary concept in our days.

Therefore, the rise of electronic commerce and the financial crisis gradually led to the introduction as a means of payment, the idea born in 1998 by Wei Dai in the "cypherpunks" email list, where he proposed the idea of a new type of money used by cryptography to control its creation and transactions (see table below).

## ARTICULATION OF THE CSR 3.0 CASE: ADVANTAGES OF THE SOCIAL-BUSINESS DIGITAL CURRENCIES (SBDC)

Consider the development of digital currencies, then how do they relate to the economy of happiness and talent, and how can they serve as a case for CSR 3.0? To respond, allow yourself a brief clarification on the future of CSR and its three stages, to then give an account of examples of CSR 3.0, and finally record the advantages and benefits of SBDC in this regard. The world consecration of CSR (beyond the business sphere, reaching all types of corporations, including NGOs or the public sector) took place with convergent initiatives of the United Nations (e.g. the *millennium agenda* of its General Secretariat, the *future of the work* of the ILO), harmonizing all this with the *global compact* (announced by K. Annan in his speech on January 31, 1999 in Davos, during the *World Economic Forum meeting,* and formally constituted on July 26, 2000). Since then, minimum global standards have been set in relationships between people, communities and the environment. In addition, a network of local support networks has been established to deepen, broaden and disseminate this commitment. This has made it possible to generate a collaborative intelligence that has given rise to new concurrent and reinforcing initiatives (e.g. the wellbeing economics alliance by *World Economic Forum,* the surveys and good practices of *Great Place To Work*). In accordance with this collection, it is possible to establish the following evolutionary categories of CSR (in the transition towards the happiness and talent economy model):

a) CSR 1.0: characteristic of incipient organizations only oriented to results and in which the hygienic measures of the workers are hardly taken care of (e.g. working risks prevention, adequate wages and payment of overtime). As such, CSR is understood in a marketing way (out-door advertising) so it is outsourced to consultants or is directly replicated by others, but does not correspond to its own business culture. It is detected by his pretentious speech, his abuse of barbarisms (linguistic loans), and commitments that are difficult to verify (e.g. reducing the carbon footprint, helping a remote town).

b) CSR 2.0: visible in consolidated organizations, in terms of their market share, but who wish to make improvement changes, going beyond hygienic measures and initiate the promotion of motivational measures (those that stimulate workers to improve and increase their productivity and your commitment). Their CSR accounts for *compliance* local (eg. equality plans, ethical codes, recycling programs), is supported by international quality certifications (such as those of ISO standards), and they begin to participate in global transformation forums (e.g. *Global Compact*-UN.). In this way, one begins to become aware of the importance of corporate culture, so that it can be lived and participated in a sustained way, with verifiable impacts and shared with others.

c) CSR 3.0: mature organizations are produced, not by seniority, but by focus, since they are companies prepared for the new corporate culture, oriented towards people and their motivation. Its CSR is local and easily measurable and verifiable, as it is based on measures that affect its social and natural environment. Thus, CSR ceases to be something from outside doors (as a mere attempt to improve the business brand, or diligent and transparent regulatory compliance), becoming something from inside (thought by and for employees, together with their families: a culture to feel part of and celebrate).

Temporarily speaking, CSR 1.0 dominated until the 2000s (although it survives in those incipient organizations - regardless of their seniority, as it is a matter of aptitude and attitude towards the economy of happiness and talent); Since the 2000s, thanks to international organizations and transnational forums, CSR 2.0 has been promoted. For its part, CSR 3.0 is the result of the creative destruction of the 2008 crisis of values, as the companies that survived and improved were due to their orientation towards talented employees and their involvement in the new corporate culture, based on a mission, vision and values with which to identify and give the best of each one.

As required by CSR 3.0 in its succinctness, just consider the following example, which already links to SBDC. In the Basque Country (in Spain), on March 5, 2019, Fagor Industrial (household appliances manufacturer) signed an agreement with Orbea (bicycle company), by which a bonus of 200 euros was offered to workers for the acquisition of sports equipment and went to work without a car. In this way, Fagor achieved the following positive results from CSR 3.0: a) it looked after the well-being and health of its employees, when they came to work by bicycle; b) it cared for the environment, by reducing emissions with the reduction of cars at work; c) it improved the natural and social environment, since as new parking spaces were not required (even some of the existing ones could be dispensed with), a larger green recreational area was available; d) increased rest space for employees and the venue for business meetings, as well as with the families of employees, etc. And all this at no cost, only increasing profits: there was no need to spend on an extension of the car park, or on future places for holding meetings; the share of health insurance was reduced; It was not even necessary to pay the 200 euros of bonus, as it was part of the discount agreement with Orbea, which thus manages to increase its sales and release stock. So simply, Fagor had created money from RSC 3.0. These practices are very common in insurance companies, which count the steps taken each day through an app on their mobile, which translates into discounts on fees and gifts.

Thus, corrected and increased, more and more companies create their own social currency, being granted for good coexistence practices and production results, being valid for the company cafeteria and surrounding businesses, or for the purchase of reduction of working hours, or any other consideration for flexibility of work (this is not only done by the leading companies of *the GAFA model - Google, Amazon, Facebook & Apple-* but also those that have gone through a process of conversion, Kodak type, even good part of the companies ranked by GPTW).

The practice of rewarding virtual tokens for environmentally friendly behavior is called "eco-friendly activities". It is recalled that CSR also affects public sector organizations, since more and more municipalities reward their fellow citizens with social currencies for their good practices: for example, Viladecans (a municipality near Barcelona), it was returned to the neighbors part of the energy savings achieved in local currency (Vilawatt) to be spent in local shops (Viladecans, 2020). This has also been done in other places, such as Brussels and its Eco Iris, and other cases that arise later (when dealing with the paradox of social currencies).

Therefore, the resource of digital social currencies is something on the rise (despite its prediction in 1976 by Hayek or in 1988 by *The Economist*), present in all types of organizations, which reports benefits not only to direct collaborators (being possible higher and better remuneration, as their purchasing power always increases, without the risk of higher tax pressure or inflation), but, as a matter of CSR 3.0, it also results in the environmental, social, etc. common good. (It is a positive externality, until its standardization in markets, which will be when the forecasts of Hayek and the editors of *The Economist*).

**DISCUSSION AND CONCLUSIONS**

At the beginning of this paper, the exposition of a sample of business practice was assumed that could serve as a refutation of the proposals by academic *mainstrean* (of welfare state economics-WSE) and the *new-luddite* activists (contrary to technological advances), who consider that there is an inversely proportional relationship between technology and job well-being. They oppose technological advances because they believe that they violate working conditions, leaving people without work and increasing social inequalities. However, it turns out that the relationship between technological advances (such as digital currencies) and work well-being (by increasing remuneration, but from motivation, by undertaking from gamification to achieve it and the commitment to help with CSR) is not proportionally inverse (dismantling the fallacy that the more machinery and programming, the less work available to people), but is exponentially convergent (the more technological advances are produced, the more suitable work is for human beings, since they can dedicate themselves to exploit your personal talent).

The real paradox (if not directly a discursive contradiction in the form of ideological cognitive dissonance), occurs with the double standard: local social currencies are beneficial if promoted by the public sector (as they are a small complement to national money), however, they become suspicious (of lack of transparency - with accusations of money laundering, including pyramid scam, see below BCE) if they appear in electronic format and, even more, if they are the result of private initiative., the collaborative intelligence shared in international forums such as the *Global Compact*-UU.NN. or *Wellbeing Economics*

*Alliance*-WEF, prove the opposite: unlike the state welfare economy, which is based on the redistribution of scarcity, on the other hand, the digital economy is based on the constant and diverse generation of abundance, thanks to creativity, talent and entrepreneurship (a question developed in other publications).

It turns out that regarding the use of alternative local currencies or social currencies (Cortés, 2008. Corrons, 2017), if it is carried out by local entities of the public sector (such as the case of *Bristol Pound* in Bristol, *SoNantes* in Nantes or the most recent , of 2018, *Citizen Economic Resource*-REC in Barcelona), is considered an example of a social and solidarity economy –even, *money with values* (Corrons, 2017) -, despite the fact that companies are conditioned to participate for their operation; However, the appreciation varies (becoming speculative) if it is an initiative of the companies themselves (in Spain, since the Rumasa case in 1983, companies were prevented from having their own banks, something that facilitated their own financing) . Faced with such prejudice, one of the most successful antecedents to date should be mentioned, as is the case of Wir (short for *Wirtschaftsring*, which means economic circle), the currency of the Wir Cooperative Bank in Switzerland (since 1934, under the postulates of the economist Gesell on free money). This system has helped finance almost 100,000 Swiss SMEs (reaching an accumulated value of operations close to 1.5 trillion euros), proving very useful especially in periods of crisis (when there has been a lack of liquidity, such as the crisis in 2008). Going back to the so-called social currencies, already in the 2000s, cases such as the French *Sol-Violette* or the German *Chiemgauer* (each one existing in multiple municipalities, with more than half a thousand of participating companies and with operations worth several million euros per year). In Spain, after the financial crisis of 2008, there have been cases of social currencies now only electronic (via mobile app), such as *Real* de Vila Real in the Valencian Community; However, reluctance increases due to transferring social currencies to electronic support and their employability in the digital economy (as happened with the Mexican Túmin, created by professors Castro and López from the Intercultural Veracruzana University, who were charged with violation of the peso and the impulse of illegal currency).

In short, as leading companies in digital transformation and in the implementation of the talent and happiness economy model (such as those ranked by GPTW) have been proving, the resource of digital socio-business currencies has the benefit of CSR 3.0 practices (helping companies, collaborators, communities and the environment), but it has many more possibilities, which will soon be discovered after the great lockdown and work stoppage by Covid19 and its associated economic depression (as already seen after the 2008 stock crisis, with Bitcoin and the subsequent boom in digital currencies, given the lack of liquidity and financing, the loss of purchasing power and incentives, etc.).

Finally, keep in mind that in the last half century alone, there have been almost 150 *bank crashes*, more than 200 monetary, and 75 sovereign debt crisis. This means that a world average of a failure of the traditional monetary system (of national currency) is fulfilled every month and a half (shortening the terms in this period of depression just started). If to this is added the aggravation of the debt crisis at the beginning of the 2019 recession and the post-Covid depression, it is obvious that the use of alternative instruments that favor the financing of companies is indispensable (introducing new fluidity), the remuneration of collaborators (including, if necessary, their hiring via alternative billing), etc. At bottom line, digital socio-business currencies balance seems quite positive, taking into account that it is something incipient and whose possibilities will start to emerge in the cycle of economic depression that is opening after the Covid19 crisis (as will already happened with blockchain and Bitcoin after the 2008 stock crisis).

# REFERENCES

Amagasa, T., et al. (2005). Karojisatsu in japan. *Journal of Occupational Health*, 47(2), 157–157.

Andreessen, M. (2014). "Why Bitcoin Matters", New York Times (URL: https://dealbook.nytimes.com/2014/01/21/why-bitcoin-matters/; consulted: May, 2020).

Andreu, A., Sánchez-Bayón, A. (2019). *Claves de Administración y Dirección de Empresas en la Posglobalización,* Madrid: Delta Publicaciones.

Bailey, B.J. (1998). *The luddite rebellion.* New York: NYU Press.

Berentsen, A. (2018). *A Short Introduction to the World of Cryptocurrencies*. St. Louis: Federal Reserve Bank of St. Louis.

Conley, J.P. (2017). *Blockchain and the Economics of Crypto-tokens and Initial Coin Offerings,* Nashville: Vanderbilt University Press.

Cortés, F. (2008): *Las monedas sociales*, Almería: Cajamar.

Corrons, A. (2017): "Monedas complementarias: dinero con valores", *Revista Internacional de Organizaciones*, 18, 109–34.

Cubeiro, J.L. (2012). *Del capitalismo al talentismo,* Bilbao: Univ. Deusto.

Easterlin, R.A. (1974). Does Economic Growth Improve the Human Lot? en David, P., Reder, M. (eds.). *Nations and Households in Economic Growth*, New York: Academic Press Inc.

Easterlin, R.A., et al (2010) The happiness-income paradox revisited. *Proceeding of the National Academy of Sciences*, 107(52), 22463-68.

Edelman (2020). *Introduction*. Edelman (URL: https://www.edelman.com/, consulted: May, 2020).

Enciso, E. (2018). *Las criptomonedas, el futuro de la economía mundial,* UTADEO (URL:https://www.utadeo.edu.co/es/noticia/destacadas/expeditio/264566/las -criptomonedas-el-futuro-de-la-economia-mundial; consulted: May, 2020).

Frank, R. (2014). Karoshi: death from overwork. *Sherwood Park News*, 24, 24.

Frey, B. (2018). *Economics of Happiness*, Basel: University of Basel.

Gallardo, I., et al. (2019). *Análisis del anonimato aplicado a las criptomonedas*. Río Cuarto: XXV Congreso Argentino de Ciencias de la Computación.

Glaser, F., et al. (2014). *Bitcoin. Asset or currency? Revealing users´ hidden intentions*. Tel Aviv:

ECIS.

González, E., Sánchez-Bayón, A. (2019). *Nuevas tendencias de RR.HH. y desarrollo de talento profesional.* Porto: Ed. Sindéresis.

Kurzweil, R. (2005). *The singularity is near. New York: Penguin Group.*

Lakomski-Laguerre, O., Desmedt, L. (2015). L'alternative monétaire Bitcoin: une perspective institutionnaliste. *Revue de la régulation. Capitalisme, institutions, pouvoirs,* 18, 1-22.

Lansky, J. (2018). Possible state approaches to cryptocurrencies. *Journal of Systems integration.* 9(1), 19-31.

Leitaer, B. (2005): *El futuro del dinero* (trad.). Buenos Aires: Longsheller.

Leitaer, B. (2010) "El futuro ya está aquí", en VV.AA.: *El futuro del dinero,* Madrid: Fundación de la Innovación Bankinter,

Mougayar, W., (2016). *The Business Blockchain.* Hoboken: John Wiley & Sons Inc.

Rojas, M. (2014). *El estudio científico de la felicidad*, México DF: FCE.

Ruwitch, J., Sweeney, P. (2013). *China prohíbe a los bancos realizar transacciones con bitcoin.* London: Reuters.

Sale, K. (1996). *Rebels against the future: the luddites and their war on the industrial revolution: lessons for the computer age.* New York: Basic Books.

Sánchez-Bayón, A. (2020). "Renovación del pensamiento económico-empresarial tras la globalización: Talentism & Happiness Economics", *Bajo Palabra*, 24, 293-318.

Sánchez-Bayón, A. (2019a). Una historia crítica de sociología del trabajo y de las organizaciones: de "trabajadores de cuello azul y blanco" a "Knowmads & freeriders". *Miscelánea Comillas,* 77(151), 431-51.

Sánchez-Bayón, A. (2019b). Transición a la Economía GIG. *Encuentros multidisciplinares,* 21(62), 1-19.

Sánchez-Bayón, A. (2016). *Problemas y retos para alcanzar la sociedad del conocimiento.* Madrid: Delta Publicaciones.

Sánchez-Bayón, A., et al. (2020). "The Spanish B-Schools trouble in digital economy", *Journal of Entrepreneurship Education,* 23(5), 1-8.

Valero, J., Sánchez-Bayón, A. (2018). *Balance de la globalización y teoría social de la posglobalización.* Madrid: Dykinson.

Viladecans (2020). *¿Qué es vilawatt?* Viladecans (URL:

https://www.viladecans.cat/es/vilawatt, consulted: May, 2020).

VV.AA.a (2020). *Defining a New Economic Paradigm: The Report of the High-Level Meeting on Wellbeing and Happiness*. UU.NN. (URL: https://sustainabledevelopment.un.org/index.php?page=view&type=400&nr=617&menu=35, consulted: May, 2020).

VV.AA.b (2020). World Happiness Report 2020. UU.NN. (URL: https://s3.amazonaws.com/happiness-report/2020/WHR20.pdf; consulted: May, 2020).

Zheng, Z., et al. (2017). An Overview of Blockchain Technology: Architecture, Consensus, and Future Trends, en VV.AA.: *2017 IEEE International Congress on Big Data*, Boston: IEEE.

# Intention to Use M-Government Services: Age, Gender and Education Really Matter?

Annie Ng Cheng San[1], Choy Johnn Yee[2], Krishna Moorthy[3], Alex Foo Tun Lee[4]

## Abstract

With the highest level of mobile penetration rate, the globe is now moving rapidly into mobile government (M-government). Despite its benefits, the acceptance of m-government services in Malaysia is still not widespread. This study attempts to incorporate the Unified Theory of Acceptance and Use of Technology (UTAUT) model with perceived risk theory (security risk and privacy risk) to explore its impact towards the intention to use m-government services. Age, gender and education level were also adopted as moderator variables to provide in-depth understanding of citizens' preference in m-government services. Partial Least Square (PLS) Structural Equation Modelling method was conducted. The results indicate that the facilitating condition, performance expectancy, social influence, and security risk can be used as the predictor of m-government services adoption. These findings confirm the application of theory in the m-government context, which provide valuable insight to the government, citizens as well as future researchers to implement a successful m-government for a better communication between government and citizens.

**Keywords:** Mobile Government, UTAUT, Security risk, Privacy risk

**Jel Codes:** 032, 033

## INTRODUCTION

With the advancement of mobile technology, it changes the form of public services globally. Today, mobile technology is no longer used for the purpose of communication and entertainment only, but it is also used to improve the competence, intelligibility, and accountability of government services. With its unique characteristics of mobility and wireless, mobile technology plays essential and growing roles in the government position to deliver reliable information and services anytime, anywhere and on any devices to meet the needs of people (Thunibat, Mat Zin, & Sahari, 2010). Mobile government (hereinafter called M-government) is the integral components of electronic government. It focuses on the use of mobile platforms in government operation and services (Al-Hujran, 2012). Effective M-government practices aid the government to advance the sustainable development and communication with citizens by providing better access to public services in health, education, labor as well as environment (Ohme, 2014; Waller & Genius, 2015). The influence of m-government has been witnessed in many developed and developing countries including Malaysia.

---

[1] Lecturer, Department of Commerce and Accountancy, Universiti Tunku Abdul Rahman, Perak Campus, Malaysia. Email: ncsan@utar.edu.my Orcid: 0000-0003-3553-550X
[2] Lecturer, Department of Marketing, Universiti Tunku Abdul Rahman, Perak Campus, Malaysia. Email: choyjy@utar.edu.my Orcid: 0000-0003-4527-0111
[3] Assistant Professor, School of Economics and Management, Xiamen University Malaysia, 43900 Sepang, Selangor, Malaysia.Email: krishna.manicka@xmu.edu.my Orcid: 0000-0003-0431-0957
[4] Lecturer, Department of Commerce and Accountancy, Universiti Tunku Abdul Rahman, Perak Campus, Malaysia. Email: godspeed.alex@gmail.com, Orcid:0000-0001-8248-5336

## M-government services in Malaysia

With the increasing number of mobile users and high mobile penetration rate of 150%, it is a good opportunity to promote and implement M-government in Malaysia (Malaysian Communications and Multimedia Commission, 2015).

**Table 1.** Types of m-government services provided in Malaysia

| M-services | Number | Functions |
|---|---|---|
| mySMS | 15888 | Received short message service (SMS) on: <ul><li>License application status</li><li>Alert on renewing road tax, driving license expiry and income tax return due date</li><li>Information about government housing loan balance, road safety tips, traffic summons and train schedules</li></ul> Send SMS complaints to the government agencies such as Social Security Organization (SOCSO) or (PERKESO) |
| myUSSD | *158# | Request respond via Unstructured Supplementary Service Data (USSD) such as provide a menu to check: <ul><li>Status or result of public exam monthly pension payment</li><li>Application status of myKAD</li><li>Operation hour for marriage counter registration</li><li>Credit loan status</li></ul> |
| myMMS | 15888 | It is an enhancement of SMS, which allow public to share the Multimedia Messages (MMS) comprising the text, images, and video clips. It provides broadcast functions such as: <ul><li>Sending alert message on missing child or</li><li>Complaints of traffic or vandalism.</li></ul> |
| myApp | | It provides mobile application download. The current applications offered are: <ul><li>myHealth - provide tips for health issues</li><li>myJakim – provide Solah information such as prayer time, mosque and Qibla locator</li><li>myKPDNKK – provide domestic trade information from Ministry of Domestic Trade, Co-operatives and Consumerism</li><li>mySPAD – provide public transport terminals, routes, schedules and fare such as rapid public bus routes, KTM commuter schedule, and Kuala Lumpur Monorail schedule.</li></ul> |
| myPay | | It provides payment channel through mobile for the government service such as: <ul><li>Traffic summons</li><li>Utility bills</li><li>Income tax payment</li></ul> |

Thus, a number of innovative initiatives have been launched by the government such as the No Wrong Door Policy in the Tenth Malaysia Plan and Mobile Community Transformation Centres to expand and strengthen the service quality of M-government to reach more people. To date, there are more than 77% of public services provided through the mobile technology such as mySMS, myMMS, myAPP, myUSSD and myPay. Table 1 shows the current mobile government service provided in Malaysia.

## Challenges of M-government services in Malaysia and Research Questions

Despite the benefits, convenience and a number of initiatives adopted for m-government services, the acceptance rate is still far from the enormous utilization in Malaysia (Abdullah, Mansor, & Hamzah, 2013). Out of the 27.3 millions of mobile users in 2015, yet only 335,768 logins were recorded (Performance Management & Delivery Unit, 2016). Besides, Malaysia has a dramatic drop from 40 to 52 in the ranking of world e-government development in the latest United Nations E-government survey (2014). As such, the Performance Management and Delivery Unit (2014) urged that the awareness activities remain essential to educate and alert the citizen about the significance of m-government services.

To realize the vision 2020 to transform Malaysia into a fully developed country, the key challenge for m-government is to ensure its quality and service delivery to transform a successful citizen centric of m-government (Abdulla, Mohd Noor, & Ibrahim, 2016). The success of government initiatives always heavily depends on the end users (Sharma, 2015). Hence, it is essential to understand the needs of citizens towards the m-government services to strengthen the service quality and delivery model which best suited the citizens' expectation.

The high acceptance of mobile devices for daily activities might not guarantee the acceptance of using this technology for government service (Venkatesh, Chan, & Thong, 2012). The past literatures on m-government services' acceptance mainly focused on Technology Acceptance Model (Belanche, Casalo, & Flavian, 2012; Wang, 2014; Liu, Lim Kostakos, Goncalves, Hosio, & Hu, 2014) and Theory of Planned Behavior (Hung, Chang, & Kuo, 2013). Yet, majority of all the above said models are not specifically designed for m-government and the past literature often faced difficulty in choosing the best model for government services. Moreover, Shafinah, Sahari, Sulaiman, Mohd Yusoff and Mohd Ikram (2013) claim that by solely grounding on theory or model, it is incomplete to understand the technology acceptance. To overcome it, Unified Theory of Acceptance and Use of Technology (UTAUT) unified eight popular models to examine the acceptance based on organizational context. It consists of four main determinants, which are Performance Expectancy (PE), Effort Expectancy (EE), Social Influence (SI), and Facilitating Condition (FC) to explain the technology adoption. It is a highly validated model, which tailors on the acceptance of technology.

Moreover, risks are major problems faced in the government services. 97% of Malaysians think that security and privacy risks are the major obstacles for them not to use the m-government services (Thunibat, Mat Zin, & Sahari, 2010; Shafinah et al., 2013). The authors highlighted that security risk and privacy risk are the most crucial determinants in government services. However, majority studies have neglected the contribution of perceived risk (Shafinah et al., 2013). Hence, this study serves to narrow the research gaps by adopting the UTAUT and perceived risk theory (Perceived Security Risk (SR) and Perceived Privacy Risk (PR)) to investigate the determinants of m-government services. It will also investigate the moderator impact of age, gender and education level on each of the causal relationship of all constructs. Accordingly, the research questions are:

RQ1: Can the determinants in UTAUT model be confirmed in M-government services context?

RQ2: Do the SR and PR have an impact on the ITU of M-government services?

RQ3: Do age, gender and education level moderate the causal relationship between these determinants?

## LITERATURE REVIEW AND HYPOTHESES DEVELOPMENT

### Unified Theory of Acceptance and Use of Technology (UTAUT)

In the past decade, several theoretical models have been developed to explain the user's acceptance of technology usage. Among the models proposed, UTAUT is one of the popular models with most encompassing theory which provides a high explanation power of the intention of mobile technology adoption. Yet, Ohme (2014) highlighted limited studies that adopted UTAUT model in m-government services.

### Intention to use of M-government Services

Intention is the individual's subjective probability to accept certain action. It is immediate antecedent of behaviour to indicate the individual's readiness to perform the said services (Lin, Tzeng, Chin, & Chang, 2010). In this study, the m-government services' intention is used as the proxy to measure the actual behaviour as it is the best indicator for actual user behaviour (Hew, Lee, Ooi, & Wei, 2015). The prior empirical studies related to the influence of the proposed constructs with the intention are presented in the following part.

### Performance Expectancy (PE)

It is defined as one's belief on the usage of technology will assist him or her in achieving goal in the job performance (Venkatest et al., 2003). When the user perceives that the technology is able to enhance their goal and performance, it will cause a favourable preference towards the acceptance intention (Dwivedi, Shareef, Simintiras, Lal, & Weerakkody, 2015). PE has been in the limelight in the past literatures. Among the past studies, it is the strong determinant of acceptance intention in organization context (Tung, Chang, Chou, 2008; Ng & Choy, 2013). Consistently, the following hypothesis was tested:

H1. PE significantly influences the intention to use M-government services.

### Effort Expectancy (EE)

EE is one's perception of the degree of effortlessness associated with the use of technology (Venkatest et al., 2003). The users will intend to use m-government service if they believe that the technology is easy to control, operate and understand. Leong, Hew, Tan and Ooi (2013) also purported the significant association between EE and PE. This is supported in the empirical finding in Hew, Lee, Ooi and Wei (2015), which agreed that when the users found that the technology are convenient and easy to accessible, he or she will have high possibility to adopt the said technology and have a perception that the technology is useful. Thus, the following hypotheses are posited:

H2. EE significantly influences the intention to use M-government services.

H3. EE significantly influences the PE of M-government services.

### Social Influence (SI)

SI is the degree of individual's perception on those who are important to them who think that he or she should accept the technology (Yi, Jackson, Park, and Probst (2006). The past studies concluded that it is an important indicator for acceptance intention of technology. One might learn and intend to adopt the technology when he or she observes the other carries the similar behaviour in the social group. According to Singh et al. (2010), individual will intend to adopt the mobile commerce service when they are influenced by friends, family or colleagues. The findings in Venkatest et al. (2003), Venkatest, Thong and Xu (2012), as well as Abdelghaffar and Magdy (2012) have proven that no matter in mandatory or voluntary setting, SI is the most significant determinant for acceptance intention of technology. Thus, the hypothesis is proposed as below:

H4. SI significantly influences the intention to use M-government services.

### Facilitating Condition (FC)

FC is the degree of technical support or training that supports the use of technology (Venkatest et al., 2003). According to Venkatest, Thong and Xu (2012), FC significantly affects the acceptance intention as well as the user behavior towards technology. For instance, an individual is willing to accept the M-government service if there is a favourable set of FC such as training, tutorial or vendor support on the m-government services. Besides, Marshall, Mills and Olsen (2008) also found that with sufficient supports, users will be able to perceive the said technology as useful and easy to use/practical. However, limited studies have focused on the association of FC with the PE and EE. In other word, the complex tri-dimensional between FC, PE, and EE remains ambiguous in the literature. Thus, this study proposes the following hypotheses:

H5. FC significantly influences the intention to use M-government services.

H6. FC significantly influences the PE.

H7. FC significantly influences the EE.

### Perceived Risk Theory

Perceived risk refers to the uncertainty that potentially affects the users' confidence towards the services in a long term (Im, Kim, & Han, 2008). The authors further commented that perceived risk and anxiety are different whereby "anxiety can be mitigated" but perceived risk will remain unchanged for a long time. Government services are dissimilar with the m-commerce activities as it often involves highly sensitive and personal information (Venkatest, Chan, & Thong, 2012). Yu (2005) argued that citizens are highly concerned about the security and privacy risks when they are enjoying the government services. Hence, security and privacy are among the major aspects that require extra surveillance in the m-government services in order to foster the usage. To narrow down the literature gap, the perceived security risk (SR) and privacy risk (PR) are incorporated to test their significant influence towards the intention to use m-government services in Malaysia.

**(a)** Perceived Security Risk (SR)

According to Ohme (2014), SR is the uncertainty of users towards the said technology due to the concern about the technical fraud by unauthorized third parties such as system hacking. When the citizens have a high doubt on the security of system, it is unlikely for the citizens to use the system. Venkatest, Chan, and Thong (2012) also agreed that when the users adopt the government services such as e-filling, they have a greater concern about the security since it involves high volume of personal and sensitive information. Hence, when the SR concern is high, the citizen will have lack of confidence towards the service which then contribute to the negative impact towards the usage of technology. Hence, the following hypothesis is put forward:

H8. SR significantly influences the intention to use M-government services.

**(b)** Perceived Privacy Risk (PR)

Majority of past studies often combine the PR and SR in their research. Yet, PR and SR are different. SR focused on the technical fraud done by unauthorized third parties; while PR emphasized on the uncertainty arises from the misuse of personal information by the authorized parties (Ozdemir, Trott, & Hoecht, 2008). Ohme (2014) further commented that PR is the uncertainty or doubt on the authority or government handling the data transmitted during the m-government services. When the PR is high, the loss of trust towards the system and the technology in general is highly likely to occur. Based on the concern on privacy, the hypothesis is formulated:

H9. PR significantly influences the intention to use M-government services.

**Moderating Effects: Age, Gender and Education level**

From the observation of past literature reviews which adopted the UTAUT model, it can be concluded that the past studies have overlooked the moderating impact towards the constructs influence (Venkatesh, Thong, & Xu, 2012). To narrow the literature gaps as well as to enrich the UTAUT model in m-government context, this study proposes individual differences such as age, gender and education level as the moderating variables. Differences of individual characteristics such as age, gender, and education background have contradicted opinions in the technology intention (Venkatesh, & Morris, 2000). For example, Yang (2005) discovered that male is more inclined towards technology as compared to female. Mature adults will be more likely to concern about the risk of adopting the technology as compared to the young users. Young users tend to be risk takers, who are willing to use the new technology (Lian & Yen, 2014). Hew, Lee, Ooi and Wei (2015) highlighted that the past studies have neglected the moderating effects of individual's education level. Individual with high education level will tend to encounter fewer barriers to adopt new technology (Chopra & Rajan, 2016). Hence, with the supportive evidences, the following hypothesis is formed:

H10. Age, gender and education level moderate all relationships among the proposed constructs and intention of M-government services.

**Proposed Conceptual Framework**

With the above discussion of prior empirical studies, there are a total of six determinants: PE, EE, SI, FC, SR and PR to gauge the influence towards the behavioural intention of m-government services. To simplify the study, use behaviour was removed in this study. Figure 1 below presents the proposed conceptual framework for this study:

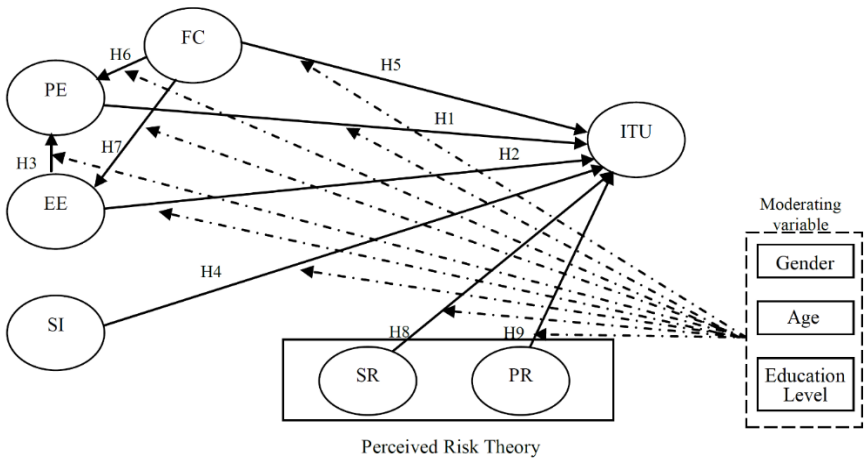

**Figure 1:** Proposed conceptual framework

Note: PE=Performance Expectancy; EE=Effort Expectancy; SI=Social Influence; FC=Facilitating Condition; SR=Perceived Security Risk; PR=Perceived Privacy Risk; ITU=Intention to use.

**Sampling technique and data collection method**

Questionnaire survey was adopted for the purpose of data collection in this study. With national mobile users as the targeted population, quota sampling technique was used by narrowing our sampling target to four states with the highest number of mobile users in Malaysia (MCMC, 2015). They are Selangor (20.9%), Johor (12.7%), Kuala Lumpur (8.9%), and Perak (8.5%), hence making up a sum of 51% out of total population. Pre-test of the questionnaire was conducted with six dominant researchers in the M-commerce field in order to ensure its feasibility and validity.

**Instrument Development**

Then, 300 self-administrated questionnaires were distributed, and 265 complete responses were collected which were usable for the data analysis. The medium of language is English. Self-administration was incorporated with the hope to reduce interviewer's bias in terms of influencing the response of the questionnaire. The survey questionnaire was made up of two sections, i.e., demographic profile as well as 26 items responsible in measuring mobile user's perception on the constructs of the study. The items in the questionnaires were adopted from the past literature, in which each construct is measured by three to four items in a 7-point Likert scale format (as provided in Table 3.1). The reliability of the instrument is attested to be within acceptable threshold, and is hence satisfactory to proceed to descriptive and inferential analysis.

**DATA ANALYSES**

**Demographic Profile of respondents**

Table 4.1 shows all the demographic profiles of the 265 target respondents. It is reported that majority of the respondents are female with the age group of 20-30 years old. Majority of them achieved Bachelor degree with the monthly income within the range of RM 2001 to RM 3,000. All of the target respondents owned at least one mobile device.

**Results of statistical analyses**

Smart-PLS 2.0 of partial least square (PLS) structural equation modelling (SEM) method was used to analyse the said model in Figure 1. According to Santhanamery and Ramayah (2015), PLS is a pertinent technique for analyzing predictive models with multiple-item constructs. Hair et al (2014) also commented that the variance-based SEM provides better efficiency in parameter estimated which is manifested in advance statistical power than the covariance based SEM. For such rationales, the PLS-SEM is the one that fits for this study. Two-step analytical procedures suggested by Santhanamery and Ramayah (2015) and Anderson and Gerbing (1988) are used to analyze the empirical findings. The reliability and validity measurement model were first evaluated, followed by the structural model assessment and hypotheses testing. According to Wu and Chen (2014), the desired sample size for PLS is ten times of the number of indicators associated or the highest number of antecedent constructs. Thus, the sample size of 265 is considered adequate for the study.

**Table: 3.1** Survey Instruments

| Latent constructs | Indicators | Coding | Source |
|---|---|---|---|
| Performance Expectancy (PE) | I would find the m-government services useful in my daily life. | PE1 | |
| | Using m-government services increases my chances of achieving things that are important to me. | PE2 | |
| | Using m-government services helps me accomplish things more quickly. | PE3 | Venkates, Thong, & Xu (2012) |
| | Using m-government services increases my productivity. | PE4 | |
| Effort Expectancy (EE) | Learning how to use m-government services is easy for me. | EE1 | |
| | My interaction with m-government services is clear and understandable. | EE2 | |
| | I find m-government services easy to use. | EE3 | |
| | It is easy for me to become skillful at using m-government services. | EE4 | |
| Social Influence (SI) | People who are important to me think that I should use m-government services. | SI1 | |
| | People who influence my behavior think that I should use m-government services. | SI2 | |
| | People whose opinions that I value prefer that I use m-government services. | SI3 | |
| Facilitating Condition (FC) | I have the resources necessary to use m-government services. | FC1 | |
| | I have the knowledge necessary to use m-government services. | FC2 | |
| | M-government services are compatible with other technologies I use. | FC3 | |
| | I can get help from others when I have difficulties using m-government services. | FC4 | |
| Security Risk (SR) | I fear that while I am using m-government service, I might make mistakes since the correctness of the inputted information is difficult to check from the screen. | SR1 | |
| | I fear that while I am using m-government services, the battery of the mobile phone will run out or the connection will otherwise be lost. | SR2 | Rakhi & Mala (2014) |
| | I fear that while I am using m-government services, I might tap out the related information wrongly. | SR3 | |
| | I fear that my password may be lost and end up in the wrong hands. | SR4 | |
| Privacy Risk (PR) | I think m-government service providers could provide my personal information to other companies without my consent. | PR1 | |
| | I think subscribing to m-government services increases the likelihood of receiving spam/ spam SMS. | PR2 | |
| | I think m-government service providers endanger my privacy by using my personal information without my permission. | PR3 | |
| | I think m-government service providers will send SMS advertisement without user's consent. | PR4 | |
| Intention to Use (ITU) | I intend to use m-government services in the future. | ITU1 | Venkatesh et al. (2012) |
| | I will always try to use m-government services in my daily life. | ITU2 | |
| | I plan to use m-government services frequently. | ITU3 | |

**Table: 4.1** Demographic profiles of respondents

|  |  | Total | Percentage |
|---|---|---|---|
| **Gender** | Female | 149 | 56.2 |
|  | Male | 116 | 43.8 |
| **Age** | Below 20 | 0 | 0 |
|  | 20 to 30 years old | 202 | 76.2 |
|  | 31 to 40 years old | 48 | 18.1 |
|  | Above 40 years | 15 | 5.7 |
| **Highest education completed** | Secondary School | 68 | 25.7 |
|  | Diploma | 29 | 10.9 |
|  | Degree | 128 | 48.3 |
|  | Others | 40 | 15.1 |
| **Income level** | Below RM1000 | 32 | 12.1 |
|  | Between RM1000 and RM2000 | 49 | 18.5 |
|  | Between RM2001 and RM3000 | 100 | 37.7 |
|  | Above RM3000 | 84 | 31.7 |
| **Own mobile phone** | Yes | 265 | 100 |

### *The Measurement Model Evaluation*

The measurement model is presented in Figure 4.1 as below:

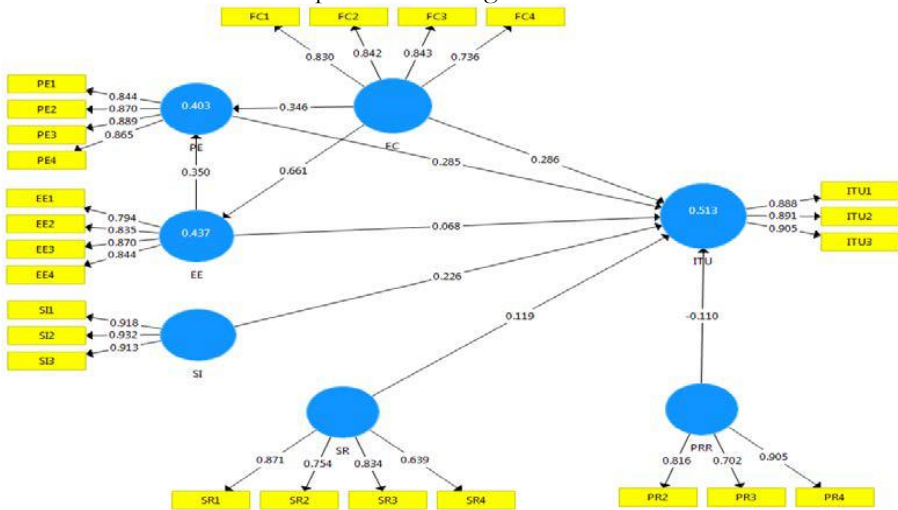

**Figure: 4.1** Measurement model

The assessment of measurement model focuses on both reliability and validity of the study. Firstly, to examine the reliability for all constructs, Cronbach's alpha and composite reliability (CR) are assessed. Hair et al. (2014) advised that the CR and Cronbach's alpha's value should be greater than 0.7 in order to demonstrate a high internal consistency of scales used in each constructs. Table 4.2 showed that both the CR and Cronbach's alpha value for all constructs are looking good.

Next, validity for constructs was evaluated by the convergent validity and discriminant validity. Convergent validity denotes as the degree to which two or more attempts share a high proportion of variance in common and it can be confirmed by the factor loadings and average variance extracted (AVE). According to the rules of thumb suggested by Hair et al. (2014), the factor loading should exceed 0.70 and AVE should be greater than 0.50. Table 4.2 presents the factor loading and AVE for each construct which has been attained. For the PR1 under the perceived privacy risk, as the factor loadings is smaller than 0.7, hence the item PR 1 will be dropped.

On the other hand, discriminant validity serves as the extent to which a construct is truly discrete from others constructs. As suggested by Fornel and Lacker (1981), discriminant validity test is conducted to determine the correlation between the constructs and square root of AVE for that construct. Referring to Table 4.3, the square root of AVE as presented in the bolded value for each construct is higher than the correlations between constructs. Hence, discriminant validity is achieved. As recorded in Table 4.2 and 4.3, we can conclude that all constructs exhibited acceptable reliability and validity.

**Table: 4.2** Indicator factor loadings, Cronbach's alpha, composite reliability, and AVE of constructs

| Constructs | | Factor Loadings | Cronbach's Alpha | Composite Reliability (CR) [a] | Average Variance Extracted (AVE) [b]: Covergence validity |
|---|---|---|---|---|---|
| **PE** | PE1 | 0.841 | 0.890 | 0.924 | 0.752 |
| | PE2 | 0.873 | | | |
| | PE3 | 0.887 | | | |
| | PE4 | 0.866 | | | |
| **EE** | EE1 | 0.791 | 0.857 | 0.903 | 0.699 |
| | EE2 | 0.835 | | | |
| | EE3 | 0.870 | | | |
| | EE4 | 0.847 | | | |
| **FC** | FC1 | 0.830 | 0.829 | 0.887 | 0.663 |
| | FC2 | 0.842 | | | |
| | FC3 | 0.843 | | | |
| | FC4 | 0.736 | | | |
| **SI** | SI1 | 0.918 | 0.911 | 0.944 | 0.848 |
| | SI2 | 0.932 | | | |
| | SI3 | 0.913 | | | |
| **SR** | SR1 | 0.871 | 0.805 | 0.859 | 0.607 |
| | SR2 | 0.754 | | | |
| | SR3 | 0.834 | | | |
| | SR4 | 0.639 | | | |
| **PR** | PR1 | 0.541 | 0.749 | 0.831 | 0.559 |
| | PR2 | 0.818 | | | |
| | PR3 | 0.720 | | | |
| | PR4 | 0.869 | | | |
| **ITU** | ITU1 | 0.888 | 0.876 | 0.923 | 0.801 |
| | ITU2 | 0.891 | | | |
| | ITU3 | 0.905 | | | |

The structural model shows the relationship between the findings of the hypotheses testing proposed in this research model. It was assessing by running bootstrap procedure using five thousand samples in SmartPLS and the findings for structural model are presented in Table 4.5 and Table 4.6. All the hypotheses proposed are supported except for H2 and H9. Firstly, the antecedents to intention to use M-government services in Malaysia were evaluated. The path coefficient for direct effects model indicated that PE ($\beta$ =0.285, p<0.001), SI ($\beta$ =0.226, p<0.001), FC ($\beta$=0.286, p<0.001) and SR ($\beta$=0.119, p<0.05) has a direct positive influence on the intention to use the m-government services. Whereby, the influence of EE and PR are not statistically significant to the intention to use, thus not supporting H2 and H9. The R square (R²) revealed that 51.30 percent of the variation in intention to use M-government services can be explained by explanatory constructs. Besides, the results show that the FC has a direct and positive influence towards the PE ($\beta$=0.346, p<0.001, R² =0.403) and EE ($\beta$=0.661, p<0.001, R²=0.437). All these R² are higher than the rule of thumb 0.35 suggested by Cohen (1988) and the FC is found as the strongest predictor of intention to M-government services.

**Table: 4.3** Discriminant Validity Test

| Discriminant Validity | PE | EE | FC | SI | SR | PR | ITU |
|---|---|---|---|---|---|---|---|
| PE | 0.867 | | | | | | |
| EE | 0.578 | 0.836 | | | | | |
| FC | 0.578 | 0.661 | 0.814 | | | | |
| SI | 0.471 | 0.522 | 0.533 | 0.921 | | | |
| SR | 0.192 | 0.098 | 0.096 | 0.048 | 0.779 | | |
| PR | 0.238 | 0.182 | 0.235 | 0.090 | 0.333 | 0.748 | |
| ITU | 0.594 | 0.532 | 0.603 | 0.545 | 0.184 | 0.091 | 0.895 |

Note: PE=Performance Expectancy; EE=Effort Expectancy; SI=Social Influence; FC=Facilitating Condition; SR=Perceived Security Risk; PR=Perceived Privacy Risk; ITU=Intention to use. The diagonals (bolded) represent the square root of the AVE while the off-diagonals are correlations among constructs. Diagonal elements should be larger than off-diagonal elements in order to establish discriminant validity.

### The Structural Model Evaluation

The structural model is shown in Figure 4.2 below:

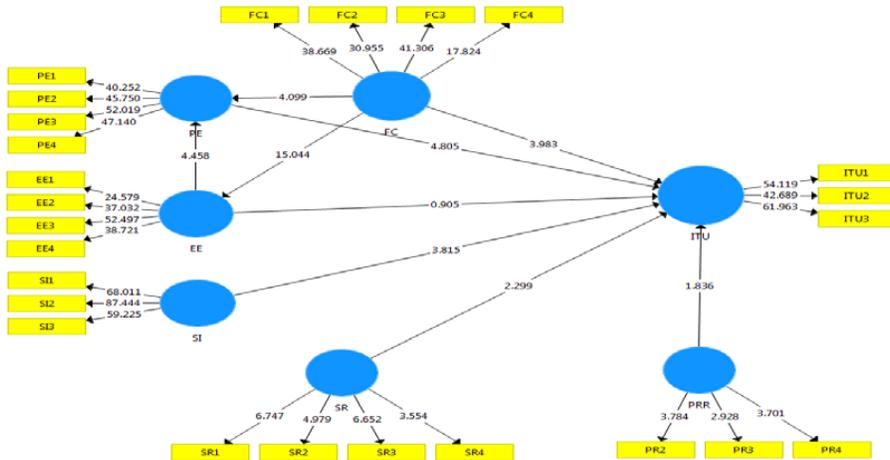

**Figure: 4.2** Structural model

### Moderating Effects of gender, age and education level

Multigroup analysis via PLS-SEM was carried out to examine the moderating effects of gender, age and education level in all paths of the proposed research model (Figure 1). Firstly, the female is coded as 1 and male is coded as 2 for gender perspective. Age and education level was re-grouped into two main categories; younger and lower education level was coded as 1 and eldest and with higher education level was coded as 2 according to the median.

**Table: 4.4** Moderate effects: multiple group analysis

| | Gender | Male | Female | T-statistics | p-value | Significant |
|---|---|---|---|---|---|---|
| H1 | PE ➜ ITU | 0.438082 | 0.170333 | 4.387 | 0.000 | **Yes** |
| H2 | EE ➜ ITU | 0.04326 | 0.078158 | 0.314 | 0.754 | No |
| H3 | EE ➜ PE | 0.222675 | 0.436938 | 2.443 | 0.015 | **Yes** |
| H4 | SI ➜ ITU | 0.162829 | 0.307243 | 2.169 | 0.031 | **Yes** |
| H5 | FC ➜ ITU | 0.187162 | 0.312181 | 1.488 | 0.138 | No |
| H6 | FC ➜ PE | 0.545462 | 0.192236 | 4.368 | 0.000 | **Yes** |
| H7 | FC ➜ EE | 0.677596 | 0.657721 | 0.342 | 0.732 | No |
| H8 | SR ➜ ITU | 0.136482 | 0.054425 | 1.087 | 0.278 | No |
| H9 | PR ➜ ITU | -0.124079 | -0.0309 | 1.117 | 0.265 | No |

| | Age | Younger | Older | T-statistics | p-value | Significant |
|---|---|---|---|---|---|---|
| H1 | PE ➜ ITU | 0.309938 | 0.334822 | 0.256 | 0.798 | No |
| H2 | EE ➜ ITU | 0.152808 | 0.175122 | 0.228 | 0.820 | No |
| H3 | EE ➜ PE | 0.336113 | 0.097844 | 3.398 | 0.001 | **Yes** |
| H4 | SI ➜ ITU | 0.21182 | 0.287115 | 0.891 | 0.374 | No |
| H5 | FC ➜ ITU | 0.490458 | 0.496003 | 0.090 | 0.928 | No |
| H6 | FC ➜ PE | 0.525503 | 0.754231 | 2.682 | 0.008 | **Yes** |
| H7 | FC ➜ EE | 0.61812 | 0.79478 | 3.737 | 0.000 | **Yes** |
| H8 | SR ➜ ITU | 0.125106 | 0.091875 | 0.650 | 0.516 | No |
| H9 | PR ➜ ITU | -0.028922 | -0.277952 | 2.358 | 0.019 | **Yes** |

| | Education Level | Low | High | T-statistics | p-value | Significant |
|---|---|---|---|---|---|---|
| H1 | PE ➜ ITU | 0.315244 | 0.267973 | 0.676 | 0.500 | No |
| H2 | EE ➜ ITU | 0.178751 | 0.218448 | 0.508 | 0.612 | No |
| H3 | EE ➜ PE | 0.311505 | 0.395437 | 1.011 | 0.313 | No |
| H4 | SI ➜ ITU | 0.30468 | 0.20721 | 2.520 | 0.012 | **Yes** |
| H5 | FC ➜ ITU | 0.450446 | 0.513521 | 1.096 | 0.274 | No |
| H6 | FC ➜ PE | 0.422665 | 0.647364 | 3.419 | 0.001 | **Yes** |
| H7 | FC ➜ EE | 0.534193 | 0.712939 | 3.293 | 0.001 | **Yes** |
| H8 | SR ➜ ITU | 0.087081 | 0.139668 | 0.554 | 0.580 | No |
| H9 | PR ➜ ITU | -0.225269 | -0.052213 | 2.207 | 0.028 | **Yes** |

Notes: PE=Performance Expectancy; EE=Effort Expectancy; SI=Social Influence; FC=Facilitating Condition; SR=Perceived Security Risk; PR=Perceived Privacy Risk; ITU=Intention to use.

The method is consistent with the technique recommended in Hew, Lee, Ooi, and Wei (2015) and Rahman, Taghizadeh, Ramayah, and Ahmad (2015) studies. The result in Table 4.4 reported that the gender was found to have significant moderating effect on PE ➔ ITU; EE ➔PE, SI ➔ ITU, and FC ➔ PE. Whereby, age was found to be significantly moderate in the relationship between EE ➔PE, FC ➔ PE; FC ➔ EE and PR ➔ ITU. Lastly, education level also found to have significant moderating effect in the path between SI ➔ ITU, FC➔ PE; FC➔ EE and PR ➔ ITU. Hence, H10 is partially supported.

*Predictive Relevance and Effect Size*

Sullivan and Feinn (2012) urged that the statistical significance (p-value) does not disclose the statistical power of the research model (substantive significant, effect size). According to Cohen's (1988) rules of thumb, the magnitude of effect size ($f^2$) of 0.02, 0.15 and 0.35 indicate the small, medium and large effect size respectively. Based on the result in Table 4.5, it can be observed that all the relationships have a passable effect size. Further to that, Table 4.6 also reported the predictive relevance of the endogenous latent variable via blindfolding procedures. Cohen (2013) suggested that $Q^2$ should be greater than the value of 0 and 0.02, 0.15 and 0.35 denoting small, medium and large predictive relevance. Table 4.6 reported that the ITU ($Q2=0.40$) has large predictive relevance with $Q^2$ more than 0.35, whereby the PE ($Q^2=0.298$) and EE ($Q^2=0.299$) which have medium predictive relevance. Therefore, it is concluded that the research model proposed has a material predictive power in explaining the ITU to use M-government services.

**Table: 4.5** Results of structural model analysis

| Hypotheses | Structural path | Path coefficients | T-statistics | Decision | $f^2$ |
|---|---|---|---|---|---|
| H1 | PE ➔ ITU | 0.285 | 4.805*** | Supported | 0.09 |
| H2 | EE ➔ ITU | 0.068 | 0.905 | Not supported | 0.01 |
| H3 | EE ➔ PE | 0.350 | 4.458*** | Supported | 0.12 |
| H4 | SI ➔ ITU | 0.226 | 3.815*** | Supported | 0.07 |
| H5 | FC ➔ ITU | 0.286 | 3.984*** | Supported | 0.08 |
| H6 | FC ➔ PE | 0.346 | 4.099*** | Supported | 0.11 |
| H7 | FC ➔ EE | 0.661 | 15.044*** | Supported | 0.78 |
| H8 | SR ➔ ITU | 0.119 | 2.299** | Supported | 0.03 |
| H9 | PR ➔ ITU | -0.110 | 1.836 | Not supported | 0.02 |

Notes: ***p<0.001; **p<0.05; PE=Performance Expectancy; EE=Effort Expectancy; SI=Social Influence; FC=Facilitating Condition; SR=Perceived Security Risk; PR=Perceived Privacy Risk; ITU=Intention to use.

**Table 4.6** Predictive relevance and effect size of the endogenous latent

| | $R^2$ | $Q^2$ |
|---|---|---|
| ITU | 0.513 | 0.400 |
| PE | 0.403 | 0.298 |
| EE | 0.437 | 0.299 |

Notes: PE=Performance Expectancy; EE=Effort Expectancy, ITU=Intention to use.

## DISCUSSIONS

With the PLS-SEM analysis, the empirical results found that all constructs have positive and significant relationship with the ITU m-government services, except the constructs of EE and PR. Among the constructs proposed, FC is seen to be the most important determinant which has stronger influence on ITU. Fundamentally, all paths from the UTAUT have been confirmed in this study except the EE and PR.

### Performance Expectancy (PE)

Based on the empirical result, PE is the second strongest predictor for the ITU of m-government services. The finding is in agreement with the findings of Venkatesh, Morris, Davis and Davis (2003); Hew, Lee, Ooi and Wei (2015); and Chopra and Rajan (2016). Hence, only when the citizens viewed that the m-government services are productive and useful for their daily life, it leads to high intention to adopt the said services. This is particularly in view of the benefit brought about by the m-government services in term of fast and beneficial transaction.

### Effort Expectancy (EE)

The result shows a surprising and interesting fact, that EE has insignificant influence towards the ITU of m-government services. This opposes the findings by Venkatest et al. (2003); Venkatest, Chan, and Thong (2012); and Leong, Hew, Tan and Ooi (2013) which suggested the EE has a significant effects towards the ITU. The inconsistent finding might be due to the fact that the target respondents are techno-savvy who perceive mobile services as friendly to them (Yu, 2012). Consequently, EE would not affect the citizens' intention to adopt the m-government services. However, the significant relationship between EE and PE is confirmed. Leong, Hew, Tan and Ooi (2013) suggested that to boost the adoption and usage of technology, the high level of convenience should be highly regarded as it results in the positive perception on the usefulness, or else users might perceive it as difficult and a hassle instead. Consequently, the purpose of adopting the m-government services will be defeated.

### Social Influence (SI)

SI has been confirmed as significant determinant of ITU. It is comparable with to the majority of past studies such as Yi, Jackson, Park, & Probst (2006); and Abdelghaffar and Magdy (2012). It is believed that family, colleagues, peers and friends' recommendation or word of mouth have a powerful impact on the individual intention to use the m-government services.

### Facilitating Condition (FC)

FC is confirmed as the most significant determinant of ITU in this study. The empirical result also confirmed that there are positive association between FC to PE and EE. These results therefore support the importance of technical support in m-government services whereby it creates favourable preference among citizens towards the implementation of m-government. When there is sufficient technical support, the citizens will have trust towards the m-government services for they are useful and easy to use (Marshall, Mills, & Olsen, 2008; Venkatest, Chan, & Thong, 2012). Thus, this study shall contribute to the significant association between FC to PE as well as EE which are neglected by the majority of past studies.

## Perceived Security Risk (SR)

SR's empirical finding is in line with the past studies by Ohme (2014) and Nasir, Wu, Yago, and Li (2015). Yet, it was surprisingly found that the SR has positively influenced the ITU of M-government services as this direction of result contradicts the past literature reviewed. This finding provides relevant insight that the higher the SR concern, the higher one's intention to use the m-government services. Roger (1995) reasoned this finding as majority of the target respondents fall between 20 to 30 years old, a group that can be classified as technology savvy. This might explain why the citizens intent to adopt the m-government services, even after some reports or incident about the security violation has been reported.

## Perceived Privacy Risk (PR)

PR was not found to have any significant influence towards the ITU of m-government services. This finding is in line with Tan, Qin, Kim and Hsu (2012) and Ohme (2014), who agreed that with the improvement of m-government policy, the citizens might have sufficient confidence towards the government in handling the data. Therefore, as compared with the SR, PR is no longer the major concern for the citizens.

## Moderating effect of Age, Gender and Education Level

In this study, the moderating effects of age, gender and education level were examined in all constructs. Remarkably, the empirical result demonstrated that gender has moderating effects on relationship of PE → ITU; EE →PE, SI → ITU, and FC → PE, which is parallel with the study of Venkatesh et al. (2003). It is agreed that male tend to be more task-oriented. Thus, when they perceive the m-government services as useful (PE) and have sufficient technical support (FC), they will be more likely to adopt the services. For female, they are more sensitive to others' recommendation and focus more on the degree of effort involved (Venkatesh & Morris, 2000). Therefore, SI and EE are more prominent for them. Next, individual with from different ages are found to have significant moderating impact on the relationship between EE →PE, FC → PE; FC → EE and PR → ITU. Older citizens tend to face more difficulties in adopting new technology; they are less innovative and more concern about the risk. As a result, the older citizens tend to emphasize on the technical supports (FC) and PR in adopting the m-government services. For younger citizens, they tend to focus more of the degree of effort and usefulness, which results in a greater moderating influence in relationship between EE and PE. Lastly, difference of education level of the citizen yields moderating effect in causal relationship between SI → ITU, FC→ PE; FC→ EE and PR →ITU. It is concluded that citizens with higher education level will tend to depend more on the FC which leads them to a better control and knowledge towards the m-government services. They believe that the system is convenient and useful with sufficient resources and support. On the other hand, citizens with lower educational level are more conscious, will consider more on the details and privacy of the data and has a higher level of risk concern before making any technology adoption (Chopra & Rajan, 2016). As a result, SI and PR are more salient in the m-government services' adoption.

## IMPLICATIONS

### Theoretical Implications

Shafinah et al. (2013) suggested that prevalence of m-government services, often referred to as the users' behavioural intention and citizens' demands and needs, and are vary. However, limited studies showcased the complete constructs as well as the moderating effects proposed by the UTAUT model. This study incorporates the UTAUT and perceived risk theory to explore the conceptual connection between the proposed variables and the behavioural intention of the m-government services among the citizens. From the theoretical point of view, all the constructs proposed have been successfully justified in this study. All of the constructs such as perceived expectancy, social influence, facilitating condition and perceived security risk have been examined and found as the salient predictors for intention to use m-government services. Furthermore, this study serves to narrow the existing literature gaps by incorporating the moderating effects of age, gender and education level in the proposed model to yield better insight in explaining the intention to use m-government services. Successful incorporation of UTAUT model and perceived risk theory (PR and SR) is another accomplishment made by this study which explained 51.3 percent of the variances in intention to use m-government services in Malaysia. The new research model would serve as a baseline for future researchers in the m-government service study.

### Managerial Implications

With the vision 2020, Malaysian government is aiming for a better, more effective and higher quality government services to the citizens. With the unique characteristic of mobile devices and high level of mobile penetration, there are clear opportunities for the SMART (Social, Mobile, Analytics, Radical Openness and Trust) m-government in Malaysia. The managerial implications of this study are crucial and comprehensive to the m-government service providers. The government should be aware of the impact of performance expectancy, social influence, facilitating condition, and security risk upon the implementation of the m-government services. In order to improve the communication between government and citizen, Malaysian government must ensure that the service is in the highest level of usefulness, with sufficient technical support, and limited security flaws.

Out of all predictors, facilitating condition contributes the highest influence power towards the intention to use. Technical support is always important. Therefore, close attention should be given to the technical support and resource in m-government services in Malaysia. Government should provide demonstration, animated tutorial as well as real time assistance to facilitate the citizens in using the m-government services. Next, performance expectancy is the second strongest predictor leading to the intention to use the m-government services. The m-government services should be able to cater the daily lives of citizens and see how it could be assimilated with the citizens' routines. Survey should be conducted from time to time to understand the citizens' needs to improve the usage of m-government services. Inter-government agencies as well as private sectors at the national or global level should be invited and engaged to provide an extensive m-government services. Given with the unlimited access to media nowadays, it is very important to promote the usage of m-government services via mass media and social network channels.

Next, citizens are worried about the increasing use of m-government services which will result in the increasing vulnerability of sensitive information as well as the security flaws. By recognizing the significant contribution of security risk, the m-government service providers in Malaysia must safeguard security principles and review the existing regulations in order to address the security concerns. Due diligence must be implied to ensure that the citizen sensitive information is protected in a secured system access, user identification and other advanced security measures. The service providers must always bear in mind that insufficient data security will greatly pose an influence on the citizen's uptake of m-government service.

The moderating effects of age, gender and education level indicates the need for market segments to provide explicit consideration for different citizens' characteristics. One size suits all strategies is not applicable in the m-government services. Male and older citizens with high education are more concerned about the usefulness and technical resource or support of m-government services. Female are more sensitive with the words of mouth and the degree of efforts in which they want the system to be more convenient and easy to use. All these findings can serve as momentum for the Malaysian government to be aware of the citizens' preferences and needs to develop a successful m-government service.

## LIMITATIONS AND RECOMMENDATIONS

Even if the study contributes a new insight into the m-government services adoption, it is limited in which the study was carried out only in Malaysia. Hence, the findings can only been transferable to country with similar culture and m-government infrastructure. A comparison with cross country or cross culture study on m-government services should be conducted in future research. On a side note, this study did not integrate mediator into the research model. It will be exciting to examine the mediating effects in the constructs to contribute deeper insight for the service provider.

## CONCLUSION

The study has successfully examined the predictors of m-government services by incorporating the UTAUT and risk theory. The effects of FC, SR, and EE are quite surprising because the results differed from the expected outcomes. FC has been proven as the most important predictor of m-government services which leads the citizens to perceive that the services are useful and convenient to use. Moreover, instead of having negative effects on ITU of m-government services, SR was found to have positive influence towards the ITU. For EE, the findings also showed that the convenience of m-government services is no longer a major impact towards the ITU. This might be due to the targeted citizens who are technology savvy and risk takers with high level of innovation. Even though there might be a high security concern, citizens are still willing to use the m-government services. As a conclusion, citizens prefer useful m-government services with adequate technical assistance and support and zero security flaw to benefit them.

# REFERENCES

Abdelghaffar, H. & Magdy, Y. (2012). The adoption of mobile government services in developing countries: the case of Egypt. International Journal of Information and Communication Technology Research, Vol.2, No.4, pp. 333-341.

Abdullah, N. R. W., Manson, N. B. & Hamzah, A. (2013). Keeping ahead of the game: Innovations and challenges in e-government in Malaysia. The Economic and Labour Relations Review, Vol.24, No.4, pp. 549-567.

Al-Hujran, O. (2012). Toward the utilization of m-government services in developing countries: A qualitative investigation. *International journal of Business and social science*, Vol.3, No.5, pp.155-160.

Anderson, J. C., & Gerbing, D. W. (1988). Structural equation modeling in practice: A review and recommended two step approach. *Psychological Bulletin*, Vol. 103, No. 3, pp. 411–423.

Belanche, D., Casalo, L. V., & Flavian, C. (2012). Integrating trust and personal value into Technology Acceptance Model: the case of e-government services adoption. *Cuadernos de Economía y Dirección de la Empresa*, Vol. 15, No. 4, pp. 277-288.

Chopra, S. & Rajan, P. (2016). Modeling intermediary satisfaction with mandatory adoption of E-government technology for food distribution. *Information Technologies & International Development,* Vol, 12. No. 1, pp. 15-34.

Cohen, J. (1988). Statistical power analysis for the behavioral sciences. Mahwah, New Jersey: Lawrence Erlbaum.

Dwivedi, Y. K., Shareef, M. A., Simintiras, A. C., Lal, B., & Weerakkody, V. (2015). A generalised adoption model for services: A cross-country comparison of mobile health (m-health).

Faaeq, M. K., Alqasa, K., & Al-Matari, E. M. (2014). Technology adoption and innovation of E-Government in republic Iraq. Asian Social Science, Vol.11, No. 3, pp. 135-145.

Hair, J. F., Hult, G. T. M., Ringle, C. M. & Sarstedt, M. (2014), A Primer on Partial Least Squares Structural Equation Modeling (PLS-SEM), Sage Publications, Los Angeles, CA.

Hew, J. J., Lee, V. H., Ooi, K. B., & Lin, B. (2016). Mobile social commerce: The booster for brand loyalty? *Computers in Human Behavior,* Vol. 59*,* No.2016, pp. 142-154.

Hew, J. J., Lee, V. H., Ooi, K. B., & Wei, J. (2015). What catalyses mobile apps usage intention: An empirical analysis. Industrial Management & Data Systems, Vol. 115, No.7, pp. 1-18.

Hung, S. Y., Chang, C. M., & Kuo, S. R. (2013). User acceptance of mobile e-government services: An empirical study. *Government Information Quarterly,* Vol. 30, No. 1, pp. 33-44.

Im, I., Kim, Y., & Han, H, J. (2008). The effects of perceived risk and technology type on users' acceptance of technologies. *Information & Management,* Vol. 45, No. 2008, pp. 1-9.

Leong, L. Y., Hew, T. S., Tan, W. H. G., & Ooi, K. B. (2013). Predicting the determinants of the NFC-enabled mobile credit card acceptance: A neural networks approach. *Expert Systems with Applications,* Vol. 40, No. 2013, pp. 5604-5620.

Lian, J. W., & Yen, D. C. (2014). Online shopping drivers and barriers for older adults: age and gender differences. *Computers in Human Behavior,* Vol. 37, No. 2014, pp. 133-143.

Lin, C. S., Tzeng, G. H., Chin, Y. C., & Chang, C. C. (2010). Recommendation Sources on the Intention to Use E-Books in Academic Digital Libraries. *The Electronic Library,* Vol. 28, No. 6, pp. 844-857.

Liu, Y., Li, H., Kostakos, V., Goncalves, J., Hosio, S., & Hu, F. (2014). An empirical investigation of mobile government adoption in rural China: A case study in Zhejiang province. *Government Information Quarterly,* Vol. 31, No. 3, pp. 432-442.

*Malaysian Communications and Multimedia Commission*. (2015). Communications & Multimedia Pocket Book of Statistics Quarter 1, 2015.

Marshall, B., Mills, R., & Olsen, D. (2008). The role of end-user training in technology acceptance. *Review of Business Information Systems,* Vol. 12, No. 2, pp. 1-8.

Nasir, M.A., Wu, J. J., Yago, M., & Li, H.H. (2015). Influence of psychographics and risk perception on internet banking adoption: Current state of affairs in Britain. *International Journal of Economics and Financial Issues*, Vol.5, No. 2, pp. 461-468.

Ng, C. S. A., & Choy, J. Y. (2013). The Modified Technology Acceptance Model for Private Clinical Physicians: A Case Study in Malaysia, Penang. *International Journal of Academic Research In Business and Social Sciences,* Vol. 3, No.2, pp. 380-403.

Ohme, J. (2014). The acceptance of mobile government from a citizens' perspective: Identifying perceived risks and perceived benefits. *Mobile Media & Communication*, Vol. 2, No. 3, pp. 298-317.

Ozdemir, S., Trott, P., & Hoecht, A. (2008), Segmenting internet banking adopters and non-adopters in the Turkish retail banking sector. *The International Journal of Bank Marketing*, Vol. 26, No. 4, pp. 212-236.

Performance Management & Delivery Unit. (2014). Economic Transformation Programme Annual Report 2014. Retrieved July 1, 2015, from

http://etp.pemandu.gov.my/annualreport2014/upload/ETP2014_ENG_full_version.pdf

Rahman, S. A., Taghizadeh, S. K., Ramayah, T., & Ahmad, N. H. (2015). Service innovation management practice in the telecommunications industry: what does cross country analysis reveal. *SpringerPlus,* Vol. 4, No. 810, pp. 1-25.

Rakhi, T. & Mala, S. (2014). Adoption readiness, personal innovativeness, perceived risk and usage intention across customer groups for mobile payment services in India. *Internet Research,* Vol. 24, No. 3, pp. 369 – 392.

Rogers, E.M. (1995). *Diffusion of Innovations*, 4th ed. The Free Press, New York, NY.

Santhanamery, T., & Ramayah, T. (2015). Understanding the effect of demographic and personality traits on the E-filing continuance usage intention in Malaysia. *Global Business Review,* Vol. 16, No. 1, pp. 1-20.

Shafinah, K., Sahari, N., Sulaiman, R., Mohd Yusoff, M. S., & Mohd Ikram, M. (2013). Determinants of user behavioural intention (BI) on mobile services: A preliminary view. *Procedia Technology,* Vol. 11, No. 2013, pp. 127-133.

Singh, S., Srivastava, V., and R. K. Srivastava. (2010). Customer acceptance of mobile banking: A conceptual framework. *SIES Journal of Management,* Vol. 7, No. 1, pp. 55-64

Tan, X., Qin, L., Kim, Y.B., & Hsu, J. (2012). Impact of privacy concern in social networking web sites. *Internet Research*, Vol. 22, No. 2, pp. 211–233.

Thunibat, A. A., Mat Zin, N. A., & Sahari, N. (2010). Mobile government service in Malaysia: Challenges and opportunities. Information Technology International Symposium (ITSim), Vol. 3, No. 1, pp. 1244-1249.

Tung, F. C., Chang, S. C., & Chou, C. M. (2008). An extension of trust and TAM model with IDT in the adoption of the electronic logistics information system in HIS in the medical industry. International Journal of Medical Informatics, Vol.77, No. 5, pp. 324-335.

Venkatesh, B., Thong, J. Y. L., & Xu, X. (2012). Consumer acceptance and use of information technology: extending the unified theory of acceptance and use of technology. *MIS Quarterly,* Vol. 36, No. 1, pp.157-178.

Venkatesh, V. & Morris, M. G. (2000), "Why men don't ever stop to ask for directions? Gender, social influence, and their role in technology acceptance and usage", *MIS Quarterly,*Vol. 24, No. 1, pp. 115-139.

Venkatesh, V., Chan, F. K. Y., & Thong, J. Y. L. (2012). Designing e-government services: Key service attributes and citizens' preference structures. *Journal of Operations Management,* Vol. 30, No. 1-2, pp. 116-133.

Venkatesh, V., Morris, M. G., B. Davis, G., & D. Davis, F. (2003). User Acceptance of Information Technology: Toward A Unified View. *MIS Quarterly,* Vol. 27, No. 3, pp. 425-478.

Waller, L., & Genius, A. (2015). Barriers to transforming government in Jamaica: Challenges to implementing initiatives to enhance the efficiency, effectiveness and service delivery of government through ICTs (e-Government). *Transforming Government: People, Process and Policy,* Vol.9, No. 4, pp. 480-497.

Wang, C. (2014). Antecedents and consequences of perceived value in mobile government continuance use: An empirical research in China. *Computers in Human Behavior,* Vol. 34, No. 2014, pp. 140-147.

Yang, K. C. (2005). Exploring factors affecting the adoption of mobile commerce in Singapore, *Telematics and Informatics,* Vol. 22, No.3, pp. 257-277.

Yi, M. Y., Jackson, J. D., Park, J. S., & Probst, J. C. (2006). Understanding information technology acceptance by individual professionals: toward an integrative view. *Information & Management,* Vol. 43, No. 3, pp. 350-363.

Yu, C.S. (2012). Factors Affecting Individuals to Adopt Mobile Banking: Empirical Evidence from the Utaut Model. *Journal of Electronic Commerce Research,* Vol. 13, No. 2, pp. 104-121.

# E-Procurement within the City of Johannesburg Metropolitan Municipality, South Africa: A gendered perspective[1]

Angelita Kithatu-Kiwekete[2], Shikha Vyas-Doorgapersad[3]

## Abstract

Municipalities in South Africa are expected to utilise their purchasing processes to promote gender equality. A key external goal of municipal procurement is to redress inequalities through economic opportunities and economic equity to the benefit of both men and women. Currently, most municipalities are transforming their services through electronic mode, resulting in the use of e-procurement processes which link business-to-business, business-to-consumer, and business-to-government via information and communication technologies. Using a Gender and Development (GAD) Approach, this chapter aims to assess the level of gender inclusivity in the municipal e-procurement processes in the City of Johannesburg as a case study. Among the questions raised in the chapter are whether gender mainstreaming is considered in the municipal procurement processes; and if there are any initiatives in place to capacitate men and women to ensure their participation in the e-procurement processes. The review of literature and official documents forms part of the desktop conceptual and theoretical analysis. Utilising qualitative, descriptive and analytical research approaches, the chapter explores the need for gender mainstreaming in the municipal e-procurement value chain processes such as e-informing, e-tendering and vendor management. The chapter then offers policy implications and suggestions for improvement.

**Keywords:** Capacity building, e-procurement, gender, information and communication technologies, municipal procurement.

**Jel Codes:** N77, H57

## INTRODUCTION

Esther Eghobamien, an interim director and head of Gender in the Social Transformation Programmes Division in the Commonwealth Secretariat (Kirton, 2013: 4) states that globally, very few "countries have designed public procurement policies which provide special derogation for competing companies based on gender (or ethnicity, for racially polarised countries)". Gender is important for public procurement policy because it can contribute positively to ensuring equitable access and provide benefits by diversifying the supply chain. Increasing the opportunities for more economic agents, particularly small and medium-sized enterprises (SMEs) to engage in the delivery of goods and services can result in improved outcomes for the alleviation of poverty and increasing gender equality, given that women-owned businesses are disproportionately located in this sub-sector of the economy (cited in Vyas-Doorgapersad & Kinoti, 2015: 97). Despite abundant scholarship on gender equality in Africa, the gendered dynamics within public procurement remains understudied (Nyeck, 2015: 28). Callerstig (2014: 53) agrees that more research is required in this field. The article therefore explores the relationship between gender and procurement and investigates the level of gender inclusiveness in procurement policies for socio-economic development.

---

[1] Kithatu A, Vyas-Doorgapersad S. (2017). Gender Based E-Procurement within the City of Johannesburg Metropolitan Municipality. International Journal of eBusiness and eGovernment Studies. 9 (1).2017. pp. 9-23.

[2] Post-Doctoral Research Fellow, School of Public Management, Governance and Public Policy, University of Johannesburg, Republic of South Africa,
Email: angelita.kiwekete@wits.ac.za, Orcid: 0000-0001-5769-9965

[3] Full Professor, School of Public Management, Governance and Public Policy, College of Business and Economics, University of Johannesburg, Johannesburg, South Africa,
Email: svyas-doorgapersad@uj.ac.za, Orcid: 0000-0002-8146-344X

## CONCEPTUAL CLARIFICATION

Public procurement refers to the purchase by governments and state-owned enterprises of goods, services and works (Organisation for Economic Co-operation and Development (OECD), 2016: 1). E-procurement is defined as a system "incorporating all purchasing activities such as purchaser request, authorization, ordering, delivery and payment by utilizing electronic means such as internet, web technology and e-commerce" (Suleiman, 2013: 1). Before the advent of the internet, procurement functions were perceived by many to be routine and repetitive processes. This perception has since been modified by the expanding capabilities of the worldwide web (www) in recent years. Various business concerns found it both appropriate and inevitable to embrace the use of internet facilities to enhance the performance of their tasks (cited in Suleiman, 2013: 2).

Underpinned by the Feminist Theory, this article adopts a Gender and Development (GAD) Approach because of its proposed policy implications. The GAD "addresses inequalities in women's and men's social role in relation to development (March, Ines and Mukhopadhyay, 1999: 9). This approach argues for "an integrated gender planning perspective" that concentrates "on the power relations between men and women" to challenge "the assumptions between traditional planning methods" (March, *et al.*, 1999: 55). Furthermore, it links the GAD with the Empowerment Approach to mainstream gender in institutional processes. Razavi and Miller (1995: 2) see this collaboration as a "strategy of relevance" which challenges established institutional dynamics and makes gender equality a key part of the development dialogue.

## GENDER AND PROCUREMENT

Public procurement has great potential to promote gender equality (European Institute for Gender Equality (EIGE), 2016: 1). It has therefore been suggested that "whenever possible … gender equality should be incorporated in the subject of the contract itself" and that his will mean "the incorporation of gender equality clauses requiring gender technical competence … as well as the inclusion of gender criteria for the evaluation of the submitted proposals and for further implementation." (EIGE, 2016: 1).

In the African context, as cited in Nyeck (2015: 22), procurement opportunities raise several concerns, such as "future of human and gender security given outsourcing in key sectors such as health, education, the environment and security"; and that "money lost through procurement is opportunity delayed for gender equality in the financing of development". Djan (2015: 6) takes up the debate and mentions that there are those who argue that there is no "correlation between the processes of public procurement and gender equality" and that "public procurement itself lies outside the field of relevance …[to] actualize social objectives (respect for equality between men and women among them)". In other words, when public procurements are viewed solely in relation to economic development, "then their wider social impact is commonly disregarded". This article furthermore stresses that in todays' public administration, public procurement is not only linked with economic development but also requires technological understanding to utilise the technological aspect of procurement (known as e-procurement). This viewpoint is substantiated by Barahona and Elizondo (2012: 109) who argue that "new disruptive technologies – Internet services, social media, collaborative platforms, cloud computing – enable the development and diffusion of new disruptive models to provide services", but caution that these must also be used to design and manage innovation and enhance the abilities and skill sets of public leaders and administrators. In addition, appropriate training must be provided for both men and women entrepreneurs to utilise the e-procurement system.

## GENDER INCLUSIVENESS IN PROCUREMENT PROCESSES: THE CASE OF SOUTH AFRICA

By acknowledging the economic disparities entrenched by apartheid, the South African Constitution, 1996, requires that national legislation be enacted to ensure that public procurement provide for categories of preference in the allocation of contracts as well as the "protection or the advancement of persons" who have been "disadvantaged by unfair discrimination" (Republic of South Africa, 1996, section 217).The contestation for economic redress that drives the agenda for economic transformation in South Africa provides the leverage for enhancing gender equality. The result is an evolving dynamic legal framework for procurement that governs all state agencies and spheres of government. Legislation that recognizes the need to include previously excluded groups, such as all categories of women, offers the opportunity to promote gender inclusiveness through procurement. Laws such as the Preferential Procurement Policy Framework Act, 2000; and the Broad Based Black Economic Empowerment (BBBEE) Act, 2003 with its corresponding Codes of Good Practice (2007), stipulate a preferential point system that encourages the use of women-owned enterprises to benefit from preferential procurement of all state organs.

A desktop study was conducted by Vyas-Doorgapersad and Kinoti in 2015. The researchers reviewed the Department of Public Works (DPW) Strategic Plan 2012-2016 and deduced that "at the institutional structure level, the DPW, under its sub-programme: Corporate Services, set a Strategic Objective 6 that emphasises mainstreaming of gender, disability and youth development in the core business of both DPW and its related industry (Construction and Property)" confirming that "the gender aspect of DPW Strategic Plan 2012-2016 only incorporates 'people with disabilities'" (Vyas-Doorgapersad & Kinoti, 2015: 104). This viewpoint is supported by Nyeck and Benjamin (cited in Vyas-Doorgapersad & Kinoti 2015: 100) emphasising that even "today, the theoretical and pragmatic rationales for complete outsourcing, privatization, or a combination of both" has implications for women in the public services supply chain. Furthermore, "shifts in the role of the state as an employer of women in the service and caring occupations around the world have not received sufficient attention. The role and impact of new public-private partnerships compared to other forms of privatization for the delivery of public services for women and by women also remains under researched". In order to address this challenge, the Ministry of Women, Children and People with Disabilities (2012) has formulated the Women Empowerment and Gender Equality Bill (WEGE) and this was published in the *Government Gazette* of 29 August 2012 for public comment. The objective is the "monitoring and the setting of targets for women empowerment to achieve equal representation of women" in the public procurement sector (Frontier Advisory, 2013: 18). The Bill is still under consideration and its impact will only be assessed in the coming years. The Presidency acknowledges that constraints on the gendered implementation of these laws still persist (The Presidency, 2009: 29). Gender integration will thus require that proactive steps be taken in the implementation of procurement and that such steps should be made more visible to service providers that are owned by women and men. One of the ways of improving visibility is by modernising procurement.

In 2015, national treasury implemented a centralised and computerised procurement system for the three tiers of government, state departments, agencies and entities through the Office of the Chief Procurement Officer (National Treasury, 2015). E-procurement makes the process more transparent and enhances accountability. This online platform utilises the e-procurement value chain that comprises e-informing, e-tendering, e-auctioning; and also vendor management, purchase order integration, e-invoicing, e-payment, and contract management. These e-procurement phases should adhere to the supply chain processes through which government purchases goods and services. This makes the recent introduction of e-procurement to be of particular importance to gender mainstreaming. In the metropolitan areas, women entrepreneurs have often been isolated from municipal procurement particularly in the larger contracts. It then becomes important for the larger municipalities to interrogate their supply chain management particularly e-procurement which encapsulates the municipal process in its entirety. The City of Johannesburg is therefore being used as a case study below.

# E-PROCUREMENT IN THE CITY OF JOHANNESBURG

Johannesburg retains the premier position as economic hub for the Gauteng region, South Africa, as well as the southern Africa region. This is emphasised in the municipality's vision and medium term strategy as aligned to the National Development Plan (NDP). The current integrated development plan (IDP), 2016-2021, sets "economic growth, job creation, investment attraction, poverty reduction, informal economy and small, medium and micro-sized enterprises (SMMEs) support" as key strategic objectives (City of Johannesburg, 2016: 9). Five economic transformation priorities have been identified, namely: i) industrial transformation to alter the present dominance of mining and service industries; ii) spatial transformation to restructure spatial patterns embedded by apartheid through efforts such as the corridors of freedom; iii) global positioning for the country in the international value chains; iv) competitive transformation particularly for SMMEs; and v) institutional transformation to support national development objectives. The city is using a fifteen-point economic development plan to realise these goals (City of Johannesburg, 2016: 46). This political economy places the municipality "at the coalface of facilitation of local economic development and delivery of utilities and other services necessary for sustainable communities, economic development and growth" (City of Johannesburg, 2016: 46). Being a municipality bound by international and national commitments for gender equality and the empowerment of women in local communities, compels Johannesburg to mainstream gender in its policies and programmes for economic growth and development. Municipal procurement thus increases the significance for realising economic development as determined by Nijaki and Worrel (2012: 135). Initiatives such as the Soweto Empowerment Zone; the local EPWP projects; and the Jozi Equity Fund have seen improved support for SMME and women-owned businesses (City of Johannesburg, 2013: 19). The next external role that local government can play in procurement is for the local sphere to realise economic equity for enterprises owned by women and other previously disadvantaged groups. Municipal procurement may be "specifically crafted as a tool to mediate equity concerns by targeting economic opportunities" for particular categories of people (Nijaki & Worrel, 2012: 140). This deliberate inclusion of enterprises that are on the periphery is important for gender mainstreaming because integrating gender into the municipal procurement enables women-owned enterprises to benefit and enhance their participation in Johannesburg's local economic development.

In looking to take advantage of competition between suppliers as well as "streamline the municipality's purchase of goods and services", efforts must also be made to ensure that gender equality is enhanced through procurement (City of Johannesburg, 2013: 19; also refer Gildenhuys, 2000: 187). The city's procurement is managed by a Supply Chain Management policy (SCM) that is prescribed by national legislation including the Local Government: Municipal Finance Management Act, 2003. Johannesburg's Supply Chain Management policy ascribes to a procurement system that supports Broad-Based Black

Economic Empowerment (BBBEE) and the Preferential Procurement Policy Framework Act (PPPFA) regulations. The policy also commits the city to make a deliberate effort to empower women-owned enterprises and enhance their gendered participation in the local economy. Johannesburg should therefore use its municipal procurement policies and processes to realise economic development as well as economic equity. It is critical that the city be able to assess, measure and integrate gender into procurement because regular evaluation provides the means to "integrate gendered data into the policy cycle" (Ambe & Badenhorst-Weiss, 2012: 252). The gender strategies presented above should "lead to better responsiveness to purchasing needs [and] a better understanding of unique local needs" because local purchasing is closer to suppliers which in turn will facilitate the inclusion of women owned enterprises in municipal procurement (Ambe & Badenhorst-Weiss, 2012: 253). Therefore, the performance advantages of utilising e-procurement should assist in increased municipal productivity, offer access to more suppliers, enhance transparency in purchasing, reduce costs as well as to promote the use of one interface that manages the municipal bid process. These benefits should in turn provide the opportunity for women owned businesses to partake in municipal procurement. More importantly, e-procurement must highlight critical areas of concern as will be examined below where women entrepreneurs encounter obstacles in increasing their share of the municipal procurement objectives for economic empowerment and economic equity. Obstacles include timeous access to electronic platforms as well as the mandate for preferential treatment of women entrepreneurs throughout the municipal bidding process Addressing these obstacles should enable Johannesburg to address gender bias that is inherent in its procurement processes that would otherwise not be highlighted.

To achieve the above objectives, the municipality's gender policy provides specific guiding strategies to mainstream gender into its municipal purchasing, namely: i) to ensure that 25% of all procurement contracts in non-traditional areas are granted to women and youth; ii) to develop systems and mechanisms to identify women involved in the informal economy and SMME level; iii) to create a data registration for SMMEs and traders in the informal sector; iv) to put in place a programme to capacitate women who run SMMEs and enable successful tenders for city projects; v) to strengthen links with entrepreneurial institutions to benefit women entrepreneurs; vi) to make available funding for women entrepreneurs through a community development bank to improve their capacity to deliver on tenders; vii) to develop a programme for women in the informal sector to enable them to participate in the mainstream economy; viii) to review the procedures of the payment system specifically for SMME because current procedures disempower women; ix) to disaggregate data on the Expanded Public Works Programme (EPWP); x) to monitor procurement trends and patterns in the city with a focus on gender; xi) to develop a strategy for women's access to credit and capital; xii) to review (with the goal to increase) the tender point system for the women's category; and xiii) to ensure that there is regular reporting on the awarding of contracts to women business owners and suppliers of services (City of Johannesburg, 2013: 19). The city must make use of online platforms for e-procurement processes that should be gender sensitive.

The following analysis on specific aspects of the city's e-procurement gives an indication of the level of gender integration on the online platforms. Three of the strategies from the above list are used, namely: reviewing the e-procurement tender process with a focus on gender; monitoring the city's procurement trends and patterns; and the disaggregation of data on Johannesburg's purchasing processes. The process of e-tendering and vendor management should provide information on: the pre-bid phase where potential service providers are invited to register on the city's database; the bidding process that includes the bidding period, evaluation of bids, adjudication period and finally the bid award to selected service provider. Information should also be available on the post-bid phase which involves the contract management by the Supply Chain Management (SCM) within the municipality (City of Johannesburg, 2009). E-informing involves communication to current and potential service providers on procurement processes and tenders. The SCM policy requires the city to use its official website, www.joburg.org.za as a platform for e-informing. As an example, it provides details on tender and bid documents. In addition, it must be acknowledged that the city offers free Wi-Fi in the Braamfontein area via the Braamfontein wireless mesh. Wi-Fi is also available in municipal libraries and clinics. This means the public is able to access the internet and furthermore, at municipal libraries computer facilities are available to check for tenders (City of Johannesburg, 2015). Johannesburg's e-procurement platform is available on the city's website https://joburg.org.za/index.php?option=com_content&id=309&Itemid=152. A screenshot is presented in Figure 1.

Through the review of official documents (City of Johannesburg: Point Claim forms, Undated a,b), it can be emphasised that the online invitation to register on the city's supplier database is given periodically. This call for interested service providers makes specific mention of SMMEs owned by women to make submissions. Despite the city's commitment to collect gendered data on procurement, there is no evidence of gendered data in the tender forms that should be completed by potential service provides and included in their tender submissions. The forms that regulate the promotion of SMMEs during service provider registration illustrate this deficiency. Interested SMMEs should supply information in compliance to the PPPFA which is the verification on the SMME status and business location within the municipal jurisdiction of Johannesburg.

Bids/Proposals

**Contents:**
**Bids/Proposals**
| Bid Opening Registers
| Cancellation of Proposals
| List of Awarded Contracts
| Supply Chain Management
| Previous Tenders
| Additional Important Information (Policies and Procedures)
| Pikitup Tenders
| Joburg Water Tender Awards
| Metropolitan Trading Company Tenders

Invitation to register on Supplier Database
The City invites interested businesses, particularly SMMEs owned by women, youth and
people with disabilities to apply for registration to be included on the City's list of
approved prospective suppliers. Read more
Supplier Database of the City of Joburg
Registered and accredited suppliers on the City's Supplier Database are hereby required to
complete and submit to Group Strategic Supply Chain Management all of the Municipal
Bidding Document (MBD) forms, by not later than 31st January 2017. Read More
Supplier Application Form Download Application Form
Bids/Proposals
**2017**
A724 **-** 528/17
527/17
….

Tender Forms
MBD 4 **-** Declaration of interest [PDF, 108.91kb]
MBD 5 **-** Declaration for procurement above R10 million (vat included) [PDF, 10.8kb]
MBD 6.3 **-** Regulation 2001 Promotion of small businesses [PDF, 18.9kb]
MBD 6.11 **-** Regulation 2001 Enterprises located in a specific Municipal area [PDF,
13.8kb]
MBD 8 **-** Declaration of bidder' past supply chain management practices [PDF, 16kb]
MBD 9 **-** Certificate of Independent bid determination [PDF, 83,9kb]

Declaration on State of Municipal accounts [PDF, 8.2kb]

**What is the procedure for City of Johannesburg Tenders?**
Copies of the tenders are available only in printed form, at a cost of R100, or, in certain
cases, an amount as specified on the tender document. Tender documents are available
from the contracts section on the 15th floor, Metropolitan Centre, Braamfontein.
**Who do I contact about tenders?**
**NO LATE TENDERS, COPIES OR FAX COPIES OF THE TENDER DOCUMENT
WILL BE ACCEPTED.**

**Is there a fee for tenders?**
Copies of tenders …….

**Figure 1.** Johannesburg's e-procurement platform

The forms do not include a gender component that the SMME can provide when completing the application. The absence of this requirement constrains the ability of the city to capture gendered data for the interested enterprises registering on the database. The website gives information on the bid process through the bid registers and reports going back at least five years; on cancelled proposals; on awarded contracts and previous tenders, including municipal entities such as Johannesburg Water and the Metropolitan Trading Company. The SCM also manages the contracts for successful tenders. In addition, the city's website provides reports by the SCM on the contract management for approved projects. The SCM reports indicate which companies bid for the tenders; specific details of the company that was awarded the contract; the work provided by the service provider; as well as the value and duration of contract including any amendments to the contract project deliverables via status reports. Reports provided online comply with MFMA regulations and show the status of current tenders that are implemented by the city including any changes in prices or duration of the tenders.

The data on the city's reports on the contracts and awarded bids also highlight the motivation for selecting the particular bidders by citing specific reference to PPPFA requirements on the point system. However, the reports give no indication of SMME empowerment in these projects. In addition, there is no mention of whether or not the service providers are women-owned enterprises.

Moreover, from Johannesburg's e-procurement value chain highlighted here, service providers do not provide evidence of gender inclusion in their operations. Tenders are still increasingly awarded to larger, more established companies than to women-owned enterprises. As a result, the "present BEE model benefits a relatively small number of individuals" and a rather skewed implementation of the BEE, whereby "ownership and senior management receive disproportionate emphasis". This is evident in the preferential point system in the awarding of municipal bids while the empowerment regulations "do not incentivise employment creation" or deliberately provide support for SMMEs (Ambe & Badenhorst-Weiss, 2012: 253).Therefore, while the city's procurement trends and patterns can be monitored in terms of type of contracts, corresponding values awarded and whether these projects are aligned to local economic development goals, it is difficult to determine whether service providers are women-owned enterprises and/or women SMMEs. The absence of gender disaggregated data throughout the e-procurement value chain impedes the city's ability to track and to improve gender thresholds through its procurement.

## POLICY RECOMMENDATIONS

The City of Johannesburg operates its procurement and e-procurement as per its SCM policy which is guided by a national framework that acknowledges the need to proactively implement measures that enhance gender equality and the empowerment of women-owned businesses within its municipal jurisdiction. This article finds that gender mainstreaming in the implementation of the city's

SCM has yet to be done adequately. Although the city has a gender policy to guide internal operational procedures as well as municipal service delivery, the e-procurement value chain for Johannesburg does not reflect any gender integration measures in the procurement process; there are no measures that may be used by women- owned enterprises to enhance their participation and visibility. While it is important to create a space for women-owned enterprises, it is also critical to transform the procurement space that large enterprises occupy so that gender mainstreaming can also be meaningfully implemented with the larger municipal contracts. Decuyper (Undated: 2) recommends three main approaches that can be used to integrate gender in public sector procurement, namely: in the selection criteria (exclusion of discriminating companies); the contract award criteria (by including gender as a sub-criterion when evaluating the quality of the offer); and in the contract performance conditions (the obligation to take the gender perspective into account when executing the commissioned tasks). Firstly, the service providers should be required to mainstream gender in their operations and gendered data should be collected during the bid process and implementation of the procured contracts. Johannesburg should therefore include gender criteria in the procurement processes such as e-registration and e-tender submission, whereby service providers are required to give "details concerning the promotion of equal opportunities" for women and men in their operations and support to women-owned SMMEs (Weewauters, 2007: 14). The reporting on the e-procurement value chain should also encapsulate gendered data to enhance gender mainstreaming. Therefore, going forward the city's SCM should collect gender disaggregated data from service providers throughout the e-procurement value chain to enhance the participation of women entrepreneurs in its municipal e-procurement.

**CONCLUSION**

This article finds that municipal procurement is critical in enhancing local economic development as well as attaining economic equity particularly for women entrepreneurs. Women are often marginalised from supplying goods and services to Johannesburg. E-procurement offers the opportunity to enhance gendered reporting of the process, but fails to do so. Johannesburg must adopt far-reaching strategies in its municipal procurement practices to ensure that gender is mainstreamed and that women-owned businesses, SMMEs as well as large enterprises adopt gender sensitive measures to win contracts with the municipality. It is important that the collection and provision of gendered data throughout the city's e-procurement process be made visibly available for women SMMEs as well as potential service providers.

# REFERENCES

Ambe, Intaher, E. and Johanna A. Badenhorst-Weiss (2012), Procurement Challenges in the South African Public Sector, *Journal of Transport and Supply Management*, Vol. 6, No. 1, pp. 242–261.

Barahona, Juan Carlos and Andrey Elizondo (2012), The Disruptive Innovation Theory Applied to National Implementations of E-Procurement, *Electronic Journal of E-Government,* Vol. 10, No 2, pp.107 – 119.

Callerstig, Anne-Charlott (2014), Can Public Procurement be an Instrument for Policy Learning in Gender Mainstreaming? *Scandinavian Journal of Public Administration*, Vol. 18, No. 4, pp. 51–71.

City of Johannesburg (Undated a), *MBD 6.3: Reference Points Claim Form in terms of the Preferential Procurement Regulations 2001 Promotion of Small Business,* http://www.joburg-archive.co.za [Accessed 17.11.2016].

City of Johannesburg (Undated b), *MBD 6.11: Reference Points Claim Form in terms of the Preferential Procurement Regulations 2001 Promotion of enterprises located in a specific municipal area,* http://www.joburg-archive.co.za [Accessed 17.11.2016].

City of Johannesburg (2009), Supply Chain Management Policy Adopted in Terms of Section III of the Municipal Finance Management Act, No. 56 of 2003, Johannesburg: City of Johannesburg.

City of Johannesburg (2013), *Gender Policy Framework,* http://www.joburg.org.za/images/stories [Accessed 16.11.2016].

City of Johannesburg (2015), *Braamfontein Gets Free Blanket Wi-Fi,* http://www.joburg.org.za [Accessed 19.11.2016].

City of Johannesburg (2016), *IDP 2016/21,* http://www.joburg.org.za [Accessed 17.11.2016].

City of Johannesburg (2017), *Johannesburg's e-procurement platform*, https://joburg.org.za [Accessed 02.05.2017].

Decuyper, Jeroen (Undated), Position Paper: The Introduction of the Gender Dimension in the Public Budgets, Public Procurement Contracts and Subsidies. http://www.oecd.org [Accessed 02.12.2016].

Department of Trade and Industry (2007), BBBEE Codes of Good Practice, Pretoria: Government Gazette No. 29617.

Djan, Aurelija (2015), Public Procurement and Gender Equality: Its Impact on Women in the Security Sector in Serbia, Belgrade: Belgrade Centre for Security Policy.

European Institute for Gender Equality (2016), Gender Procurement, http://eige.europa.eu [Accessed 24.09.2016].

Frontier Advisory (2013), The Current Status of Policies, Practices, Measures and Barriers Regarding Women-owned Businesses in Government Procurement: Report Prepared for the Businesswomen's Association of South Africa (BWASA), http://bwasa.co.za [Accessed 28.11.2016].

Gildenhuys, J.S. Hans (2000), Introduction to Local Government Finance: A South African Perspective, Pretoria: J.L. Van Schaik Publishers.

Kirton, Raymond Mark (2013), Gender, Trade and Public Procurement Policy: Kenya, Australia, Jamaica, http://assets.thecommonwealth.org [Accessed 25.10.2016].

March, Candida, Innes Smyth and Maitrayee Mukhopadhyay (1999), *A Guide to Gender Analysis Frameworks.* Oxford: Oxfam GB.

Ministry of Women, Children and People with Disabilities (2012), *Women Empowerment and Gender Equality Bill*, Pretoria: Government Printers.

National Treasury (2015), *Central Procurement Portal for Government,* https://eprocurement.csd.gov.za [Accessed 24.11.2016].

Nijaki, Laurie Kaye and Gabriela Worrel (2012), Procurement for Sustainable Local Economic Development, *International Journal of Public Sector Management,* Vol. 25, No. 2, pp.133–153.

Nyeck, S.N. (2015), (Out)bidding Women: Public Procurement Reform Diffusion and Gender Equality in Africa,

http://webhost1.cortland.edu [Accessed 23.11.2016].

Organisation for Economic Co-operation and Development (OECD) (2016), Public Procurement, http://www.oecd.org [Accessed 28.11.2016].

Razavi, Shirin and Carol Miller (1995), From WID to GAD: Conceptual Shift in the Women and Development Discourse, United Nations Research Institute for Social Development, Occasional Paper 1, Geneva: Switzerland.

Republic of South Africa (1996), Constitution of the Republic of South Africa, Pretoria: Government Printer.

Republic of South Africa (2000), Preferential Procurement Policy Framework Act. No. 5 of 2000, Pretoria: Government Printer.

Republic of South Africa (2003), Broad Based Black Economic Empowerment Act. No. 53 of 2003, Pretoria: Government Printer.

Republic of South Africa (2003), Local Government: Municipal Finance Management Act. No. 53 of 2003, Pretoria: Government Printer.

Suleiman, Mohammed A. (2013), Adoption of e-procurement and value addition to Tanzanian public institutions a case of Tanzania public institutions, Tanzania: Mzumbe University.

The Presidency (RSA) (2009), South African CEDAW Report: Progress Made on the Implementation of the Convention 1998 – 2008, Pretoria: The Presidency.

Vyas-Doorgapersad, Shikha and Abel Kinoti (2015), Gender-based Procurement Practices in Kenya and South Africa, *African Journal of Public Affairs*, Vol. 8, No. 3, pp. 96–108.

Weewauters, Marijke (2007), Equal Opportunities for Women and Men in Public Procurement Contracts: A Few Recommendations, Brussels: Institute for the Equality of Women and Men.

# Impediments to the use of eLearning technology in an applied sciences and technology at a university in South Africa

Anthony Kiryagana Isabirye[1], Nobukhosi Dlodlo[2], Lydia Mbati[3]

## Abstract

This chapter examines the impediments that derail the intensive uptake of eLearning programmes in a particular higher education institution. The study adopted an inductive research paradigm that followed a qualitative research strategy. Data were collected by means of one-on-one in-depth interviews from selected faculty members at a nominated institution of higher learning. Data were iteratively and reflexively analysed, leading to the emergence of four themes. Notably, the scepticism towards the implementation of transformative eLearning was ascribed to complex initiation procedures, inadequate training and support, an incoherent e-policy at the institution as well as resistance to change. In lieu of this, the paper advocates for the incremental adoption of fully-fledged eLearning strategies and policies among academic institutions as well as the effusive use of blended learning approaches. Thus, as opposed to merely enabling academic faculty to refine their teaching, eLearning strategies could possibly alter the manner in which faculty members conduct their teaching and assessment activities.

**Keywords:** eLearning, implementation, transformative.
**Jel Codes:** 031, 032, 033

## INTRODUCTION

If current university structures have to embrace innovative teaching and learning strategies, they must be flexible enough to adapt to the contemporary teaching and learning approaches. Without such flexibility, students' entrance to the worldwide knowledge repositories could be impeded. Within the same vein, an array of transformational enablers exists to provide academics with adequate motivation for re-thinking curricula. These may include among others; globalisation, commercialisation and internationalisation of higher education (Zakaria, Janjua & Fida, 2016); the inevitable shift from product based economies to knowledge based economies together with the changing student profiles and learning styles (Engelbrecht, 2003). The discussion is no longer about whether to introduce digital technologies into mainstream teaching and learning but rather, how to use the technology and skills students already have to create meaningful learning experiences (Ng'ambi, Brown, Bozalek, Gachago & Wood, 2016). Similarly, Garrison and Kanuka (2004) maintain that curriculum transformations that cater for emerging technologies play a pivotal role in the global

---

[1] Dr., Department of Human Resource Management, Vaal University of Technology, South Africa, Email: anthonyi@vut.ac.za, Orcid: 0000-0003-3601-2241

[2] Dr., Department of Marketing, Vaal University of Technology, South Africa, Email:nobukhosid@vut.ac.za, Orcid: 0000-0002-4727-5453

[3] Dr, Open Distance Learning Research Unit, College of Education, University of South Africa, South Africa, Email: mbatilsa@unisa.ac.za, Orcid: 0000-0002-1182-2654

competitiveness of universities. Consistent with this view is the evident phenomenal growth in the integrated usage of information communication technologies (ICTs) in South African higher education institutions (HEIs). This is because the 'traditional' lecture is no longer an appealing product to the digital natives who are leading a 'wired,' anytime, anywhere lifestyle (Czerniewicz & Brown, 2009). Relatedly, universities' competitiveness in the global higher education market will be dependent on their flexibility and ability to embrace and make use of current technological advancements to change educational and business practices. The precarious position of many HEIs is the struggle to wade off the threat of being 'left behind' by their competitors. Likewise, in the business world, new entrants continue to give innovative solutions at low cost as the markets continue to expand. This makes it difficult for the 'static' or 'complacent' higher education providers to compete (Mapuva, 2009). At the primary level, the emphasis on the part of higher education institutions is to create a learning experience that prepares the higher education student to function in the global world. Secondarily, students and academic faculty are encouraged to contribute meaningfully to the digitally connected global work environment.

Keegan (2003:1) defines electronic learning (henceforth referred to as eLearning) as the "provision of education or training electronically through the Internet" whereas Koohang and Harman (2005) portray eLearning as a confluence between Internet interfaces and software developments that produces education and learning that is ubiquitous and engaging. However, these definitions are only limited to the Internet's ability to alter the cognitive abilities of users. Other scholars argue that real learning is an activity that changes the individual's perceptions and attitudes whilst simultaneously empowering them with both cognitive and physical skills (Rekkedal & Qvist Eriksen, 2003). In this study, the authors conceptualise eLearning as all forms of authentic web-enabled teaching and learning that actively engages students in the process of knowledge construction. eLearning has been adopted in South Africa as an inevitable advancement in spite of the plethora of challenges that are consequential toward its adoption by the learner, the academic, the web developer and university management (Ravjee, 2007). This study focuses on eLearning challenges presented to full-time academics at a South African University of Technology (UoT).

## LITERATURE REVIEW

The review of literature in this section focuses on the eLearning and mobile learning (mLearning) affordances that may be effective in enhancing teaching and learning within the fields of applied sciences and technology disciplines.

### eLearning in the field of applied sciences and technology disciplines

eLearning allows for collaborative activities in disciplines that rely on practical application as a demonstration of learning. Literature on the prominent pedagogical underpinnings of applied sciences and technology education denotes the use of social constructivism (Fransen, Weinberger & Kirschner, 2013), interactive lecturing and modelling and simulation (Saraswat, Anderson & Chircu, 2014). In addition, problem-based learning is viewed as a viable pedagogical approach in applied sciences and technology education practice. While there are other pedagogical approaches that may be used in teaching applied sciences and technology education, the focus of this chapter is on the prominent approaches mentioned in this paper. in addition to eLearning, this chapter focuses on mobile learning as a subset of eLearning.

### Interactive lecturing

Interactive lectures include a strong element of interaction on multiple levels. Interaction may be between the facilitator and the students, interaction between the students themselves, as well as student interaction with learning resources, which may be facilitated through eLearning platforms. In an eLearning environment, lectures may be offered via synchronous means through video conferencing and interaction can be facilitated using synchronous online discussions. Supportive resources may also be accessed on the internet to supplement the lecture.

### Problem based learning

Problem-based learning is a learning approach in which students are expected to work (in teams), harnessing a variety of resources to solve a specific problem. Problem-based learning calls for teamwork, creativity and meta-cognition (Kumar & Natarajan, 2007). In the eLearning environment, problem-based learning may be facilitated through eLearning using applications that allow for interaction and social constructivism. Social constructivism views learning as occurring because of the social process of knowledge construction. In the realm of eLearning, social constructivism may occur through the social learning enabled applications such as blogs, discussion forums and/or wikis. These tools allow for social learning and collaboration beyond geographical boundaries.

### Modelling and simulation

A number of applied sciences require authentic practice and application, which can be expensive, impractical and even risky. Modelling and simulation could be presented using controlled eLearning platforms. As technology advances, simulated and virtual learning contexts could be harnessed to develop skills required in real life situations. These modelling techniques and simulations are available as cloud-based computing; four-dimensional (4D) computer-aided design as well as geometric software.

## Mobile Learning (mLearning) in the field of applied sciences and technology disciplines

mLearning is ordinarily assumed to mean "learning on the move", however mLearning refers to learning with the use of a mobile device, particularly a smartphone or tablet, typically a handheld mobile device. The affordances offered by mLearning to applied sciences and technology education are the focus of this section.

### Game-based Learning

Game-based learning combines virtual reality with location-based learning. These affordances allow for learning to occur within authentic settings and stimulates competition both collaboratively and individually. Motivation is enhanced in game-based learning activities. Game-based learning may serve as a means to achieving other pedagogical goals such as problem-based learning, collaborative learning and interactive learning.

### Simulations and Virtual Reality

Simulations in teaching and learning are particularly relevant to the field of applied sciences and technology. There are skills (both cognitive and motor) which are essential in certain applied disciplines. Similarly, it may be too expensive, too dangerous or against ethical norms to carry out training and skill enhancement in certain fields of study. This may be true of pilot training and medical procedures. In this fields, the use of simulations and virtual reality is a realistic, safer and cheaper option.

### Augmented Reality

The portability of mobile devices allows students to conduct field work *in situ*. Using augmented reality enabled devices, students may view plants, buildings and other physical artefacts with information in a variety of formats related to the artefact embedded and displayed on the mobile device.

### Usage of a variety of mobile apps to achieve learning

Mobile devices are capable of carrying out multiple functions due to the variety of tools and devices available in students' hands. The excerpt below provides an example of the use of multiple mobile apps to enhance students' learning experiences:

> *A mobile application was designed to create an in situ experience of geospatial concepts and representations in science. This was achieved through the use of cameras, video data logging, and*
> *QR codes to access lecturer selected web-based information…*
> *PSTs (Pre-Service Teachers) selected a number of in-application features, illustrating adaptive use with*
> *various groups of students. Primary features chosen were the camera, QR codes, plant characteristics,*
> *video and ambient data section accessible on the "collect data" page*
> *Cameras were used to take pictures*
> *Ambient data logging on mobile devices*
> *QR codes were used to give students access to information not readily available in situ*

*QR codes linked to different representations related to the same concept* (Price et. al., 2014).

## MOTIVATION FOR THE STUDY

In its Draft White Paper on E-education in South Africa (2003:44), the Department of Education recommended that innovative teaching and learning in the form of eLearning becomes a "mainstream activity" among HEIs. This recommendation is consistent with the department's (2015) development strategy of attaining the millennium goal of inclusivity through the "education for all by 2020" plan. Regardless of these evident plans and policies, Salmon (2005) notes that implementation of real eLearning beyond HEI initiated projects has so far been modest. Some institutions still strive to bring aboard the majority of students and staff onto the eLearning podium. Notwithstanding this fact, the scant technology champions who already exist are rarely appropriately guided towards the use of educational innovation. Furthermore, they are not amply motivated to effect comprehensive changes through eLearning.

The reasons why academics shun the use of eLearning is worthy of investigation. This study is therefore, structured to find out the constraints towards embracing eLearning by considering the experiences of academic staff at a tertiary institution in South Africa. The authors are of the view that consultations with the academics will assist in determining the intervening variables that influence non-adoption decisions with a view to condense these constructs whilst developing a framework for eLearning inhibitors within the contextual setting of the university.

## RESEARCH DESIGN

In accordance with Henning, Van Rensburg and Smit (2004), a qualitative, interpretive research design was adopted for the study with a view to solicit detailed information to explain the constraints to eLearning. The authors envisage that this type of information would expand knowledge and understanding beyond what is already known, consequently proffering a detailed account of the experiences of academics and providing clear explanations of the reasoning behind the decisions not to adopt eLearning in spite of its numerous advantages.

### Sample participants

The purposive sampling method was used to select the participants for inclusion in this study. The process commenced with identification of a single participant. More respondents were further traced through the snowballing technique, ensuring that only participants with the required information were included in the sample. New participants were continually brought into the study until after ten participants, where no new information was being added. This signified completeness or saturation of the data (Charmaz, 2003; Groenewald, 2004; Henning *et al.,* 2004).

### Data collection process

Semi-structured interviews were used to collect research-specific data. The process of the qualitative interviews entailed preparing the interview-guide based on the research questions, familiarising with the interviewees, the actual interview sessions and audio recording of the interviews. The interviews were conceptualised as planned social interactions between equals (interviewer-interviewee). This created a sense of relaxation and trust between the interviewee and the participants; enabling the latter to provide the best narration of their experience, thoughts and feelings with regard to eLearning constraints. The interviews were documented through audio-recordings and notes for further analysis. The field notes were used as part of the data. The field notes were also used as a measure of triangulation, whereby interviewees' facial expressions and easiness (or uneasiness) during the course of the interview sessions were captured. In view of discerning any contradiction between what the participants had said and the non-verbal signals, exhibited characteristics were collated with the responses and reconciled. The notes were also made use of during the coding process. In line with Charmaz's (2003) recommendation, the notes were used to document the products of coding, examine the codes further, establish and ascertain how the different categories were related and further explore emergent gaps in the formed categories.

### DATA ANALYSIS

Data were iteratively and reflexively analysed (Srivastava & Hopwood, 2009). Collected data were organised and re-arranged following the procedures of a qualitative investigation, as suggested by Henning *et al.* (2004) and Ezzy (2010). The audio-recorded interviews were transcribed verbatim. The researchers listened to each audio-taped interview, read and re-read the transcripts several times, line by line; ensuring familiarity with the data and further determining data quality (Holliday, 2007). Moreover, constant reference was made to the research questions in order to keep the analysis focused. The data were then compiled, labeled, separated and organised through a process called coding.

### Credibility

Maritz and Visagie (2010) indicated that research credibility is about truth-value and truth in reality. This study provides a comprehensible and justifiable connection linking each phase of the research from the data collection process right through to the reporting of findings. The authors make further attempts to present information coherently, while interpreting it in light of the empirical findings and eluding any personal assumptions and pre-conceived ideas that would possibly influence the outcomes the research.

### Ethical considerations

Ethical clearance was obtained through the Ethical Research committee of the UoT under whose auspices the study was conducted. Participation in the study was voluntary and the respondents were free to withdraw at any stage without victimisation. None withdrew, however. Informed consent was attained by revealing the purpose of the investigation to all participants in writing and verbally. Assurance was given to participants that their names would remain anonymous and the collected data would not be used for any other purposes other than to advance scholarly research and enhance scientific findings in the field.

## RESULTS AND DISCUSSION

Data were collected from ten participants' code named R1 to R10. The respondents' revealed that adopting ICTs within HEIs was unavoidable. This was based on their observation that digital communication and information models are the preferred methods of preserving, retrieving and distributing information. However, the academics' voices were beset with undertones of under-preparedness with regard to teaching within a blended learning domain, whereas eLearning platforms are used without the basic facilitating conditions. From the interviews, it emerged that the process of access registration at the institution was a cumbersome exercise, which was short of buy-in from staff members *(complex initiation procedures)*. It also emerged that a de-motivator towards eLearning adoption was a technologically illiterate academic populace *(Inadequate training and support)*. Additionally, the interviewees indicated that there were many cases when the academics themselves were unable to make use of eLearning since there was no clear e-policy to that effect *(incoherent e-policy)*. Resultantly, a majority of the academics opted to remain attached to the traditional way of teaching *(resistance to change)*. The ensuing themes are elaborated on in the ensuing subsections.

## Theme 1: Complex initiation procedures

Complex initiation procedures were cited as a big deterrent towards the implementation of eLearning at the institution. The respondents indicated that not all staff members were able to use eLearning without going through a cumbersome and time-consuming registration procedure. While the faculty members were reluctant to go through the lengthy eLearning registration process, those who were already registered were discouraged from using the system due to lack of IT support coupled with the scantily available e-support material (Childs, Blenkinsopp & Walton, 2005). To exacerbate the eLearning implementation problem even further, registering the students online did not always happen timeously and academics would be unable to access learning materials until mid-semester every academic year. The participants' concerns were aptly captured in the words of R₅ when she stated that:

> *"…there are tedious registration problems before an academic can obtain access rights on the LMS [learning management system]… as a lecturer I feel that I have limited accessibility rights on the LMS"… [and]… "some students may be omitted from the e-platform if there is incongruence between the institution's online system and student enrolment services."*

A majority of the respondents further voiced their concern with regard to the lack of collaboration between the academics and the online enrolment services personnel. Noble (2002) suggests that departmental synergies and university buy-in are necessary in order to ensure that both learners and staff members are enrolled on the online platform and obtain un-interrupted access to the LMS. Though research has proved the importance of top-down strategies in the implementation of eLearning, such strategy implementation requires the buy-in and engagement of academic staff (Cummings *et al.,* 2005). For wider adoption, there is also need for the support of senior management, among other stakeholders. In this vein, academic staff as subject matter experts could potentially shun the implementation of eLearning if they were to be left out in the process of implementation. These sentiments were strongly captured by participants $R_3$ and echoed by $R_9$:

> *"......I am comfortable with m*y (traditional) *way of teaching"*
> *"I would not want to move from one teaching method to another because the face to face contact is working very well for me."*

**Theme 2: Inadequate training and support**

Schuler and Jackson (2006) point out the importance of training and development as major tools to ensure successful acquisition of the relevant skills and knowledge to implement eLearning. Training does not only provide participants opportunities to engage in practical training sessions but it also empowers them with the cognitive, affective and psycho-motor skills and knowledge to use online tools. There was need in this case for participants to receive training in audio podcasts, script-writing, recording, editing, uploading and the use of different software. They also needed to acquires skills and knowledge vital for the use of the self-assessment and onscreen marking tools. But training alone may not yield the expected results as it should be accompanied by a supportive context. In such a context training is not only supported by participants in the training programme but also the facilitators and the institution. What this implies is that the entire University management, junior and senior lecturers should be motivated and buy into the training programme.

Therefore, for training and development to be effective academic faculty should also be motivated to learn. Indeed, Volery (2000) observes that technical proficiency (on its own) is not of great value unless the academics are encouraged and internally motivated to use eLearning. Some of the respondents admitted that they possessed limited knowledge about eLearning and its contribution. This was a very interesting finding for the study since eLearning aptitude plays a fundamental role towards providing the impetus for academics to utilise eLearning in their teaching practise (Meyer, 2001). Suffice to say; the academics who had computer proficiency demonstrated greater confidence and perceived ease of eLearning use. On the contrary, respondents who had minimal skills were reluctant to use eLearning as highlighted by $R_{10}$:

*"I am expected to use online learning tools and yet I have not been trained on what the eLearning platform can help me to achieve in terms of teaching and assessment".*

This finding is in line with Rekkedal and Qvist Eriksen's (2003) assertion that lack of skills and IT competencies significantly contribute towards the non-adoption of eLearning. According to Charlesworth (2002), academic faculty are neither resistant to training nor to the use of technology in their teaching. On the contrary, the entire process is obfuscated by a lack of training regarding the implementation and incorporation of technology in their daily teaching. Such perceptions inadvertently become impediments in the process of implementing an innovation, causing problems in perception, application and technology usage (Volery, 2000). Training of staff should therefore, be used as an invaluable motivational tool for augmenting the confidence of academics towards various eLearning initiatives. It is indeed against this backdrop that Shapiro (2000) advocates for proficient training that should include both technical and conceptual issues. Relatedly, Macpherson *et al.* (2005) observe that appropriate skills and ability to use eLearning platforms generates increased user' satisfaction. Such satisfaction is closely connected to active participation and devotion to the innovation. Thus, if lecturers do not realise the importance of a particular technology and its contribution towards the achievement of teaching goals, they are likely to be deprived of any commitment towards using the technology, rendering it impossible to integrate the technology into teaching practise (Meyer, 2001).

## Theme 3: Incoherent e-policy

Generally, the university training policy, needs to acknowledge the importance of staff development. Such a policy should ensure that faculty take part in professional development activities that are related to their work. The policy should not only support e-learning, like in this context, but also other staff development related activities like professional conferences and sabbatical leave. Furthermore, the policy needs to make allowance for the university to accord enough human and physical resources to ensure the success of the e-learning intervention. Resources like appropriate presentations, handouts, venues, materials and refreshments during the period of training are as vital as the provision of human resources. Furthermore, the need for a staff development budget, support and guidance for all staff participating in e-training should be well embedded in the e-policy. All this calls for efficient management as managers should always act as the champions of professional development. Leu and Ginsburg (2011) note in this regard that as their leadership role has the potential to influence change as they direct academics towards professional development

Mapuva (2009) discusses the absence of institutional leadership that channels the modus operandi of HEIs towards the fruitful adoption of eLearning. In this regard, the successful implementation and use of the technology is dependent upon created institutional structures that are designed to improve the effectiveness of pedagogical methods to disseminate educational material through technological innovations. E-policy documents usually act as indispensable tools through which institutions can avoid a laissez-faire proliferation of eLearning (Czerniewicz & Brown 2009). These documents range from systematic teaching and assessment e-documents, strategic documents, e-quality assurance documents and manuals that guide university processes towards uptake of ICTs (Department of Education, 2015). The institution under review is currently in the process of establishing eLearning policy documents for the first time since its re-organisation from a former technikon to a university in 2004.

The need to appoint faculty-based eLearning managers dedicated at tailoring the eLearning packages to discipline-specific needs was emphasised. The university is currently pursuing a strategic mission that integrates ICT usage in teaching and assessment but does not necessarily have core institutional polices on ICT usage or appointed faculty-based eLearning managers. Some participants have highlighted that this apparent absence of frameworks governing the use of eLearning has often acted as an impediment towards the adoption of eLearning technologies by academics at the university. Moreover, some academics are of the opinion that the decision by management to exclude them from the policy development process leads to a contestation of ideas, seemingly contrary to the adherence to eLearning (Sesemane, 2008). Academics feel that they are being coerced to implement eLearning owing to commands from management, rather than facilitator or learner progression. Resultantly, academic faculty implement flawed pedagogical practises upon servicing the eLearning technology (O'Neill, Singh & O'Donoghue, 2004). This has left staff sceptical about the likelihood not only of successful implementation of the innovation, but also of realising teaching and learning objectives. The absence of e-policies and the corresponding exclusion of staff members from the development thereof; have almost certainly resulted in the unplanned and *ad hoc*, fragmented and uncoordinated adoption of eLearning at the institution. In line with this notion, the frustration of respondents were captured in R$_7$'s responses which epitomised the rest of the respondents' views. The respondent stated thus:

> *"Students and staff members are thirsty for eLearning but the haphazard (lack of e-policy) implementation and support structures are a problem"… [and]… "I wish we could have a dedicated eLearning manager to help us within our faculty."*

## Theme 4: Resistance to change

Several academics seemed to have a negative attitude towards eLearning. Whilst some felt comfortable using traditional ways of teaching, others felt that it was time consuming and cumbersome to learn new ways and methods of teaching. This resistance to the uptake of innovative teaching methods was mostly noted among academics who lionised the traditional teaching methods. Giving his reasons for not embracing eLearning, R10's response epitomised the responses of all those who did no readily embrace eLearning technologies, stating that:

*"I feel comfortable with the way I teach. I need to be in class and physically face my students other than posting assignments on the eLearning platform".*

## RECOMMENDATIONS AND FUTURE RESEARCH AVENUES

Against the findings of this study, it is recommended that institutional leadership be directly driven towards eLearning solutions and where possible, support blended learning strategies, e-skills training and development in order to empower academics. In this vein, there is need to launch eLearning awareness programs. Such programmes should be implemented in line with management driven e-quality assurance strategies. Care and caution should be exercised upon developing institutional policies that recommend eLearning interventions. It is also the authors' view that IT support personnel can play an important role towards the development of eLearning. This can be done when they render support to academic faculty as and when it is required. Ultimately, synergy among content, pedagogy and technology is fundamental prior to the complete integration of eLearning across the applied sciences curriculum. Future studies can identify the importance of institutional leadership as a key driver towards eLearning uptake by academic staff. In this respect, the focus of institutional policies falls squarely on the circumstances upon which eLearning can be utilised, with consideration for the needs of both academics and students alike.

## CONCLUSION

This paper explored the eLearning adoption constraints faced by academics at a university. While it is envisaged that successful adoption of eLearning will transform teaching and learning to meet the increasing demands for change and modernisation in higher education, faculty members alluded to a number of factors that impede technology adoption. Primarily, the stated barriers include inter alia; complex initiation procedures, inadequate teacher training and support, absence of a coherent e-policy in the institution as well as general staff resistance. Based on this, it was recommended that institutional leadership plays a supportive and pro-active role to counter the identified constraints.

# REFERENCES

Charmaz, K. (2003). Qualitative interviewing and grounded theory analysis. In J.F. Gubrium & J.A. Holstein (eds.). *Handbook of Interview Research: Context and Methods* (pp. 675 - 694). London: Sage.

Childs, S. Blenkinsopp, H.A. & Walton, G. (2005). Effective eLearning for health professionals and students; barriers and their solutions: A systematic review of the literature, findings from the HeXL project. *Health Information and Libraries Journal,* 22, 20-32.

Czerniewicz, L. & Brown, C. (2009). A study of the relationship between institutional policy, organisational culture and eLearning use in four South African universities. *Computers & Education,* 53(1), 111-121.

Department of Education (2015). Strategic plan for the fiscal years 2015-2020. http://www.dhet.gov.za/Strategic%20Plans/Strategic%20Plans/Department%20of%20Higher%20Education%20and%20Training%20Strategic%20Plan%202015-16%20-%202019-20.pdf. Accessed2016/04/21.

Draft White Paper on E-education (2003). Transforming Learning and Teaching through ICT. Department of Education, (2003). http://www.info.gov.za/whitepapers/2003/e-education.pdf. Accessed 2016/08/27.

Engelbrecht, E. (2003). A look at eLearning models: investigating their value for developing an eLearning strategy. *Progressio,* 25(2), 38-47.

Ezzy, D. (2010). Qualitative interviewing as an embodied emotion performance. *Qualitative Inquiry,* 16(3), 163-170.

Fransen, J., Weinberger, A. & Kirschner, P.A. (2013). Team effectiveness and team development in CSCL. *Educational Psychologist,* 48(1), 9-24.

Garrison, D.R. & Kanuka, H. (2004). Blended learning: Uncovering its transformative potential in higher education. *Internet and Higher Education,* 7(1), 95-105.

Groenewald, T. (2004). A phenomenological research design illustrated. *International Journal of Qualitative Methods,* 3(1), 1-26.

Henning, E. Van Rensburg, W. & Smit, B. (2004). Finding your way in qualitative research. Pretoria: Van Schaik.

Holliday, A.R. (2007). Doing and writing qualitative research, 2nd ed. London, UK: Sage Publications.

Keegan, D. (2003). Ireland Ericsson competence solutions. Working Paper in ZIFF Papiere 121. Hagen, Germany: Fern Universität, University of Hagen.

Koohang, A. & Harman, K. (2005). Open Source: a Metaphor for eLearning. *Information Science Journal*, 8(1), 1-12.

Kumar, M. & Natarajan, U. (2007). A problem-based learning model: Showcasing an educational paradigm shift. *The Curriculum Journal,* 18(1), 89-102.

Macpherson, A. Homan, G. & Wilkinson, K.H. (2005). The implementation and use of eLearning in the corporate university. *Journal of workplace learning,* 17(1/2), 33-48.

Mapuva, J. (2009). Confronting challenges to eLearning in higher education institutions. *International Journal of Education and Development using Information and Communication Technology,* 5(3), 101-114.

Maritz, J. & Visagie, R. (2010). Methodological rigour and ethics of accountability within a qualitative framework. Paper presented to academic staff at UNISA. Pretoria, South Africa.

Meyer, S.M. (2001). The adoption of technology in higher education. *Curationis,* 24, 32-36.

Ng'ambi, D., Brown, C., Bozalek, V., Gachago, D. & Wood, D. (2016). Technology enhanced teaching and learning in South African higher education: A review of a 20-year journey. *British Journal of Educational Technology,* 14, 30-44.

Noble, D.F. (2002). Digital diploma mills: The automation of higher education. Albany, NY: Monthly Review Press.

O'Neill, K. Singh, G. & O'Donoghue, J. (2004). Implementing eLearning programmes for higher education: A review of the literature. *Journal of Information Technology Education*, 3, 314-323.

Price, S., Davies, P., Farr, W., Jewitt, C., Roussos, G. & Sin, G. (2014). Fostering geospatial thinking in science education through a customisable smartphone application. *British Journal of Educational Technology, 45*(1), 160-170.

Ravjee, N. (2007). The politics of eLearning in South African higher education. International *Journal of Education and Development using Information and Communication Technology,* 3(4), 27-41.

Rekkedal, T. & Qvist Eriksen, S. (2003). Internet-based eLearning, pedagogy and support systems. Working Paper in ZIFF Papiere 121. Hagen, Germany: Fern Universität, University of Hagen.

Salmon, G. (2005). Flying not flapping: a strategic framework for eLearning and pedagogical innovation in higher education institutions. *Research in Learning Technology,* 13(3), 201-218.

Saraswat, S.P., Anderson, D.M. & Chircu, A.M. (2014). Teaching business process management with simulation in graduate business programs: an integrative approach. *Journal of Information Systems Education,* 25(3), 221-232.

Sesemane, M.J. (2008). E-Policy and higher education: from formulation to

Implementation. *South African Journal of Higher Education,* 21(6), 643-654.

Shapiro, L. (2000). Evolution of Collaborative Distance Work at ITESM: structure and process. *Journal of Knowledge Management,* 4(1), 44-55.

Srivastava, P. & Hopwood, N. (2009). A practical iterative framework for qualitative data analysis. *International Journal of Qualitative Methods*, 8(1), 76-84.

Volery, T. (2000). Critical success factors in online education. *The International Journal of Educational Management,* 14(5), 216-223.

Zakaria, M., Janjua, S.Y. & Fida, B.A. (2016). Internationalization of higher education: Trends and policies in Pakistan. *Bulletin of Education and Research,* 38(1), 75-88.

# The Role of Knowledge Management Systems On the Export Performance of Manufacturing Firms: Evidence from Zimbabwe

Edmore Tarambiwa[1], Chengedzai Mafini[2]

## Abstract

There is a general acceptance that Knowledge Management Systems (KMS) are a primary source of value and have taken a center stage in the definition, operation and performance of most business organisations. However, their use within the manufacturing sector in developing countries remains inconsistent. This chapter investigated the role of KMS in enhancing the export performance of firms operating within the manufacturing sector in Zimbabwe. The study used a quantitative approach in which a survey questionnaire was distributed to 555 managers drawn from 185 manufacturing firms based in Harare. Data analyses involved the use of descriptive statistics, Spearman correlations and regression analysis. The results of the study showed that combined IT/social driven KMS exerted the greatest impact on export performance. The availability of both information technology centered and social centered KMS influences export performance by improving the firm's export strategy, export commitment, export orientation, export growth, export sales, export profits and export market share.

**Keywords:** Knowledge management systems, export performance, manufacturing firms, Zimbabwe

**Jel Codes:** M10

## INTRODUCTION AND BACKGROUND

Despite being endowed with a wide array of natural resources, most developing countries continue to face economic challenges (Vijil & Wagner, 2012). As noted in a report by the World Economic Forum (2013) most countries in Asia, South America and Africa have remained as net importers of finished and capital goods. The report further indicates that 31 percent and 34 percent of exports from the European and American manufacturing sector respectively, end in developing countries. According to the Bertelsmann Stiftung's Transformation Index (2016), these imports normally result in a high trade deficit amounting to billions of United States dollars within most developing countries. The report further reveals that in Southern Africa, Zimbabwe had a trade deficit of USD3.9 billion in the 2015-2016 fiscal year whilst Zambia had a trade deficit of USD1.24 billion within the same year. Likewise, Botswana had a trade deficit of P176 million in 2016 (United States Census Bureau, 2017) and South Africa had a trade deficit of ZAR9.5 billion by May 2017. Another report by the Namibia Statistics Agency (2016) indicates a N29.8 billion trade deficit in 2016 for Namibia (Trading Economics, 2017).

[1] Mr, PHD Student, Vaal University of Technology, South Africa
Email: tarambiwae@gmail.com, Orcid: 0000-0002-6175-846X
[2] Professor, Logistics, Vaal University of Technology, South Africa
Email: chengedzaim@vut.ac.za, Orcid: 0000-0002-9426-0975

According to the United States Census Bureau (2017), most of the trade deficits in the above countries were as a result of poor export performance. To counter their unsatisfactory export performance, some countries have resorted to economic integration by becoming members of regional economic blocs (Hartzenberg, 2011). Regional economic integration is aimed at improving on export performance through market expansion (Felix, 2007). An example of a country that has resorted to the adoption and implementation of regional economic integration as a strategy of boosting its export performance is Zimbabwe. The country became a signatory to a number of regional and international economic blocs such as the Common Market for Eastern and Southern Africa (COMESA), Southern Africa Development Committee Preferential Trade Area (SADC PTA) and World Trade Organization (WTO) (Mapuva & Muyengwa-Mapuva, 2014). However, this strategy has not yielded any positive results, particularly in the manufacturing sector. In 2013, the Zimbabwean government acknowledged that the country had been turned into an import-based economy and attributed this development to global competition which had increased in response to the country's enlarged bloc membership (Bimha, 2013). The situation calls for the implementation of other strategies that augment current efforts to turnaround the economic fortunes of the country.

Both individual corporates and countries of today cannot avoid global competition, which may in part, be linked to the increased use of Information Communication Technologies (ICT) all over the world (Kotler, 2011)). Even most notable multi-national corporations such as Nestle, Coca Cola and Toyota have embraced ICT models based information technology (IT) driven knowledge management systems (KMS) as tools for improving performance (Edward & Alves, 2009). Various authors (Argote & Ingram, 2000; Malik & Malik, 2008; Pawlowski & Bick, 2012) support the use of KMS as vehicles for the improvement of corporate performance at micro level, which translates to economic performance at macro level. A study conducted by Man Li (2012) concluded that there were great gains in competitive advantage to be realised by corporations utilising ICT infrastructure as KMS to manage knowledge. Malik and Malik (2008) also found out that there is a general acceptance that knowledge management is a primary source of value, which is an indication that knowledge has taken a center stage in the definition, operation and performance of corporates. Pawlowski and Bick (2012) suggest that as corporates develop globally, their need for KMS increases. Other authors (Barney, 1991; Singer & Czinkota, 1994; Coff, 1997; Shamsuddoha, Ali & Ndubisi, 2009) have also highlighted the importance of KMS in export marketing. These researches seem to suggest the existence of a relationship between KMS and export performance, although not explicitly.

**PURPOSE AND EXISTING RESEARCH GAPS**

The aim of the current study is to test the relationship between KMS and export performance from a context of manufacturing firms in Zimbabwe. A literature search shows that most previous studies on KMS included one focusing on definitions (Edwards, 2011) challenges and benefits (Alavi & Leidner, 1999); practices and theories (Dalkir, 2005). Other studies focused on limitations (Swan, Newell & Robertson, 2000); formulation of theoretical frameworks (Gallupe, 2001; Maier & Lehner, 2003); evolution (Halverson, Ericson & Ackerman, 2004); KMS in product development (Hidiyanto & Efendy, 2010); KMS in Business (Thierauf, 1999); and requirements of a KMS (Mau & Mau, 2008). In addition, Plessis and Boon (2004) examined the role of knowledge management in customer relationship management in South Africa. Kaniki and Mphahlele, (2002) and Ngulube (2002) focused on knowledge management related approaches to the preservation of indigenous knowledge. Jain (2007) conducted a survey to establish the level of knowledge management practices in east and southern Africa. Although these studies gave an insight into the subject of knowledge management, none of them investigated its link to export performance, which leaves an important research gap. This study suggests that the adoption and implementation of KMS could be a vehicle for the improvement of export performance by manufacturing firms in Zimbabwe. Hence the study investigates the relationship between KMS and export performance.

**LITERATURE REVIEW**

The review of literature discusses export performance and knowledge management systems.

**Export Performance**

Export performance is defined as either the relative success or failure of the efforts of an entity to sell its goods and services in other nations (Lages & Lages, 2004). There are several reasons why superior export performance is important for firms. Through exporting, firms are able to increase their sales potential by ensuring that their markets have been expanded beyond national borders (Lages, Silva & Styles, 2009). Since the average orders from international customers are often larger than those from domestic buyers, exporting can be a useful way of increasing firm profits (Sousa & Bradley, 2008). Exporting is also an important approach to diversification, which assists in avoiding risks or exposures due to fluctuations in local markets (Carneiro, da Rocha & da Silva, 2011). In addition, exports are essential in putting redundant production capacity to work, leading to more efficient utilisation of the existing factories, equipment and employees (Freeman, Styles & Lawley, 2012). Exporting may further be a useful means to offset seasonal fluctuations in sales (Boehe & Cruz, 2010). For instance, when an unfavourable season in one country begins, certain product sales take a knock. However, in the same period, the same products can be exported to markets in another country where the season is favourable to sales. Still, some domestic markets are either too small or saturated, creating the need for expansion to other untapped markets (Brouthers, Nakos, Hadjimarcou & Brouthers, 2009). These reasons, amongst others, demonstrate the importance of maximising export performance to both firms and the economy.

**Knowledge Management Systems**

Several authors (Thierauf, 1999; Alavi & Leidner, 2001; Hidayanto & Efendy, 2010; Assegaf & Hussin, 2012) define KMS as the IT technology that supports or facilitates knowledge management. There are various categorisations of KMS.

However, in this study, a categorisation of KMS developed by Nielsen and Michailova (2007), which divided KMS into three classes namely IT driven KMS, social driven KMS and combinations of IT driven and social driven KMS, was adopted. IT driven KMS are based on information technologies whereas social driven KMS are based on interactions of people. Examples of IT driven KMS include decision support systems, data mining and warehousing, simulations, intranet and the internet. Examples of social driven KMS include organisational structure, organisational culture and communities of practice. To counteract the strengths and weaknesses of IT and social driven KMS, the two can be combined, creating a robust and often more effective hybrid system (Hidiyanto & Efendy, 2010).

According to Malik and Malik (2008) KMS are an important tool for driving export performance. In support, Pawlowski and Bick (2012) adds that KMS manage the intangible asset of intellectual capital within organisations thus creating distinct competencies. Lowry (2014) reports that the European Union and the USA have embraced KMS as tools for improving export marketing. This contributed significantly to the European Union and USA's success in exporting to international markets. In Zimbabwe, the National Trade Development and Promotion Organisation of Zimbabwe (ZimTrade) was established to provide the relevant knowledge and support structures to stakeholders at national level (Chigumira, 2013). ZimTrade implemented IT driven KMS by launching a website in 2007 to enhance national exports as suggested by Malik and Malik (2008). However, regardless of having implemented KMS strategies that have worked elsewhere, Zimbabwe's export promotion reports from Zimbabwe National Statistics Agency indicate a continuously downward trend. Based on the literature review, the following hypotheses were formulated and put forward to guide this investigation:

> *H1: There is a positive relationship between IT driven KMS and export performance*
> *H2: There is a positive relationship between social driven KMS and export performance*
> *H3: There is a positive relationship between combined IT driven and social driven KMS and export performance*

**RESEARCH DESIGN**

The research adopted a quantitative survey design, based on the need to generalise the study to other environments of manufacturing firms in developing countries. In addition, a review of previous literature showed that previous studies on both KMS and export performance (Zou, 1998; Alavi & Leidner, 2001; Kautz & Mahnke, 2003; Abdullah, Selamat, Sahibudin & Alias, 2005; Pawlowski & Bick, 2012) were conducted using quantitative surveys.

## Sample Design

The target population in this study was composed of firms operating within the Zimbabwean manufacturing sector. This included eleven industries, namely food, drink, textile, wood, clothing, paper, chemicals, metals and automotive, as categorised by Zimbabwe National Statistics Agency (2016). The names of the firms were drawn from the Confederation of Zimbabwe Industries (CZI) and ZimTrade databases.

To select the sample, a combination of the cluster and some purposive techniques were used. Firms were clustered according to their respective industries. Thereafter, within each cluster three key professionals with the relevant information were selected using the purposive sampling technique. The professionals that were considered as respondents were marketing managers, human resources managers and information technology. The purposive sampling technique was used since the field of study was a technical one which required individuals possessing the required information in each situation. The final sample was composed of 555 respondents drawn from 185 firms.

## Instrumentation and Data Collection Procedures

Data were collected by means of a questionnaire. The questionnaire was divided into four sections. Section A elicited information on the demographic profile of respondents and their firms. Section B sought responses on three KMS sub-elements, namely IT driven KMS, Social driven KMS and Combined IT and Social driven KMS based on measures developed by Nielsen and Michailova (2007) and Malik and Malik (2008). Section C sought information on export performance based on measures developed by Zhou, Taylor and Osland (1998). Response options in Section B of the questionnaire were presented on Likert-type scales anchored by 1=strongly disagree and 5= strongly agree.

Data were gathered from manufacturing firms between June and December 2015. Questionnaires were either emailed to respondents, or administered in person by the principal researcher. Out of a total of 410 questionnaires emailed to respondents, 271 usable questionnaires were retained after the process of screening the questionnaires. Moreover, out of a total of 145 questionnaires that were administered using the drop and collect method, 96 were retained after the screening of the questionnaires. This culminated in a total of 555 questionnaires that were used in the final data analysis. Respondents were given a period of two weeks to complete the questionnaire. During the process of data collection, several ethical considerations, namely participant's rights to anonymity, voluntary participation, confidentiality and protection from victimisation were followed.

## Data Analysis

Data were analysed using the Statistical Packages for the Social Sciences (version 22.0). The strengths and direction of associations between KMS and export performance were measured using Spearman Correlation analysis, whilst predictive relationships between constructs were measured using regression analysis.

## Validity and Reliability

To establish face validity, the questionnaire was reviewed by four faculty members at a Zimbabwean military university who are experts in ICT. Three staff members of ZimTrade who are experts in export marketing were also given the opportunity to review the questionnaire. Feedback obtained from the two panels was used to modify the questionnaire in order to establish face validity. To establish content validity, a pilot study was conducted using a conveniently selected sample of 50 respondents. Further modifications were made to the questionnaire, using feedback obtained from the pilot sample. The pilot sample was excluded from the main survey. To establish construct validity, Spearman's correlations were used. The results of the correlation analysis as shown in Table 2 showed positive correlations between the constructs, thereby providing evidence of acceptable construct validity. Predictive validity was tested using regression analysis. The results of the regression analysis showed statistically significant relationships between the constructs and are illustrated in Table 3, which attests to satisfactory predictive validity within the scales. Reliability was tested using the Cronbach alpha coefficient. All measurement scales attained alpha values above the recommended threshold of 0.7 as indicated in Table 1, thereby providing evidence of satisfactory reliability in the study.

## RESEARCH RESULTS

### Demographic Profile of Respondents

An analysis of the demographic profile of the respondents shows that 29.7 percent of the respondents were marketing professionals, 48.5 percent were IT professionals and 21.8 percent were HR professionals. With respect to their age groups, 54.2 percent of respondents were aged between 31 and 49 years of age. The racial profile showed that all but three respondents who were of the mixed race, were black. At least 72.6 percent of the respondents were male. In terms of the distribution of respondents per manufacturing industry, 22.1 percent of the respondents were in the chemical industry, 42.8 percent in the beverages industry and 35.1 percent in the metals industry. Further analysis revealed that 39.8 percent of the firms had been in operation for up to 15 years; 52.3 percent had been in operation for periods ranging between 16 and 30 years and 7.9 percent had been in operation for more than 45 years.

**Mean Scores and Reliabilities**

The mean-scores and reliabilities of the measurement scales used in the study are reported in Table 1.

Table 1: Mean scores and Reliabilities

| Dimension description | Number of items | Cronbach Alpha | Mean | Standard Deviation |
|---|---|---|---|---|
| IT driven KMS | 6 | 0.724 | 4.01 | 1.05 |
| Social driven KMS | 5 | 0.703 | 4.33 | 1.53 |
| Combined IT and Social driven KMS | 4 | 0.741 | 4.01 | 1.22 |
| Export Performance | 9 | 0.842 | 3.94 | 1.02 |
| Scale 1=strongly disagree; 2= disagree; 3=neutral; 4= agree; 5=strongly disagree | | | | |

Mean scores for the four scales ranged between 3.94 and 4.44. These values depict an inclination towards the 'agree' point in the Likert-type scale. This implies that most respondents perceived that implementation of KMS was satisfactory within their firms. Respondents considered the implementation of combined IT and social driven KMS to be more important than implementing them separately. Cronbach alpha values ranged between 0.703 and 0.842, which were above the recommended minimum threshold of 0.7 (Malhotra, 2011), which confirms that the scales used in the study were reliable.

**Correlation Analysis**

Correlation analysis shows the strength and direction of association amongst the constructs under consideration in a research study (Genest, Kojadinovic, Neˇslehov´a & Yan, 2011).

**Table 2:** Correlation Analysis

| Constructs | N | IT driven KMS | Social driven KMS | Combined IT and Social driven KMS | Export Performance |
|---|---|---|---|---|---|
| IT driven KMS | 555 | 1.000 | 0.241* | 0.473* | 0.599* |
| Social driven KMS | 555 | 0.241* | 1.000* | 0.372* | 0.569* |
| Combined IT and Social driven KMS | 555 | 0.473* | 0.372* | 1.000 | 0.623* |
| Export Performance | 555 | 0.599* | 0.587* | 0.623* | 1.000 |
| ** Correlations are significant at the 0.01 level (2-tailed) | | | | | |

The correlation analysis results are indicated in Table 2. In this study, the constructs were IT driven KMS, social driven KMS, Combined IT and Social driven KMS, and export performance. A two-tailed Spearman Correlation Analysis was undertaken at a significance level of p<0.01 to establish the level of association between the hypothesised associations. In this study, positive inter-factor correlations were observed between the constructs under consideration. The strongest correlation was observed between combined IT and social driven KMS and export performance (r = 0.623; p < 0.01) while the weakest correlation was observed IT driven KMS social driven KMS (r = 0.41; p < 0.01). This indicates that when one of these constructs either increases or decreases, the other constructs either increase or decrease correspondingly.

**Regression Analysis**

Since positive associations existed between KMS dimensions and export performance, it was necessary to establish whether KMS dimensions predicted export performance. This was achieved through application of the regression analysis procedure as illustrated in Table 3. Regression analysis is a statistical process for estimating predictive relationships amongst variables (Armstrong, 2012). To test predictive relationships, IT driven KMS, Social driven KMS, Combined IT and Social driven KMS were used as independent variables and export performance was used as a dependent variable. The results of the regression analysis are reported in Table 3.

**Table 3:** Regression Model Summary

| Model summary Dependent variable- Export Performance | Beta | T | Sig. | Tol | VIF |
|---|---|---|---|---|---|
| IT driven KMS | 0.267 | 3.414 | 0.000 | 0.326 | 2.418 |
| Social driven KMS | 0.143 | 3.012 | 0.002 | 0.712 | 1.732 |
| Combined IT and Social driven KMS | 0.577 | 3.124 | 0.000 | 0.442 | 3.411 |
| R² 0.168        p<0.01 | | | | | |

Multicollinearity tests were conducted by calculating the tolerance value and variance inflation factor (VIF) associated with each independent variable. According to Tabachnick and Fidell, (2001) thresholds for testing for multicollinearity include a minimum of 0.1 for tolerance and a maximum of 10 for VIF. In the current study, tolerance and VIF values were within the recommended thresholds, indicating that multicollinearity did not constitute a problem in the study and the independent variables are not highly correlated ($r$= 0.90 and above). The regression analysis showed an R² of 0.168 which demonstrates that nearly 17 percent of the variation in manufacturing firms export performance is attributable to adoption and implementation of KMS.

**DISCUSSION OF RESULTS**

The purpose of the study was to investigate the relationship between KMS and export performance in manufacturing firms in Zimbabwe. To achieve this purpose, three hypotheses were put forward. The first hypothesis (H1) suggested that there is a positive relationship between IT driven KMS and export performance. This hypothesis was accepted in this study because as revealed in Table 2, there was a strong positive correlation between IT driven KMS and export performance (r = 0.599; p <0.01). Moreover, in the regression analysis, IT driven KMS were statistically significant in predicting export performance ($\beta$ = 0.267; t= 3.414; p=0.000). The second hypothesis (H2) indicated that there is a positive relationship between social driven KMS and export performance. This hypothesis was supported because a strong positive correlation was observed between social driven KMS and export performance (r = 0.587; p < 0.01). Also, analysis of the regression model shows that social driven KMS were statistically significant in export performance ($\beta$ =0.143; t=3.012; p=0.002). The third hypothesis (H3) stated that there is a positive relationship between combined IT and social driven KMS and export performance. This hypothesis was supported because there was a strong positive correlation existed between combined IT and social driven KMS and export performance (r= 0.623; p<0.01). Regression analysis indicates that combined IT and social driven KMS problems were statistically significant in predicting export performance ($\beta$ =0.577; t=3.124; p=0.000). These results illustrate that export performance is likely to increase with an increase in the use of KMS in Zimbabwean manufacturing firms. It is important then for manufacturing firms intending enhance their export performance to, amongst other things, adopt and implement effective KMS along the three dimensions proposed in this study.

## CONCLUSIONS

The study concludes that firms in the Zimbabwean manufacturing sector could improve their export performance by adopting and effectively implementing KMS as part of their strategic ethos. The best model would be to use a combination of IT driven KMS and social driven KMS as this exerts a greater impact on export promotion when compared to applying the two systems separately. However, implementation of both IT driven and social driven KMS requires reliable ICT infrastructure backed by relevant information security policies. It is thus imperative that effective ICT policies and infrastructure be put in place to support the transfer and utilisation of knowledge by the firms in the manufacturing industry. It would also be useful for the Zimbabwean government, through its trade-agency: ZimTrade, to embark on a nationwide KMS awareness program aimed at educating firms on the importance of adopting and implementing KMS.

# REFERENCES

Abdullah, R., Selamat, M.H., Sahibudin, S. & Alias, RA. (2005). A framework for knowledge management systems implementation in collaborative environments for higher learning institutions. *Journal of Knowledge Management Practices,* 2(1), 45-86.

Alavi, M. & Leidner, D.E. (1999). Knowledge management and knowledge management systems: Conceptual foundations and research issues. INSEAD Working Paper: Oxford University.

Alavi, M. & Leidner, D.E. (2001). Review: Knowledge management and knowledge management systems conceptual foundations and research issues. *MIS Quarterly,* 25(1), 107-136.

Argote, L. & Ingram, P. (2000). Knowledge transfer: A basis for competitive advantage in *firms. Organizational Behavior and Human Decision Processes,* 82(1), 150-169.

Armstrong, J. (2012). Illusions in regression analysis. *International Journal of Forecasting,* 28(3), 689-694.

Assegaff, S. & Hussin, A.R.C. (2012). Review of knowledge management systems as socio-technical system. *International Journal of Computer Science Issues* (IJCSI), 9(5), 120-450.

Barney, J.B. (1995). ìFirm Resources and sustained competitive advantage. *Journal of Management,* (17), 99-120.

Bertelsmann Stiftung's Transformation Index. 2016. *Zimbabwe Country Report* . https://www.btiproject.org/fileadmin/files/BTI/Downloads/Reports/2016/p df/BTI_2016_Zimbabwe.pdf. Accessed 2017/07/13.

Bimha, M. (2013). Review of Zimbabwe's Export Performance. ZimTrade Annual Conference, Harare. December 2013.

Boehe D.M. & Cruz, L.B. (2010). Corporate social responsibility, product differentiation strategy and export performance. *Journal of Business Ethics,* 91, 325-346.

Bostock, K. (2002). Avoiding information overload: Knowledge management on the internet. *Journal of knowledge management,* 3(3), 781-880.

Brouthers, L.E., Nakos, G., Hadjimarcou, J. & Brouthers, K.D. (2009). Key factors for successful export performance for small firms. *Journal of International Marketing,* 17(3), 21-38.

Carneiro, J., da Rocha, A. & da Silva, J.F. (2011). Determinants of export performance: A study of large Brazilian manufacturing firms. *Brazilian Administration Review*, 8(2), 107-132.

Chigumira, D.R. (2013). Role of value chains in export development and competitiveness. Harare: ZimTrade.

Coff, R.W. (1997). Human assets and management dilemmas: Coping with hazards on the road to resource-based theory. *Academy of Management Review*, 22(2), 374-402.

Dalkir, K. (2013). *Knowledge management in theory and practice*. California: Routledge.

Davies, R. (1991). Trade, trade management and development in Zimbabwe. *Trade and development in sub-Saharan Africa*, 2(2), 281-313.

Edwards, J.A. (2009). A process view of knowledge management: it ain't what you do, it's the way that you do it. *Electronic Journal of Knowledge Management*, 9(4), 297-306.

Felix, A.S. (2007). Enhancing economic performance through regional intergration. *Journal of International Marketing.* 12(1), 332-358.

Freeman, J., Styles, C. & Lawley, M. (2012). Does firm location make a difference to the export performance of SMEs? *International Marketing Review,* 29(1), 88-113.

Gallupe, B. (2001). Knowledge management system: Surveying the landscape. *International Journal of Management Reviewers,* 3(1), 61-98.

Genest, C., Kojadinovic, I., Neˇslehov´a, J. & Yan, J. (2011). A goodness-of-fit test for bivariate extreme-value copulas. *Bernoulli,* 17, 253-275.

Halverson, C.A., Erickson, T. & Ackerman, M.S. (2004). Behind the help desk: Evolution of a knowledge management system in a large organistion. Proceedings of the 2004 ACM Conference on computer supportive cooperative work, pp. 304-313. ACM-Chicago.

Hartzenberg, C. (2011). Transnational Communities: Shaping Global Economic Governance. London: Cambridge University Press.

Hidayanto, A.N. & Efendy, R. (2010). Analysis and design of knowledge management system in product development. *International Conference on Management and Information Engineering.* 2(1), 104-107.

Jain, P. (2007). An empirical study of knowledge management in academic libraries in East and Southern Africa. *Library review*, 56(5), 377-392.

Kaniki, A.M. & Mphahlele, M. K. (2002). Indigenous knowledge for the benefit of all: Can knowledge management principles be used effectively? *South African Journal of Libraries and Information Science*, 68(1), 11-85.

Katsikeas, C.S., Leonidas, C.L. & Neil A.M. (2000). Firm-level export performance assessment: Review, evaluation, and development. *Journal of the Academy of Marketing Science*, 28 (4), 493-511.

Kautz, K. & Mahnke, V. (2003). Value Creation through IT-supported knowledge management? The utilization of a knowledge management system in a global consulting company. *Journal of Information Science,* 6(1), 85-135.

Kotler, P. (2011). *Principles of marketing.* 13th ed. USA: Pearson Education.

Lages, L.F. (2004). A conceptual framework of the determinants of export performance: Reorganizing key variables and shifting contingencies in export marketing. *Journal of Export marketing,* 13(3), 202-451.

Lages, L.F. & Lages, C.R (2004). The STEP scale: A measure of short-term export performance improvement. *Journal of International Marketing,* 12(1), 36-56.

Lages, L.F., Silva, G. & Styles, C. (2009). Relationship capabilities, quality, and innovation as determinants of export performance. *Journal of International Marketing,* 17(4), 47-70.

Lowry, S. (2014). Small Business Administration Trade and Export Promotion Programs. January 2014, USA: Congressional Research Services.

Maeir, R. & Lehner, F. (2003). Perspectives on knowledge management systems theoretical framework and design of an empirical study. White Paper, November 2003, Regensburg: University of Regensburg.

Maier, R. (2007). Knowledge Management Systems: Information and Communication Technologies for Knowledge Management. 3rd ed. Berlin:Springer.

Malhotra, Y. (2005). Integrating knowledge management technologies in organizational business processes: getting real time enterprises to deliver real business performance. *Journal of knowledge management*, 9(1), 7-28.

Malik, K.P. & Malik, S. (2008). Value creation role of knowledge management: A developing country perspective. *Electronic Journal of Knowledge Management*,6(1), 41-48.

Man Li, R. (2012). Knowledge management, sharing and creation in developing countries' banking industries. *Advanced in Network and Communication.* 1(1), 10-156.

Mapuva, J. & Muyengwa-Mapuva, L. (2014). The SADC regional bloc: What challenges and prospects for regional integration? *Law, Democracy and Development,* 18, 22-36.

Mau, M. & Mau, N. (2008). Requirements of knowledge management systems according to performance and risk related issues in global supply chains, *IBIMA,*6(1), 36-40.

Moghaddam, F.M., Hamid, A.B.B.A., Rasid, S.Z.A. & Darestani, H. (2011). The influence of export marketing strategy determinants on firm export performance: a review of empirical literature between 1993-2010. *International Journal of Fundamental Psychology and Social Sciences,* 1(2), 26-34.

Namibia Statistics Agency. (2016). *Annual trade statistics bulletin 2016*. http://cms.my.na/assets/documents/Media-Release_15_March.pdf. Accessed 2017/07/13.

Ngulube, P. (2002). Managing and preserving indigenous knowledge in the knowledge management era: Challenges and opportunities for information professionals. *Information development*, 18(2), 95-102.

Nielsen, B,B. & Michailova, S. (2007). Knowledge management systems in multinational corporations: Typology and transitional dynamics long range planning. *International Small Business Journal*, 40(3), 314-340.

Pawlowski, J. & Bick, M. (2012). The global knowledge management framework: Towards a theory for knowledge management in globally distributed settings. *Electronic Journal of Knowledge Management,* 10 (1), 92-108.

Plessis, M.D. & Boon, J.A. (2004). Knowledge management in e-business and customer relationship management: South African case study findings. *International Journal of Information Management*, 24(1), 73-86.

Robertson, S.L. (2000). Re-imagining and re-scripting the future of education: Global knowledge economy discourses and the challenge to education systems. *Comparative education*, 41(2), 151-170.

Shamsuddoha, A.K., Ali, M.Y. & Ndubisi, N.O. (2009). A conceptualisation of direct and indirect impact of export promotion programs on export performance of SMEs and entrepreneurial ventures. *International Journal of Entrepreneurship,* 13(1), 87-106.

Singer, T.O. & Czinkota, M.R (1994). Factors associated with effective use of report assistance. Journal of *International Marketing.* 2(1), 53-71.

Sousa, C.M.P. (2004). Export performance measurement: An evaluation of the empirical research in the literature. *Academy of Marketing Science Review,* 22(3), 231-458.

Swan, J., Newell S. & Robertson, M. (2000). Limits of It-driven knowledge management systems initiatives for interactive innovation process: Towards a community based approach. *The Journal of Development Studies*, 5(2), 305-365.

Sousa, C.M.P. & Bradley, F. (2008). Antecedents of international pricing adaptation and export performance. *Journal of World Business,* 43, 307-320.

Tabachnick, B. G. & Fidell, L. S. (2001). Using multivariate statistics. 4th ed. Boston, MA: Allyn and Bacon.

Trading Economics. (2017). *South Africa balance of trade: 1957-2017*. https://tradingeconomics.com/south-africa/balance-of-trade. Accessed 2017/07/13.

Thierauf, R.J. (1999). Knowledge Management for Business. Connecticut: Quorum Books.

United States Census Bureau. (2017). Trade in goods with Botswana. https://www.census.gov/foreign-trade/balance/c7930.html. Accessed 2017/07/16.

Vijil, M. & Wagner, L. (2012). Does aid for trade enhance export performance? Investigating the infrastructure channel. *The World Economy*, 35(7), 838-868.

World Economic Forum. (2013). Report on World Economic Performance. January 2013, USA: World Economic Forum.

World Trade Organisation. (2013). Report on World Trade Statistics for 2013. London: WTO.

Zhou, S., Taylor, C.R. & Osland, G.E. (1998). The EXPERF scale: A crossnational generalized export performance measure. *Journal of International Marketing*, 6(3), 37-68.

# Public finance support for digitalization implementation within the SME's in Pelagonija region

Anastas DJUROVSKI[*]

## Abstract

The modern concept of SME development is based on the digital economy and its approach. Public finance as a vehicle for public policy management and its central part is widely accepting the concept on development based on several pillars- and among the others SME development. Yet most of the studies for SME development are including digitalization of the SME's. Public finance support currently goes towards various sets of measures on expenditures and revenues budget side. The theory on regional growth is widely based on the SME development. Currently there are no studies on impact of the digital SME's in Macedonia. In Pelagonija region there are 3790 active businesses out of which 32 are large enterprises and the rest are SME's. GDP in 2016 was 1,078 billion EUR and on national level SME's contribute with 64.5% of value added to the GDP. Survey was undertaken on 131 SME's from Pelagonija region in order to determine the current level of SME digitalization within the region, to compare with EU average and to make conclusions on the impact of the SME digitalization to region GPD growth as well as revenues collection.

**Keywords:** Fiscal Policies, SME's development, Digitalization
**Jel Codes:** H30, H32

## DIGITALIZATION AND SME'S FROM PELAGONIJA REGION

Digitalization is attracting widespread interest along various industries and politics. It can be defined as "ability to turn existing products or services into digital variants, and thus offer advantages over tangible product". [1] Digital SME's and its impact in regional development is widely recognized in many studies. There are examples of regions in most developed countries where regions are trying to provide greater level of incentives in order digitalization of SME's to be achieved to the highest level [2].However there are still lack of studies that are establishing link between public finance spending towards digitalization of SME's as well as measures on the revenues side like tax cut in favor of digitalization process implementation inside SME's.

Regional development processes in underdeveloped country such as Macedonia has to be one of the government public finance priorities. Investment decision making tool integrated into the regional development policy should be integral part.

---

[*] PhD, Associate Professor, Faculty of Law, University St Kliment Ohridski Bitola, North Macedonia, Email: anastas.dzurovski@uklo.edu.mk Orcid: 0000-0002-2105-744X

[1] Parviainen, Tihinen, Kääriäinen, Teppola, 2017:6-22;
[2] Randall, Berlina, Teräs, Rinne, 2018:18;

Digitalization of the SME's many times mean more cheap option as compared with Large enterprise. [3]Digitalization of the SME's Though the compelling economic factors as well as enforcement from the government in form of various schemes and programmes has led to greater adoption of digital technology by these firms. In our country there are approximately SME firms and traditionally these were exposed to an informal credit system due to lack of access to formal credit system. Productivity gap of SME's can be overcome through intensive digitalization of SME;s.[4] According to Financial Institutions Practice at Boston Consulting Group, this process will help increase the access to formal credit system among 85% of SMEs by 2023. As of now, low level of awareness, unavailability of talented human resource and cost of adoption etc. are the impeding factors in the process of digitalization. Apart from it, the absence of an understanding about the benefits that could be reaped through the use of technology, lack of guiding forces towards integration of technology and its institutionalization into the business, inhibitions towards upfront investment oriented costs have also been the causes that led to low adoption of digitalization among SMEs.[5] In the area of ecosystem, most findings are consistent with previous research. The difficulty of defining the ecosystem's borders is supported by the interviews as well as by literature [6]. Though researchers analyzed this field for more than 20 years, its practical relevance is still low. Thus, we aim to bridge the gap and highlight the relevance of the topic for SMEs. Still, it needs to be considered, that these results describe the relation between roles in ecosystems and stages for digitalization in a qualitative way based on the findings of the case studies.

Digital skills are crucial for SMEs if they are to improve their productivity and especially if they want to scale up. [7]Digital technologies can allow SMEs to improve their relationship with their customers through customer relationship management (CRM), improve and speed up accounting, resource planning and people management processes, delivering efficiencies, especially in terms of staff time.[8]

## LEVEL OF DIGITALIZATION OF PELAGONIJA REGION SME'S

Trying to approach more widely local SME's we have divided their core business into two basic segments products and services. Total of 36 production companies (that are producing physical products) and 95 services sector companies were surveyed. They were asked to provide answers about their level of digitalization as per the segments of business operations that they are dealing with. Both products and services sector are dealing with four core areas where they can implement digitalization solutions: General management, Finance, Production and Marketing. As shown within the table below level of the digitalization is higher within the General management business process and lowest level of digitalization is within the Production processes within the Production companies and for the Services companies the lowest level of digitalization can be found within the Marketing.

---

[3] Pankaj,2019:5.
[4] ASEAN, 2018:10.
[5] Pankaj, 2019:10-15.
[6] Gawer, 2009:3.
[7] Van Ark, 2014:15-20.
[8] Enterprise Research Centre, 2018:30.

**Table 1.** Survey on level of digitalization of Pelagonija SME's answers

| Production Sector | | | n=36 | Services sector | | n= 95 | |
|---|---|---|---|---|---|---|---|
| In which area do you use digitalization | | | | In which area do you use digitalization | | | |
| General Management | 32 | 88.89% | | General Management | | 85 | 89.47% |
| Finance | 36 | 100.00% | | Finance | | 91 | 95.79% |
| Production | 4 | 11.11% | | Service product preparation | | 22 | 23.16% |
| Marketing | 20 | 55.56% | | Marketing | | 50 | 52.63% |
| Production | | | | Services | | | |

Next table shows the level of digitalization within the specific within the business sectors of the companies that are SME's. At the General Management sector there is widespread use of digitalization tools. Mostly there is use of phone applications as well as e-mail (although not all of the respondents are using e-mail both within the products and within the services sector). Very small percent of the respondent companies are using the special software for general management. Also within the general management only small portion of the companies are using platforms (which are very popular and commonly used by the competition abroad). Within the finance sector only in the field of accounting companies are 100% digitalized if we can agree on it that accounting software is a digitalization tool and it is used both insourced and outsources by the companies that are part of the survey. The situation is different within the other sectors like bank correspondence and money market. In the products sector 11.11% of the respondents are using digital means such are online account checking and payment orders processing. Within the services sector such percentage is 33.68%. Yet this cannot be seen only as tool of the companies but even most tool of the banks and respondents are only passive users. In the analysis of this part of business operations also Financial markets are included. Basically in this part companies were asked to answer whether they buy foreign exchange and the definition was given broader meaning by whether they communicate with stock exchanges through buying or selling stocks (of other companies). Here the digitalization is lower that is probably both due to the lower levels of money and financial markets development but also due to the low level of knowledge of digital tools. Within the field of production, the situation can be found as more relevant for explanation of the digitalization level within the Pelagonija region SME's. These differences in the adoption of key digital technologies indicate that different needs are prioritized in each industry. For instance, the need for robotic and automated machinery is higher in the food sector than in the construction sector due to the nature of the production processes in this industry. Interestingly within EU industries adopt social media technologies at a high rate, which may imply that there is a greater need to engage with customers than to improve production processes; while on the contrary, in both industries, 3D printing technology is only adopted by a low percentage of firms in each industry. [9]. Hence, through the digitalization process within the product or services creation many of companies are building their competitive advantage and the same situation counts for the Pelagonija region SME's. As expected more companies that are within the services sector are digitalized in the field of production process. The level of use of digitalized machines is 11.11% within the services sector while such figure is only 23.16% within the services sector. Probably the situation above is due to the fact that technology for services has advanced more it is easier to be implemented and costs are lower that the production sector. However, if regional and national economy wants to be more competitive it has to improve much more within the area considered. It the situation of digitalized machines companies are also technology takers. Those company owners that are following the technology, have access to it and can finance can decide to buy it. The situation is even worse within the field of use of automated software within the production process where only 11.11% of the production companies surveyed have provided answers that they are using automated software for the production process. Situation is not better within the services sector where 14.74% of the companies are using the automated software. Again here are used available software applications (for example Archicad within the construction design or Adobe illustrator within the graphic services), tools that are widely available and can be easily adopted to the services production. Situation can be different in the field of specially developed and tailored software for company's purposes. The most symptomatic area is robotics where none of the companies surveyed responded positively – none of them are using robots although robotics are one of the key elements of the so called fifth technological revolution. Industry 4.0, which will be driven by a new generation of information technologies such as Internet of Things (IoT), cloud computing, big data and data analytics, robotics, artificial intelligence, machine learning, virtual reality and 3D printing.[10] Certainly use of robots can be key competitive advantage element both for production and services companies within the Pelagonija region.

The area of marketing is interesting not only for measurement of digitalization level but also for measurement of the overall marketing tools use for the companies surveyed. We have focused only of digitalization but we think that there is a strong correlation among the use of digital marketing tools and overall marketing tools, models and concepts implementation within the 4P concept. In this area companies are using mostly their own websites although still about 1/2 (50.00% are using) in the products area and 3/5 (40.00% are using) in the services area are not having web sites. Such situation within the products are can be related to low knowledge and need of a part of the companies while in the services sector can be related to moving towards more advanced ways for communication with stakeholders such as platforms (for example logistics companies are using platforms), business tools of social media etc. As expected companies are using free social media advertising tools (such as opening profiles under company name and posting products and services info) and they are also using paid advertising tools. Even all surveyed companies in the products sector are using free social media advertising tools and almost 9 out of 10 services companies are using the same media. But paid advertising tools such as google, facebook and Instagram ads are used to a lesser extent since there are about 1/4 of the companies that are surveyed.

[9] European Comission, 2018: 22
[10] European commission, Digitalisation Support to SMEs 2017a: 18

**Table 2.** Pelagonija SME's digitalization by business processes

| Which tools do you use mostly? | | | Which tools do you use mostly? | | | | |
|---|---|---|---|---|---|---|---|
| **Production** | | | **Services** | | | | |
| **General Management** | | | **General Management** | | | **Average** | **Management** |
| Phone apps | 36 | 100.00% | Phone apps | 95 | 100.00% | 100.00% | 49.47% |
| E mail | 31 | 86.11% | E mail | 95 | 100.00% | 93.06% | |
| Special software (company tailored) | 4 | 11.11% | Special software (company tailored) | 15 | 15.79% | 13.45% | |
| Platforms | 5 | 13.89% | Platforms | 22 | 23.16% | 18.52% | |
| Digital markets | 10 | 27.78% | Digital markets | 16 | 16.84% | 22.31% | |
| | | 47.78% | | | 51.16% | | |
| **Finance** | | | **Finance** | | | | Finance |
| Accounting software | 36 | 100.00% | Accounting software | 95 | 100.00% | 100.00% | 67.45% |
| Bank correspondence | 32 | 88.89% | Bank correspondence | 95 | 100.00% | 94.44% | |
| Financial market | 0 | 0.00% | Financial market | 15 | 15.79% | 7.89% | |
| | | 62.96% | | | 71.93% | | |
| **Production** | | | **Service product preparation** | | | | Products |
| Computerized machines | 5 | 13.89% | Computerized machines | 28 | 29.47% | 21.68% | 14.53% |
| Automated software | 1 | 2.78% | Automated software | 39 | 41.05% | 21.92% | |
| Robots | 0 | 0.00% | Robots | 0 | 0.00% | 0.00% | |
| | | 5.56% | | | 23.51% | | |
| **Marketing** | | | **Marketing** | | | | Marketing |
| Web sites | 28 | 77.78% | Web sites | 63 | 66.32% | 72.05% | 61.98% |
| Free advertizing tools | 36 | 100.00% | Free advertizing tools | 71 | 74.74% | 87.37% | |
| Paid advertizing tools | 10 | 27.78% | Paid advertizing tools | 24 | 25.26% | 26.52% | |
| | | 68.52% | | | 55.44% | | |

The graph below shows digitalization of the business operations of the products creating companies.

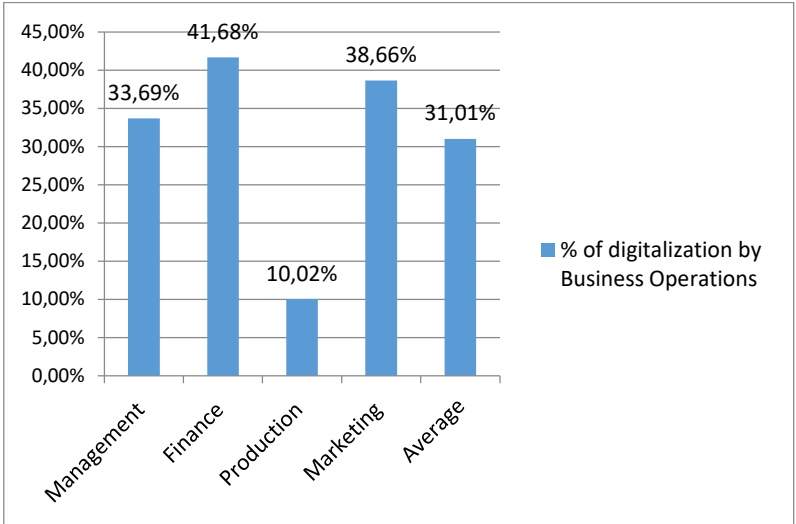

**Graph 1.** Product creating Pelagonija region SME's Digitalization of Business processes

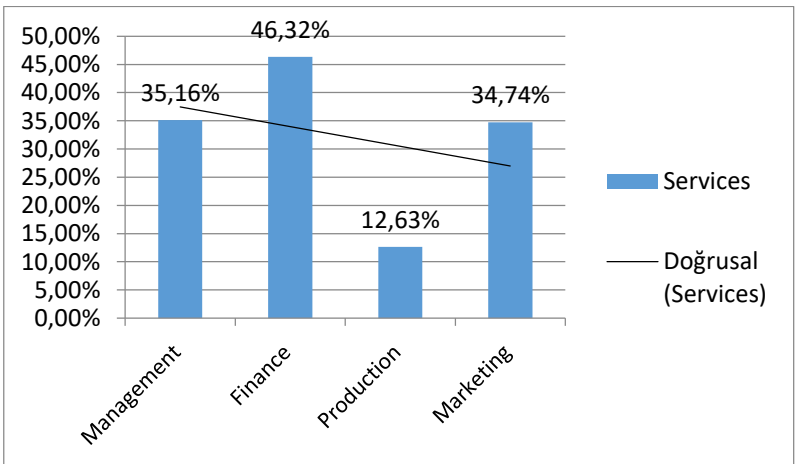

**Graph 2.** digitalization of the business operations of the services producing companies

Graph 3 shows overall level of the digitalization by business operations.

As it can be seen above the level of digitalization is highest within the finance departments. That is due to the fact that all of the companies that were surveyed are using software for accounting and even although the current law permits this part of finance to be kept in paper form the public revenue body is encouraging use of digital tools and communication more than 10 years ago and all of the companies and accountants have switched to the digital forms of accounting. The lower level of digital business operations can be found within the production processes that is 10%.

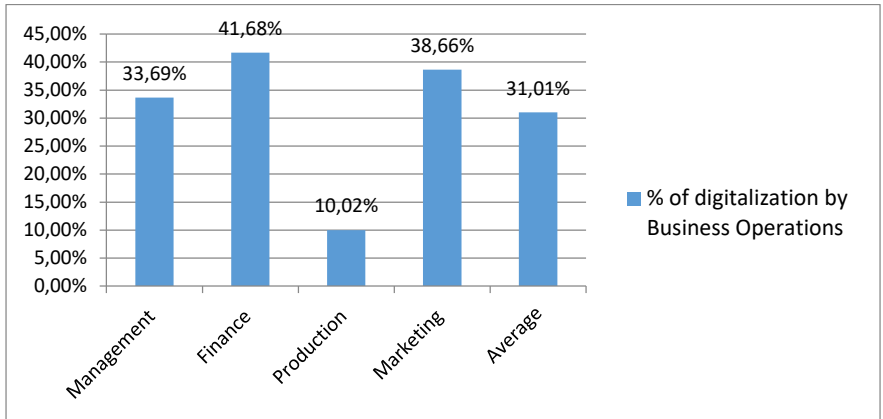

**Graph 3.** Level of digitalization by business operations among the Pelagonija region SME's

**Table 3.** Key indicators tracking digitalization processes

| Key Indicator | % of EU SME's that have adopted |
|---|---|
| Having a web site or homepage | 76.00% |
| Website has some interactive functionalities | 58.00% |
| Use any social media | 47.00% |
| >50% of persons employed use computers and internet | 40.00% |
| Fastest broadband connection is at least 30mbs | 37.00% |
| Have ERP software package to share information | 33.00% |
| Use Customer Relationship Management (CRM) | 32.00% |
| >20% of workers with portable devices for business use | 32.00% |
| Employ ICT specialist | 18.00% |
| Selling online (at least 1% of turnover) | 17.00% |
| Share electronically supply chain management data | 17.00% |
| Exploit B2C eCommerce | 7.00% |

Data as of 2017 [11] Source : European commission services based on Eurostat Data.

If comparison is to be made with Compared with the EU DESI Business Digitalization Index where mean value is above 40%, SME's Within Pelagonija region should significantly improve.

If in particular comparison is to be made with key indicators for tracking o the digitalization processes where the average fo the SME;s is 34.72 % Pelagonija region SME's are still below the EU average . There is a room for improvement and also state policy can benefit from it if proper measures are tailored and implemented in order SME's to benefit towards digitalization improvement. Our aim is to provide data on how much state budget is losing from the non-satisfactory level of digitalization of the companies.

## IMPORTANCE OF SME DIGITALIZATION FOR PELAGONIJA REGION GPD AND ITS IMPACT TO REVENUES COLLECTION

In order to show the financial cost for the state from the current non-acceptable level of the SME's digitalization level provide data on the level of the GDP's growth influence of the better digitalized companies at least to the EU average and to provide data on the tax lost due to such reasons we need data on the GPD growth with the current level of digitalization and to provide trend extrapolation of the GPD's growth. Also we need data on the tax collection. Since there is no precise statistics on the level of the value added of the SME's to GDP by region we have added the general ponder of the 64.5% level within the value added of Pelagonija region SME's to the region GDP.

**Table 4.** Level of contribution to GPD of SME's in Macedonia

|  | Million EUR | % of share |
|---|---|---|
| Micro | 818 | 21.9 |
| Small | 836 | 22.4 |
| Medium | 758 | 20.3 |
| SMEs total | 2412 | 64.5 |
| Large | 1327 | 35.5 |
| Total | 3739 | 100 |

Source: 2017 SBA Fact Sheet, Macedonia, European Comission 2017.

Table 5 shows the levels of GDP by regions in Macedonia in 2016.

**Table 5.** Level of contribution to GPD of statistical regions in Macedonia

| Region | GDP in MKD in million | GDP in EUR in million |
|---|---|---|
| **Macedonia** | **594 795** | **9671.46** |
| Vardar | 46 172 | 750.7613 |
| East | 46 975 | 763.8264 |
| Southwest | 48 810 | 793.655 |
| Southeast | 59 332 | 964.7411 |
| Pelagonija | 65 057 | 1057.835 |
| Polog | 42 487 | 690.8381 |
| Northeast | 29 655 | 482.1924 |
| Skopje | 256 308 | 4167.611 |

Source: Macedonian Statistics Office last available data for 2016

Table 6 shows an estimate of the contribution in GDP of the SME enterprises within the Pelagonija region.

**Table 6.** SME contribution to GDP in Pelagonija region

| Level | Gross domestic product (in million denars) | Gross domestic product (in million EUR) | Estimate of share of SME into GPD,% | Estimate of share of SME into GPD, millions of EUR |
|---|---|---|---|---|
| **Macedonia** | **594 795** | **9671.46** | **64.50%** | **6238.09** |
| Pelagonija | 65 057 | 1057.835 | 64.50% | 682.30 |

Source: DZS (Macedonian Statistics Office)

In 2016 the level of the taxes collected by Public revenues office in Pelagonija region was 127.561 EUR. The table below shows the level of SME contribution towards public revenues collection:

**Table 7.** Share of SME into all companies tax collection in Pelagonija region

| Total taxes collected from Pelagonija Region in Millions of MKD | Total taxes collected from Pelagonija Region in Millions of EUR | Contribution of the SME;s |
|---|---|---|
| 7845 | 127.561 | 82.27683 |

Source: Macedonian Public revenues collection office

Having considered the data below we have extrapolated trends in two fields for Pelagonija region: A) GDP growth with and without EU average level digitalized SME's and B) Public revenues growth with and without EU average level digitalized SME's. As a baseline value we have extrapolated the need for growth of digitalization. Supporting growth and development of SMEs as well as innovative policies targeted at fostering their growth should belong to the state-level priorities.[12]

The graph below shows the level of GPD prediction with the current state of SME digitalization vs level of GPD prediction growth if the level of digitalization is to be upgraded to the EU average. Current SME digitalization in Pelagonija region is estimated at the level of 31.01% while the EU average is 34.72%. That says at least there is a need for 3.71% absolute improvement or 10.68% relative improvement. If it is assumed that at least for that level the company productivity will be increased (although there is data for higher company benefits and in particular profit).

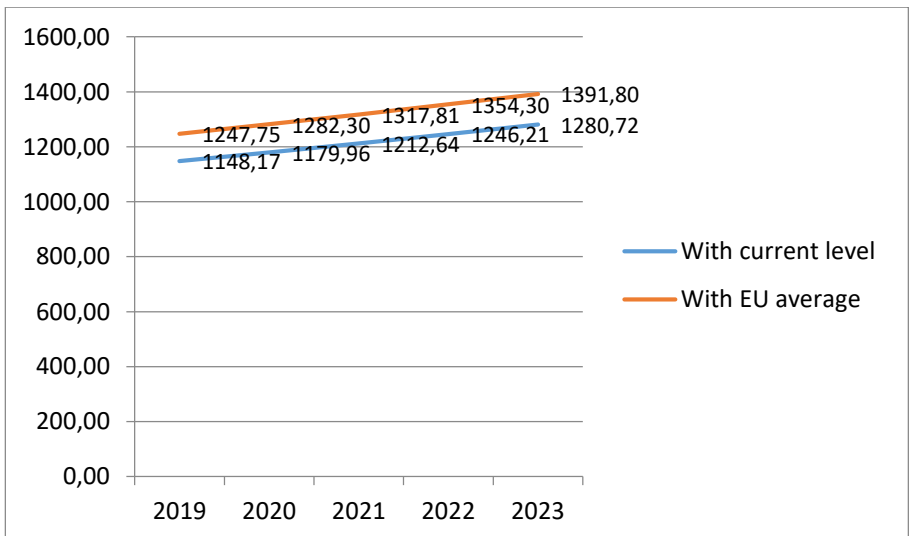

**Graph 4.** Difference in GPD among current and potential level of digitalization meeting the EU average

---

[12] Ruchkina Melnichuk, Frumina, Mentel, 2017:259-271

The graph above depends on the following basis for forecast:

- Annual growth of GPD with no digitalization of SME's of 2.7% taken as average of GPD growth of Pelagonija region for the years 2013,2014,2015 and 2016 as years for which the data is available.

- 10,68% difference in the digitalization to be improved within local SME's in order EU average of 2017 to be reached. 10.68% is current level of relative difference between level of digitalization of the Pelagonija SME's and average level of EU SME's digitalization

- Relative contribution of SME's to regions GDP at national average level of 64.5%.

Above mentioned data shows that if the EU average level of SME digitalization is achieved GPD will grow for additional 94.2 million of EUR

The table below shows the projection difference among the current contribution of SME's within Pelagonija region GDP with current level of digitalization and contribution of SME's within Pelagonija region GDP with EU SME digitalization average as of 2017 achieved.

**Table 8.** Difference in GPD projection growth in Pelagonija region with EU average SME digitalization achieved in millions of EUR

|                    | 2019    | 2020    | 2021    | 2022    | 2023    |
|--------------------|---------|---------|---------|---------|---------|
| With current level | 1148.17 | 1179.96 | 1212.64 | 1246.21 | 1280.72 |
| With EU average    | 1247.75 | 1282.30 | 1317.81 | 1354.30 | 1391.80 |
| Difference         | 99.58   | 102.34  | 105.17  | 108.09  | 111.08  |

Lost GPD is shown at the Graph 5.

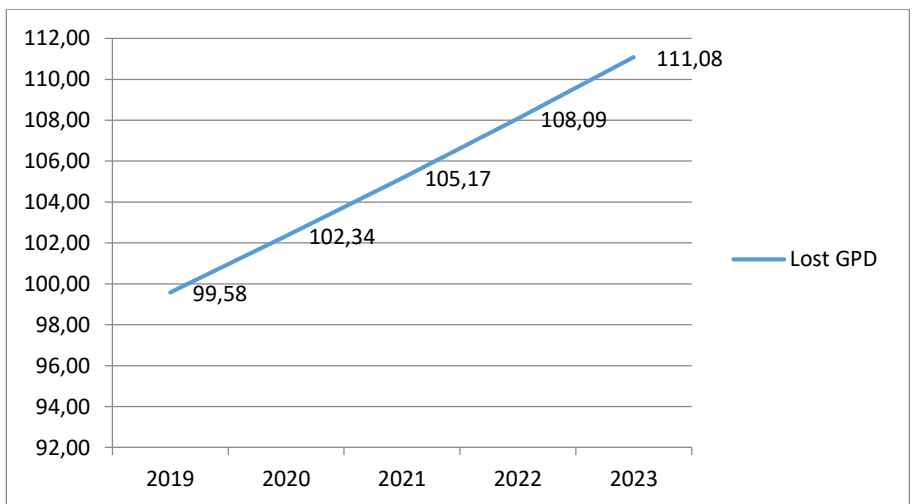

**Graph 5.** Lost GPD for Pelagonija region due to the current level of digitalization of the SME's

As per the data of Public revenue collection office the level of taxes collected from business entities is 3.778 million of MKD or 61.43 million of EUR. Following the ratio of 64.5/35.5 of percentual contribution of SME's and large companies the contribution of SME in revenues is 39.62 million of EUR that is 3.74% of the annual GPD. If we assume that this portion will not increase (by the means of higher level of tax burden), the last is taken as baseline amount that can be increased in revenues collection if Pelagonija region SME's digitalization level is upgraded to EU average. The levels of the taxes lost due to the current level of SME's digitalization are shown in Table 9.

**Table 9.** Difference in taxes collected due to not reached EU average level of digitalization of SME's in millions of EUR

|  | 2019 | 2020 | 2021 | 2022 | 2023 |
|---|---|---|---|---|---|
| Difference in GDP due to not reached EU average level of digitalization | 99.58 | 102.34 | 105.17 | 108.09 | 111.08 |
| Difference in taxes collected from digitalized SME's due to not reached EU average level of digitalization | 3.72 | 3.82 | 3.93 | 4.04 | 4.15 |

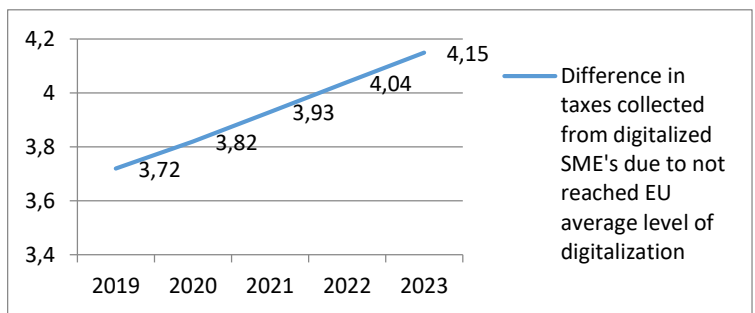

**Graph 6.** Difference in taxes collected due to not reached EU average level of digitalization of SME's in millions of EUR

**CONCLUSION**

There is great room for improvement of digitalization level of Pelagonija SME's. Current state of digitilizaion as per survey undertaken on total of 131 companies out of which 95 are from production sector and 36 are from services sector the overall level of digitalizaiton is 31.1%. Compared with EU average that is measured through EU SME digitalization index this level is lover than the average. Namely in 2017 this level was 34.72% . That shows the fact that there is a evident need for state and regional policy makers should act towards providing climate and incentives for digitalization improvement. Such incentives will not provide benefit only for SME's and GDP. They will also provide direct financial revenues for the state since it is expected level of taxes that will be collected from SME's that advanced in digitalization will bring additional 3.72 million EUR  in the year 1 and 4.15   million EUR in the year 5 if the EU average level of digitalization is going to be achieved from the SME's from Pelagonija region. That questions the debate on how incentives should be provided. Public finance incentives are to be introduced and they should be considered as investments that will be returned within the budget on medium run.

# REFERENCES

Annabele Gawer (2009). Platforms, Markets and Innovation, *Edward Edgar Publishing*, 10-31.

ASEAN (2018). Study on MSMEs Participation in the Digital Economy in ASEAN, *Nurturing ASEAN MSMEs to Embrace Digital Adoption, Economic Research Institute for ASEAN and East Asia* ,18-21.

Enterprise Research Centre (2018). *State of small Business Britain Report 2018*, Sage Group, p 30.

European commission (2017a). Digitalisation Support to SMEs, *European commission*; 18-31.

European commission (2017b). Europe's Digital Progress Report *2017 European commission*; 20-25.

European commission, (2018). Digital Transformation Scoreboard, *European Comission*;10-22.

Gierlich M., Schüritz R., Malte V., Hess T.(2019). SMEs' Approaches for Digitalization in Platform Ecosystems, *Proceedings from Twenty-Third Pacific Asia Conference on Information Systems*, 8-11.

Pankaj M (2019). Study on Impact of Digital Transformation on MSME Growth Prospects in India, *PradeshIJRAR- International Journal of Research and Analytical Reviews*, Amity Business School, 5-18.

Parviainen P., Tihinen M., Kääriäinen J., Teppola S. (2017). Tackling the digitalization challenge: how to benefit from digitalization in practice, *International Journal of Information Systems and Project Management*, 6-22.

Randall L, Berlina A, Teräs, J & Rinne T (2018). Digitalization as a tool for sustainable Nordic regional development: *Preliminary literature and policy review. Discussion paper prepared for Nordic thematic group for innovative and resilient regions*, 12-20.

Ruchkina, G., Melnichuk, M., Frumina, S., & Mentel, G. (2017). Small and medium enterprises in regional development and innovations. *Journal of International Studies*, 10(4), 259-271. 2017;

Van Ark B.(2014).Productivity and Digitalization in Europe: Paving the Road to Faster Growth, *The Lisbon Council  Think Thank*, 15-20.

# Structural Equation Modelling of Internet Banking Service Quality in South Africa: A Framework for Managers

Ephrem Habtemichael Redda[1], Jhalukpreya Surujlal[2]

## Abstract

A challenging global business environment has propelled banks across the world to be innovative and to use alternative delivery channels such as Internet banking, mobile banking and automated teller machine (ATM) banking. The purpose of this study was to develop a measuring and modelling framework/instrument of Internet banking service quality (IBSQ) for the South African banking sector. Snowball and convenience sampling, both non-probability techniques were used to recruit participants for the study. A total of 310 Internet banking customer responses were utilised in the analysis. Using exploratory factor analysis (EFA), eight determinant factors that explained IBSQ were extracted. Following this, the study determined the causal relationship amongst IBSQ, customer value, satisfaction and loyalty through correlation analysis and structural equation modelling (SEM). The proposed model indicates that IBSQ, comprising eight factors, positively influences customer value, satisfaction and loyalty. The model found customer satisfaction to be a predictor of customer loyalty in an Internet banking context. Contrary to the hypothesised model, the influence of customer value was limited to customer satisfaction. The influence of customer value on customer loyalty was found to be rather weak; it influenced customer loyalty only indirectly through customer satisfaction. Understanding the intricate relationships amongst service quality, customer value, satisfaction and loyalty will definitely enhance banks' understanding consumer behaviour and decision making in this digital era. The model may assist bankers to measure, manage and improve IBSQ. Banks could utilise this measurement model to design and improve their Internet banking services.

**Keywords:** IBSQ, customer value, satisfaction, loyalty, consumer behaviour, SEM

**Jel Codes:** M3; M30; M31

## EXECUTIVE SUMMARY

Technological developments and financial liberalisation (deregulation) are considered as the main drivers influencing the developments in the banking sector around the world. South African banks are not immune to this phenomenon. With the introduction of these service outlets banks aim to attract more customers, deliver superior service (create value) for their customers, satisfy their customers and build long-lasting relationships with their customers. Banks

[1] Associate Professor, School of Management Sciences, North-West University (Vanderbijlpark Campus), South Africa, Email: ephrem.redda@nwu.ac.za, Orcid: 0000-0002-0233-1968
[2] Professor & Deputy Dean, Faculty of Economic and Management Sciences, North-West University (Vanderbijlpark Campus), South Africa, Email: babs.surujlal@nwu.ac.za, Orcid: 0000-0003-0604-4971

ultimately aim to create loyal customers who patronise their brand for benefits that come with it such as increased sales and profit, decreased sales and marketing costs and the generation of positive word of mouth.

To achieve the afore-mentioned benefits, banks need to have a deeper understanding of the intricate relationship amongst certain crucial service marketing constructs, namely service quality, customer value, satisfaction and loyalty. The introduction of technology and machines in the delivery of Internet banking service means that traditional service quality measures such as SERVQUAL are not applicable in measuring service quality attributes in an online setting. Thus, industry specific measures of service quality are needed. Accordingly, the primary purpose of this study was to develop a measuring and modelling framework/instrument of Internet banking service quality for the South African banking sector. The study also sought to examine the relationships amongst the crucial marketing constructs mentioned earlier, viz. service quality, customer value, satisfaction and loyalty.

Using exploratory factor analysis (EFA), eight determinant factors were extracted that explained the Internet banking service quality (IBSQ). Following this, the study determined the causal relations amongst IBSQ, customer value, satisfaction and loyalty through correlation analysis and structural equation modelling (SEM). The study has proposed a model that may assist bankers to measure, manage and improve Internet banking service quality at different levels. It has identified the building blocks for improving service quality and a mechanism to create value customers, enhance customer satisfaction and ultimately attain loyal customer in the relationship.

## INTRODUCTION

A challenging global business environment has propelled banks across the world to be innovative and to use alternative delivery channels such as Internet banking, mobile banking and automated teller machine (ATM) banking. Technological developments and financial liberalisation (deregulation) are considered as the main factors influencing developments in the banking sector (Aziz, Elbadrawy & Hussien, 2014). With the introduction of these service outlets, banks aim to attract more customers, deliver superior service (create value) for their customers, satisfy their customers and build long-lasting relationships with their customers.

Over the past few decades, several research studies have been conducted on service and service quality measurements on traditional forms of businesses (Adil, 2013), with limited attention to electronic services. The concept of service quality from an electronic service perspective is described as the clients' overall evaluation and judgement of excellence and quality of electronic service offerings in the virtual marketplace (Santos, 2003). This description suggests that, unlike the evaluation of traditional service offerings, customers in an electronic environment are less likely to evaluate each sub-process in detail during a single visit to a bank's website. Clients in an electronic banking environment are likely to perceive the service as an overall process and outcome (Van Riel, Liljander & Jurriëns, 2001).

There has been long-standing debate regarding the application of existing popular models such as the SERVQUAL model across a broad range of service categories. Dabholkar, Thorpe and Rentz (1996) argue that a single measure of service quality across industries is not feasible. These authors suggest that future research on service quality should involve the development of industry-specific measures of service quality. Such arguments signal a move from attempts to adapt the SERVQUAL to the development of alternative industry-specific measures. As a result, studies have been conducted in electronic service quality across different settings. However, there is increasing evidence of variation in the outcomes of studies on the dimensions of electronic service quality that have surfaced in an attempt to address the key attributes of service quality of online services, directly or indirectly (Jun & Cai, 2001; Barnes & Vidgen, 2003; Santos, 2003; Han & Baek, 2004; Parasuraman, Zeithaml & Malhotra, 2005; Narteh, 2013). The scale developed by Parasuraman *et al.* (2005) referred to as the Electronic Service Quality instrument (E-SQ) comprises seven dimensions, namely efficiency, fulfilment, system availability, privacy, responsiveness, compensation and contact. In a study conducted in the Irish online banking sector, Loonam and O'Loughlin (2008) identified ten dimensions, namely web usability, security, information quality, access, trust, reliability, flexibility, responsiveness, self-recovery and personalisation/customisation that are focal to e-service quality delivery, with the applicability of each of the proposed dimensions to e-banking.

Currently, all major commercial banks in South Africa offer Internet banking facilities to their customers. Research in areas of service quality of Internet banking in South Africa is scant and limited in scope. Hence a research gap was identified that dealt with empirical work in the conceptualisation, measurement and modelling of the services of Internet banking. Therefore, the purpose of this study was to develop a measuring and modelling instrument of Internet banking service quality (IBSQ) for the South African banking sector. In addition, the study aimed to determine the relationships between the constructs of IBSQ, customer value, satisfaction and loyalty formed the empirical objectives of the study.

**LITERATURE REVIEW**

**Service quality**

Many early models of service quality, including the Nordic Model of Service Quality (Grönroos, 1984) and SERVQUAL (Parasuraman, Zeithaml & Berry, 1985, 1988) were based on the disconfirmation model applied in the physical goods' literature. The disconfirmation model is based on the premise that service quality is perceived through a comparison between expectations and experiences of a number of service quality dimensions (Grönroos, 2007). Cronin and Taylor (1994) were amongst the scholars who levelled serious criticism on the SERVQUAL scale, and subsequently introduced their own performance-only scale called the SERVPERF. Cronin and Taylor (1994) questioned the conceptual basis of the SERVQUAL scale and found it confusing with the customer satisfaction construct. SERVPERF was developed by conducting research in four industries, namely banks, pest control, dry cleaning and fast foods. Service quality has been measured in businesses - ranging from financial services to restaurants.

**Customer value**

By introducing alternatives to traditional banking, banks aim to create value for their customers. Zeithaml (1988) describe customer value as "the consumer`s overall assessment of the utility of a product based on perceptions of what is received and what is given". In this context value is, therefore, a trade-off between what the customer received such as quality, benefits, worth or utilities and what the customer gave up to acquire and use the product, for example, price or any other sacrifice. In literature focusing on the service industry, it is argued that customer value is the result of a customer's perception of the value received, where value equals perceived service quality relative to price (Hallowell, 1996). It is a subjective norm since it involves an evaluative judgment of what is received and sacrificed (Ruiz-Molina & Gil-Saura, 2008).

**Customer satisfaction**

Customer satisfaction is a must-achieve objective for any business. In services marketing literature there has been a long-standing debate on the concept of satisfaction (Dong, 2003). Lovelock and Wright (1999:88) define satisfaction as "the outcome of the subjective evaluation that the chosen alternative meets or exceeds expectations". This denotes that the two variables that determine satisfaction are expected and perceived service. The basis of this definition stems from the disconfirmation paradigm as a post-purchase evaluation (Torres, Summers & Belleau, 2001). Satisfaction is also considered from a perspective of cumulative satisfaction and is defined as the customers' overall experience with the service provider after a series of service encounters (Johnson, Gustafsson, Andreassen, Lervik & Cha, 2001). The majority of the past studies view satisfaction from a cumulative perspective to measure the construct (Gupta & Zeithaml, 2006).

**Customer loyalty**

Customer satisfaction is closely related with customer loyalty. Loyalty can be described as a consumer's inclination to patronise a given firm or chain of firms over time (Knox & Denison, 2000). Loyalty consists of two dimensions, namely behavioural and attitudinal aspects (Dong, 2003). The behavioural aspect of loyalty focuses on a measure of the proportion of purchase of a specific brand, while attitudinal loyalty is measured by a psychological commitment to a firm. Koo (2006) conducted a study to identify the variables that determine customer loyalty. The results revealed that customers' favourable perceptions of website design, visual appeal, well-organised hyperlinks, information quality, product assortment and after-sale services are positively associated with online store loyalty. The main advantages of having loyal customers include insensitivity of customers to price increases, possibility of cross-selling, resistance to competition, positive attitude, and of course increased revenue, profit and market share (Cronin, Brady & Hult, 2000; Iacobucci, 2016).

The relationships between service quality, customer value, customer satisfaction and customer loyalty or behavioural intentions have stimulated interest among marketing scholars both in online and in offline service settings (Parasuraman *et al.*, 1988; Cronin *et al.*, 2000; Han & Baek, 2004; Kuo, Wu & Deng, 2009; Lee, 2010). Patterson and Spreng (1997) established that each performance dimension is positively linked to perceived value and in turn perceived value is linked positively to customer satisfaction and (re)purchase intentions. Similarly, other studies have indicated that service quality as a precondition for customer satisfaction (Boshoff & Du Plessis, 2009). The literature also suggests that there is a direct link between service quality and customer loyalty (Koo, 2006). A few studies also suggest a direct link between customer satisfaction and customer loyalty (Dong, 2003). In a study on the relationships between service quality, perceived value, customer satisfaction, and post-purchase intention of mobile value-added services, Kuo *et al.* (2009) found that service quality positively influences perceived value and customer satisfaction, suggesting that when companies provide good service quality, perceived value and customer satisfaction can be improved. The results further attest that perceived value and customer satisfaction directly and positively influence customer loyalty. In light of the research objective of this study and the literature reviewed, the proposed hypothesised research model is presented in Figure 1.

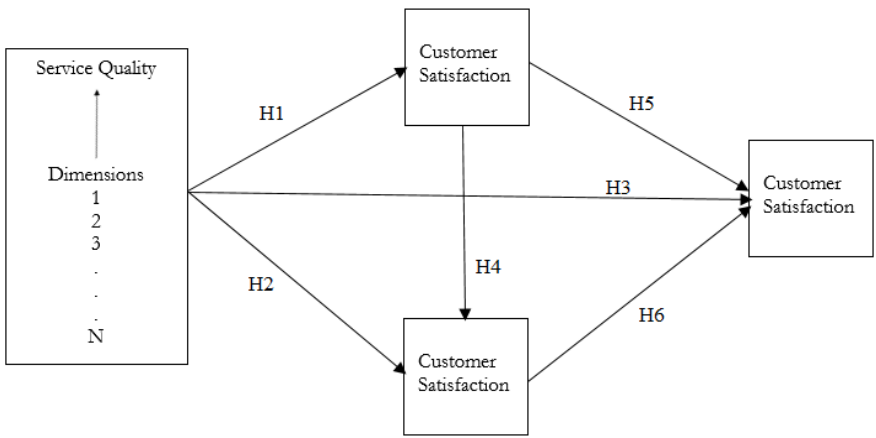

**Figure 1:** Proposed research model

The proposed research model hypothesises a set (*N*) of dimensions that determine IBSQ that positively influence customer value, customer satisfaction and customer loyalty. Furthermore, the model also hypothesises that customer value influences both customer satisfaction and customer loyalty; while customer satisfaction is perceived to be a predictor of customer loyalty.

To support the hypothesised research model, the following alternative hypotheses are formulated:

*H1: Internet banking service quality positively influences customer value.*

*H2: Internet banking service quality positively influences customer satisfaction.*

*H3: Internet banking service quality positively influences customer loyalty.*

*H4: Customer value positively influences customer satisfaction.*

*H5: Customer value positively influences customer loyalty.*

*H6: Customer satisfaction positively influences customer loyalty.*

## RESEARCH DESIGN

The study adopted a sequential-mixed method approach to achieve the formulated objectives. Qualitative data were first collected and analysed. Thereafter, quantitative research was conducted. Use of the sequential-mixed method approach is in line with the practice of the development of a service model (or a scale) where a qualitative research is first conducted followed by quantitative research (Churchill, 1979; Parasuraman *et al.,* 1988). Following a critical study of the extant literature (inductive analysis) and initial generation of a pool of items, a focus group interview comprising five participants and separate in-depth interviews with three participants were conducted to generate original items and descriptions of what constitutes service quality of Internet banking in a South African context (deductive analysis). The findings of the qualitative study (Redda, Surujlal & Leendertz, 2015) were used in compiling items for the questionnaire. This paper reports only the quantitative part of the study. The quantitative part of the research, which involved the use of a questionnaire to collect data, and refine and validate the scale through various statistical applications, was conducted in Southern Gauteng, South Africa in 2015. Since a sampling frame could not be obtained from banks for security and privacy reasons, a probability sampling could not be used in this study. Therefore, snowball and convenience sampling, both non-probability techniques were applied to conduct the study.

### Instrument and procedures

A questionnaire requesting demographic information of participants and scaled items was developed for the study. The scaled items included service quality with eight latent factors collectively comprising of 31 items, five items for customer value, four items for satisfaction and five items for loyalty. All scaled responses were recorded on a six-point Likert-type scale ranging from strongly disagree (1) to strongly agree (6). The use of even number response categories is often preferred especially when the researcher wants to eliminate the neutral effect (Garland, 1991). The questionnaire was pre-tested with three experts in the discipline to check whether any changes were required before administering it to Internet banking customers. Furthermore, to ensure reliability, the questionnaire was pilot-tested on a sample of 50 conveniently selected Internet banking customers that did not form part of the main study. Four hundred (N=400) questionnaires were administered over a one-month period through Survey monkey and self-administration to the identified sample. Of these data captured, 310 completed questionnaires were used in the final analysis. This sample size in this study (n=310) is consistent with similar previous studies conducted on Internet banking services using a non-probability sampling technique (Santos, 2003; Parasuraman, *et al.,* 2005).

## DATA ANALYSIS

The statistical programs IBM Statistical Packages for the Social Sciences (SPSS version 22.0) and the Analysis for Moment Structures (AMOS version 22.0) for Microsoft Windows, were used to perform analysis for the quantitative data. Descriptive statistics, exploratory factor analysis (EFA), correlation analysis and structural equation modelling (SEM) were computed to address the research objectives.

## RESULTS AND DISCUSSION

### Sample profile

Of the 310 respondents, 53 percent (n=163) were male and 47 percent (n=144) were female. The majority of the respondents were aged between 25-34 (34%) followed by the 35-44 (33%) age cohort and the 45-54 (13%) age cohort. The youngest age cohort (18-25) and the oldest age cohort (54-64) were in the minority representing eight percent and 12 percent of the respondents respectively. The majority (31%; n=95) of respondents earned an annual income in the category R250 001 to R350 000, followed by 29 percent (n=89), R350 001 to R450 000, and 15 percent (n=46) in the R450 001 to R550 000 category. A majority (66%) of the respondents indicated that they use Internet banking for most of their banking needs suggesting that they had adequate knowledge and experience of the service provided. In terms of how long the respondents had been using Internet banking, 65 percent (n=199) of the respondents indicated that they had been using Internet banking for more than three years, which depicts a solid experience of usage of the service.

### Exploratory factor analysis (EFA)

Exploratory factor analysis was conducted to identify the underlying service quality dimensions (factors) that influence Internet banking services and consumer decision making in South Africa. Furthermore, separate factor analyses were conducted on the other three constructs, namely value, satisfaction and loyalty to assess whether the items in each of the constructs adequately explained their respective constructs. Table 1 presents a summary pattern matrix of factors and constructs.

**Table 1: Summary pattern matrix of factors and constructs**

| Factors / Constructs | No. of items | Eigen value | % of variance | Cumulative% | Cronbach`s alpha |
|---|---|---|---|---|---|
| Factor 1 | 4 | 11.7 | 37.8 | 37.8 | 0.921 |
| Factor 2 | 6 | 3.2 | 10.3 | 48.1 | 0.940 |
| Factor 3 | 6 | 2.2 | 7.0 | 55.2 | 0.883 |
| Factor 4 | 3 | 1.9 | 6.2 | 61.4 | 0.931 |
| Factor 5 | 3 | 1.8 | 5.8 | 67.2 | 0.934 |
| Factor 6 | 3 | 1.5 | 4.7 | 71.9 | 0.827 |
| Factor 7 | 3 | 1.3 | 4.2 | 76.1 | 0.874 |
| Factor 8 | 3 | 1.1 | 3.5 | 79.7 | 0.924 |
| Value | 5 | 3.4 | 67.3 | 67.3 | 0.875 |
| Satisfaction | 4 | 3.0 | 74.2 | 74.2 | 0.883 |
| Loyalty | 5 | 3.2 | 63.1 | 63.1 | 0.847 |

The dataset was assessed for its suitability for factor analysis using sample size determination, the Kaiser-Myer-Olkin (KMO) and the Bartlett's test of sphericity. In terms of sample size, it is recommended that the size should be more than 150 and there should be ratio of at least five cases per variable (Pallant, 2013). In this study, the sample (310) yielded a ratio of ten cases for each variable. The result of the KMO measure of sampling adequacy indicated that the 31-item scale sufficiently meets the necessary threshold at 0.893 (Malhotra, 2010). Furthermore, the Bartlett's test of sphericity was significant ($p<.05$) indicating the dataset was appropriate for the factor analysis (Pallant, 2013). The factor extraction method applied was principal component analysis (PCA) with Oblimin with Kaiser Normalisation. Factor loadings greater than 0.30 were used as a threshold for the extraction of factors.

The eigenvalues and percentage of variance explained were used in determining the factors that influence IBSQ. The eigenvalue extraction indicated that the eight factors were appropriate and best fit for capturing and explaining the IBSQ construct. These eight factors are the building blocks upon which South African Internet banking customers base their decisions. The eight factors accounted approximately 79 percent of the variance (cumulative variance), which is considered more than acceptable (Hair, Black, Babin & Anderson, 2010). Thus, the IBSQ construct is explained collectively by eight factors. Furthermore, the Cronbach alpha reliability for each factor was above 0.8 portraying very good reliability (Hair *et al.*, 2010). With respect to the other three constructs, a similar method was applied. The cumulative variance for customer value, satisfaction and loyalty were approximately 67 percent, 74 percent and 63 percent respectively. The Cronbach alpha reliability for these three constructs also portrayed very good reliability as illustrated on Table 1.

After careful examination and scrutiny of the items that loaded together to describe the IBSQ construct, the factors were named and operational descriptions provided. Factor one, named *efficiency* refers to the speed at which the electronic banking facility enables customers to complete their banking transactions. Factor two, labelled *privacy and security* refers to the degree to which customers find transacting through electronic banking safe and secure. Factor three, named *contact and responsiveness* refers to the ability of the bank to be contacted to and be responsive of when customers need it, encounter problems and/or to solve problems. Factor four, *ease of use* refers to the accessing and using the bank's website for searching, navigating and transacting electronically with less effort. Factor five, named *reliability*, relates to the extent to which the bank keeps its promise with regard to electronic banking, provide dependable service and keep accurate records of transactions conducted over the Internet. Factor six named *site aesthetics* refers the extent to which customers find bank's websites and designs to be visually appealing and attractive. Factor seven, labelled *functionality,* refers to the sum or any aspect of the electronic banking as a product can do for the customer. The last factor, named *system availability,* describes the operational availability of the bank's website for electronic banking transactions.

Determining the relationships between the constructs IBSQ, customer value, satisfaction and loyalty formed the empirical objectives of the study. To address these objectives, first correlation analysis was computed.

**Correlation analysis**

Table 2 provides the correlation between eight of the dimensions, IBSQ, customer satisfaction and customer loyalty.

**Table 2: Correlation matrix**

| Factors / Constructs | F1 | F2 | F3 | F4 | F5 | F6 | F7 | F8 | IBSQ | Val | Sat | Loy |
|---|---|---|---|---|---|---|---|---|---|---|---|---|
| Factor 1 | 1 | | | | | | | | | | | |
| Factor 2 | .24 | 1 | | | | | | | | | | |
| Factor 3 | .39 | .28 | 1 | | | | | | | | | |
| Factor 4 | .24 | .41 | .41 | 1 | | | | | | | | |
| Factor 5 | .39 | .46 | .50 | .41 | 1 | | | | | | | |
| Factor 6 | .39 | .41 | .34 | .49 | .45 | 1 | | | | | | |
| Factor 7 | .40 | .40 | .42 | .47 | .49 | .46 | 1 | | | | | |
| Factor 8 | .38 | .39 | .42 | .44 | .39 | .48 | .37 | 1 | | | | |
| IBSQ | .67 | .62 | .65 | .69 | .71 | .69 | .77 | .66 | 1 | | | |
| Value | .33 | .32 | .38 | .36 | .36 | .32 | .32 | .47 | .50 | 1 | | |
| Satisfaction | .40 | .46 | .44 | .43 | .53 | .50 | .44 | .54 | .66 | .71 | 1 | |
| Loyalty | .31 | .29 | .40 | .32 | .42 | .34 | .40 | .40 | .52 | .47 | .55 | 1 |

Statistically significant at p≤0.01 (2-tailed)

A two-tailed significance level is assumed at the cut off level $p < 0.01$. On inspection of each pair of correlation, the Pearson's correlation coefficient at $p < 0.01$ level of significance indicates a positive linear association between each of the dimensions and constructs suggesting nomological validity, and none of the coefficients were above 0.9 so there were no multicollinearity issues (Hair et al., 2010). As can be seen from Table 2, the correlations between each pair of the eight dimensions that collectively explained IBSQ were significant, ranging from $r=0.242$ to $r=0.460$ at $p < 0.01$ level of significance, suggesting the existence of positive inter-factor associations. In addition, the correlations between each of the eight dimensions that collectively constitute the IBSQ construct also showed statistically significant associations with customer value, satisfaction and loyalty. The relationship between IBSQ, customer value, satisfaction and loyalty was also found to be positive and significant.

To assess the causal effect of the relationships detected through correlations analysis further, and to test the hypothesised research model specified earlier, structural equation modelling (SEM) was performed. The establishment of nomological validity and the absence of multicollinearity issues made it possible to perform SEM as a confirmatory factor analysis.

**Structural equation modelling (SEM)**

In line with the hypothesised research model, four constructs are identified in the model, namely IBSQ, customer value, satisfaction and loyalty. Following the recommendation of Malhotra (2010) and Hair et al. (2010) the measurement model was specified and identified, and the measured indicator items were assigned to latent constructs.

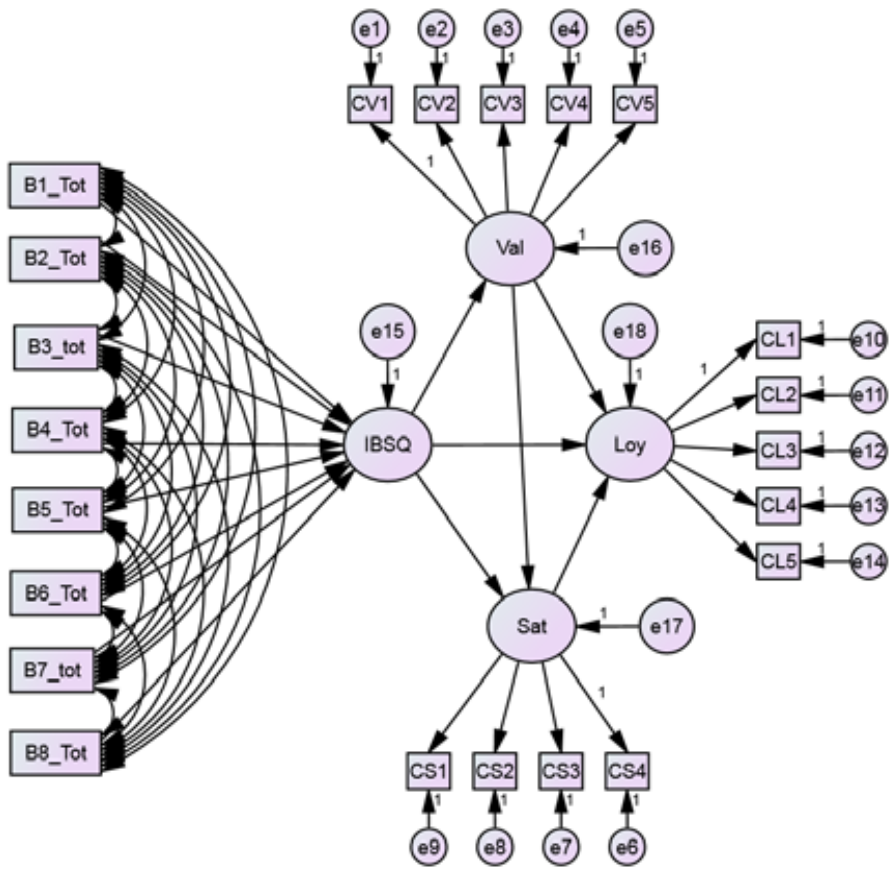

**Figure 2:** Specified measurement model

Figure 2 depicts the hypothesised specified measurement model. IBSQ, with eight latent factors, collectively comprised 31 items, with five items for customer value (Val), four items for satisfaction (Sat) and five items for loyalty (Loy). The eight latent factors are the dimensions of IBSQ (extracted through EFA), namely reliability, system availability, privacy and security, website aesthetics, ease of use, functionality, efficiency, and contact and responsiveness.

**Reliability and validity of the measurement model**

Composite reliability (CR), average variance extracted (AVE) and the correlation coefficients were computed to determine the reliability and validity of the scale. The CR results were above the cut off 0.7 level, suggesting a good internal consistency of the scale. It must be borne in mind that the Cronbach alpha reliability, of the all the factors extracted that constituted the scale, were above 0.8 portraying very good reliability (Hair et al., 2010). In terms of validity, AVE values for all constructs were above 0.50 indicating evidence of convergent validity of the scale (Malhotra, 2010). In addition, an average inter-item correlation of the scale fell within the suggested range of 0.15 and 0.50 which indicates acceptable discriminant validity (Clark & Watson, 1995). Factor loadings of the items had absolute value scores above 0.5 which further confirms that convergent validity was acceptable in this study.

**Assessment of goodness-of-fit indices**

Four competing models were identified in order to test the hypothesised research model and identify the best model fit. Table 3 reports on the goodness-of-fit indices of the competing models. Structural Model A is a three-construct model that included IBSQ, customer value and loyalty. Similarly, Structural Model B is also a three-construct model that included IBSQ, customer satisfaction and loyalty. However, Structural Model C and D are models that included four of the constructs of this study, namely IBSQ, customer value, satisfaction and loyalty.

**Table 3:** Goodness-of-fit indices for the competing models

| Indices | Acceptable level | Structural Model A | Structural Model B | Structural Model C | Structural Model D |
|---|---|---|---|---|---|
| $X^2/df$ | $\leq 5$ | 3.027 | 2.833 | 2.786 | 2.775 |
| IFI | $\geq 0.90$ | 0.899 | 0.918 | 0.903 | 0.903 |
| TLI | $\geq 0.90$ | 0.882 | 0.903 | 0.889 | 0.890 |
| CFI | $\geq 0.90$ | 0.898 | 0.917 | 0.903 | 0.903 |
| RMSEA | $< 0.08$ | 0.081 | 0.077 | 0.076 | 0.076 |

The chi-square test ($X^2$) is viewed as an overly strict indicator of model fit, given its power to detect even trivial deviations from the proposed model. Mueller (1996) suggested that the chi-square test statistic be divided by degrees of freedom where an acceptable level is observed at <3. Interpretation of the size of this value depends largely on the viewpoint of the investigator, but in practice, some interpret ratios as high as three, four or even five as still representing a good model fit (Mueller, 1996). All the competing models exhibited good model fit with regard to the chi-square test. The indices used to assess the goodness-of-fit of the structural models include incremental fit index (IFI), Tucker Lewis index (TLI) and comparative fit index (CFI). Indices values closer to one indicate a perfect fit and those closer to zero represent no fit (Malhotra, 2010; Hair et al., 2010:665). With regard to root mean square error of approximation (RMSEA), there is a good model fit if RMSEA is less than or equal to 0.05 and an adequate fit if RMSEA is less than or equal to 0.08 (Blunch, 2008). Blunch (2008) is of the view that models with RMSEA values of 0.10 and larger should not be accepted. Overall, Structural Model A exhibited poor goodness-of-fit indices while Structural Model B produced much improved and acceptable indices. Of the remaining two Structural Models (C & D) that included four of the constructs, Structural Model D provided better fit producing overall acceptable fit indices with the exception of Tucker Lewis index (TLI) missing the threshold with 0.010. On a scale of zero being no fit and one being perfect fit, Model D is still an acceptable model fit for the dataset and for purposes of testing the hypothesised research model. In the following section, analyses of the percentage of mediation effect and standardised regression weights are used to test the hypothesised research model further.

**Structural model and mediation effects**

Table 4 provides standardised regression weights of the four competing models.

**Table 4:** Standardised regression weights: IBSQ, satisfaction, value & loyalty

| Models | Constructs | | | Standardised regression weights | p-value |
|---|---|---|---|---|---|
| Structural Model A | Val | ← | IBSQ | 0.909 | *** |
| | Loy | ← | IBSQ | 0.729 | *** |
| | Loy | ← | Val | 0.310 | *** |
| Mediation effect of customer value on loyalty = **27.86%** | | | | | |
| Structural Model B | Sat | ← | IBSQ | 1.17 | *** |
| | Loy | ← | IBSQ | 0.432 | 0.01 |
| | Loy | ← | Sat | 0.512 | *** |
| Mediation effect of customer satisfaction on loyalty = **58.05%** | | | | | |
| Structural Model C | Val | ← | IBSQ | 0.601 | *** |
| | Sat | ← | IBSQ | 0.455 | *** |
| | Sat | ← | Val | 0.560 | *** |
| | Loy | ← | IBSQ | 0.285 | 0.009 |
| | Loy | ← | Sat | 0.080 | 0.492 |
| Combined mediation effect of customer value & satisfaction on loyalty = **60.16%** | | | | | |
| Structural Model D | Val | ← | IBSQ | 0.936 | *** |
| | Sat | ← | IBSQ | 0.668 | *** |
| | Loy | ← | IBSQ | 0.410 | 0.011 |
| | Sat | ← | Val | 0.534 | *** |
| | Loy | ← | Sat | 0.529 | *** |

The causal effects are read in the direction of the arrows. In Structural Model A, the results suggest that IBSQ has a statistically significant positive influence on customer value and on loyalty, and customer value in return has positive effect on loyalty. In this three-construct model, the mediating effect of customer value on loyalty was calculated to be 27.86 percent. In Structural Model B, also a three-construct model, the results indicate that IBSQ has a statistically significant positive influence on satisfaction and on customer loyalty and customer satisfaction in return has significant positive effect on customer loyalty. In this three-construct model, the mediating effect of customer satisfaction on loyalty was estimated to be 58.05 percent much higher than the percentage of mediation effect of customer value on loyalty.

In Structural Model C, two constructs (customer satisfaction and loyalty) play a mediating role together. The regression path estimates suggested that IBSQ yields positive effect on three of the constructs, namely customer value, satisfaction and loyalty. However, the regression path estimate of customer value on loyalty was reduced by two thirds of its original effect without the intervention of customer satisfaction. In this four-construct model, the combined mediating effect of customer value and satisfaction on loyalty was estimated to be 60.16 percent, where once again a diminished effect is observed with regard to the influence of customer value on loyalty. In this Model, the influence of IBSQ on value, satisfaction and loyalty is significant. In turn, value also positively influenced satisfaction while satisfaction was found not to be a predictor of loyalty.

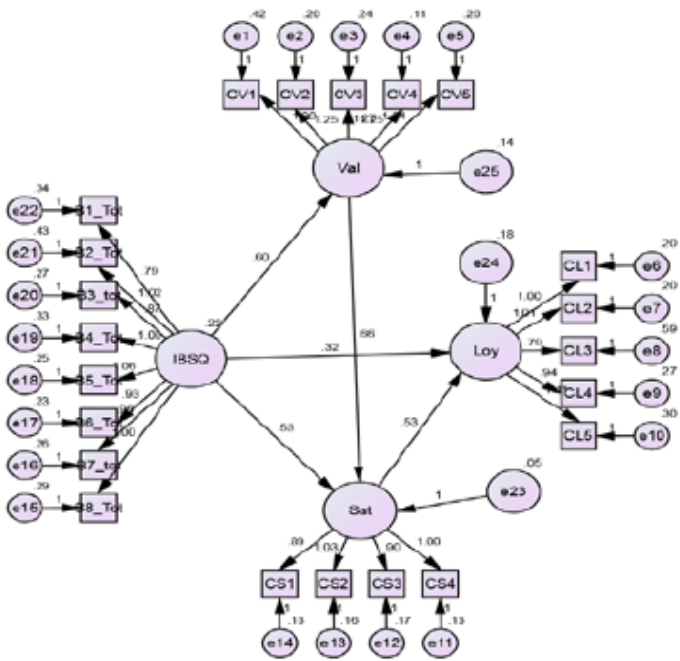

**Figure 3:** Structural Model D: IBSQ Measurement Model

Figure 3 exhibits the regression path estimates (coefficients) of the fourth competing Structural Model (D). The direct effect of customer value on loyalty was removed in light of the diminished mediation effect observed in Structural Model C. Moreover, Structural Model D did exhibit a better-fit indices compared to Structural Model C. The regression path estimates suggest that IBSQ yields positive effect on three of the constructs namely customer value, satisfaction and loyalty. Furthermore, the regression path estimates indicate that customer value in return has a positive effect on satisfaction but not on loyalty. Customer satisfaction, in return, is indicated to have a positive effect on customer loyalty. The results of standardised regression weights also support these findings. The four-construct Structural Model D indicates that the model fits well in representing the dataset. Therefore, Structural Model (D) was identified and proposed yielding a better fit for the hypothesised research model.

**Hypotheses testing**

Hypotheses testing was carried out with a significance level set at the conventional $p<0.05$ level. The causal effects read in the direction of arrows indicate that IBSQ positively influences customer value significantly (0.936) at $p<0.05$ level; IBSQ positively influences satisfaction significantly (0.668) at $p<0.05$ level and IBSQ positively influences customer loyalty (0.410) at $p<0.05$ level. Accordingly, H1, H2, and H3 are supported. This finding corroborates the results of previous studies conducted in other contexts that found service quality to be a predictor of customer value (Patterson & Spreng, 1997; Cronin *et al.*, 2000; Kuo *et al.,* 2009), customer satisfaction (Cronin *et al.*, 2000; Kuo *et al.,* 2009) and customer loyalty (Parasuraman *et al.*, 1988; Patterson & Spreng, 1997; Koo, 2006; Siddiqi, 2011). Similarly, the causal effects read in the direction of arrows suggest that customer value positively influences satisfaction significantly (0.534) at $p<0.05$ level, and customer satisfaction in turn positively influences loyalty significantly (0.529) at $p<0.05$ level. Hypotheses H4 and H6 are therefore supported. This result is consistent with other studies that established the positive influence of customer value on customer satisfaction (Anuwichanont & Mechinda, 2009), and the positive influence of customer satisfaction on loyalty (Dong, 2003; Siddiqi, 2011).

It must be borne in mind that the direct effect of customer value on loyalty was removed in light of the diminished mediation effect observed in Structural Model C. The influence of customer value on customer loyalty is only indirect through its influence on customer satisfaction. Hypothesis H5 is therefore rejected. This finding corroborates the finding of Cronin *et al.* (2000) that loyalty does not directly influence customer satisfaction. However, it must be noted that a few previous studies conducted in other contexts have found customer value to have a direct and positive link with customer loyalty (Zeithaml, 1988; Patterson & Spreng, 1997; Lewis & Soureli 2006). Ultimately, a four-construct Structural Model (D) was identified and proposed as yielding a better fit for the hypothesised research model. It is the proposed IBSQ Measurement Model that can be used for measuring and modelling of IBSQ in the South African banking sector.

## CONCLUSIONS AND RECOMMENDATIONS

Employing a sequential-mixed method approach in line with the practice of the development of a service model (or a scale) where a qualitative research is first conducted followed by quantitative research, the study has proposed a framework for internet banking service quality management in the context of South Africa. The proposed framework can easily be applied in other contexts as is or with little adaptations. Eight determinant factors that explain IBSQ (reliability, system availability, privacy and security, website aesthetics, ease of use, functionality, efficiency, and contact and responsiveness) were identified. Furthermore, the model has determined the causal relationships among four important constructs, namely IBSQ, customer value, satisfaction, and loyalty. Understanding consumer behaviour and decision making in this digital era will enhance the banks` quest to provide quality services and devise appropriate customer service solutions. The research revealed that reliability, privacy and security are the top concerns customers have with regard to Internet banking. Therefore, it is recommended that banks invest in the robustness of the websites for banking transactions by using cutting-edge technology to protect their customers from illicit criminal activity, as security and trust are of crucial importance to customers when engaging in online transactions. To enhance efficiency of Internet banking, it is recommended that banks should ensure that the service delivered through the bank's website is quick to access for transactions from any location and at any time, the bank's website loads fast all the time and the bank's website does not freeze during a transaction. With regard to contact, responsiveness and system availability, it is recommended that banks need to repair a breakdown on the website quickly as and when it occurs, promptly resolve serious problems that customer encounter, provide prompt feedback to customer requests by e-mail or other means, and improve and closely manage customer complaints.

It must be emphasised that satisfying customers is a 'must achieve' objective for any bank that wishes to remain profitable and relevant in the competitive banking sector. This must be done by providing quality services that create value for customers. Achieving loyal customers who will patronise and associate themselves with the bank is of particular significance for market growth and success. In view of the relationship of IBSQ dimensions with customer value, satisfaction and loyalty, focus must be placed on the individual building blocks of service quality, *inter alia* the factors that influence Internet banking service quality. These are the factors that influence consumer behaviour and decision making in online banking environment. Periodic measurement of the levels of Internet banking service quality through valid and reliable measuring scale should become an integral part of any bank's effort and strategy in improving service quality levels. Using the scale developed in this study, future studies could use a larger sample size to test the robustness of this scale, and obtain more exact result to draw wider generalisations.

# REFERENCES

Adil, M. (2013). Efficacy of SERVPERF in measuring perceived service quality at rural retail banks: Empirical evidences from India. *International Journal of Business Insights and Transformation*, 6(1), 52-63.

Anuwichanont, J. & Mechinda, P. (2009). The impact of perceived value on spa loyalty and its moderating effect of destination equity. *Journal of Business & Economics Research,* 7(12), 73-90.

Aziz, A.B., Elbadrawy, R. & Hussien, M.I. (2014). ATM, Internet Banking and Mobile Banking Services in a Digital Environment: The Egyptian Banking Industry. *International Journal of Computer Applications,* 90(8), 45-52.

Barnes, S.J. & Vidgen, R. (2003). Measuring website quality improvements: a case study of the forum of strategic improvement. *Industrial Management and Data Systems*, 103(5), 297-309.

Blunch, N.J. (2008). *Introduction to structural equation modelling using SPSS and Amos*. London: SAGE.

Boshoff, C. & Du Plessis, P.J. (2009). *Services marketing: a contemporary approach*. Cape Town: Juta.

Churchill, G.A. (1979). A paradigm for developing better measures of marketing constructs. *Journal of Marketing Research*, 16, 64-73.

Clark, L.A. & Watson, D. (1995). Construct validity: Basic issues in objective scale development. *Psychological assessment,* 7(3), 309-319.

Cronin, J.J., Brady, M.K. & Hult, G.T.M. (2000). Assessing the effects of quality, value, and customer satisfaction on consumer behavioural intentions in service environments. *Journal of Retailing*, 76(2), 193-218.

Cronin, J.J. & Taylor, S.A. (1994). SERVERF versus SERVQUAL: reconciling performance based and perceptions minus expectations measurement of service quality. *Journal of Marketing,* 58(1), 125-33.

Dabholkar, P.A., Thorpe, D.I. & Rentz, J.O. (1996). A measure of service quality for retail stores: scale development and validation. *Journal of the Academy of Marketing Science*, 24(1), 3-16.

Dong, M.K. (2003). Inter-relationships among store image, store satisfaction, and store loyalty among Korea discount retail patrons. *Asia Pacific Journal of Marketing and Logistics*, 15(4), 42-71.

Garland, R. (1991). The mid-point on a rating scale: Is it desirable? *Marketing bulletin*, 2, 66-70.

Grönroos, C. (1984). A service quality model and its marketing implications. *European Journal of Marketing*, 18(4), 36-44.

Grönroos, C. (2007). *Service management and marketing: Customer management in service competition.* London: John Wiley.

Gupta, S. & Zeithaml, V. (2006). Customer metrics and their impact on financial performance. *Marketing Science*, 25(6), 718-739.

Hallowell, Roger. (1996). The Relationship of Customer Satisfaction, Customer Loyalty, and Profitability: An Empirical Study, The *International Journal of Service Industry Management*, **7** (4), 27-42.

Hair, J.F., Black, W.C., Babin, B.J. & Anderson, R.E. (2010). *Multivariate data analysis: a global perspective.* 7th ed. New Jersey: Pearson Education.

Han, S. & Baek, S. (2004). Antecedents and consequences of service quality in online banking: An application of the SERVQUAL instrument. *Advances in Consumer Research,* 31, 208-214.

Iacobucci, D. (2016). *Marketing Management.* Boston: Cengage.

Johnson, M., Gustafsson, A., Andreassen, T., Lervik, L. & Cha, J., (2001). The evolution and future of national customer satisfaction index models. *Journal of Economic Psychology*, 22(2), 217-245.

Jun, M. & Cai, S. (2001). The key determinants of internet banking service quality: A content analysis. *International Journal of Bank Marketing,* 19(7), 276-291.

Knox, S.D. & Denison, T.J. (2000). Store loyalty: Its impact on retail revenue, an empirical study of purchasing behaviour in the UK. *Journal of Retailing and Consumer Services*, 7, 33-45.

Koo, D.M. (2006). The fundamental reasons of e-consumers' loyalty to an online store. *Electronic Commerce Research and Applications*, 5(2), 117-130.

Kuo, Y.F., Wu, C. & Deng, W.J. (2009). The relationships among service quality, perceived value, customer satisfaction, and post-purchase intention in mobile value-added services. *Computers in Human Behaviour*, 25, 887-896.

Lee, H.S. (2010). Factors influencing customer loyalty of mobile phone service: Empirical evidence from Koreans. *Journal of Internet Banking and Commerce,* 15(2), 2-14.

Lewis, B.R. & Soureli, M. (2006). The antecedents of consumer loyalty in retail banking, *Journal of Consumer Behaviour*, 5, 15-31.

Loonam, M. & O'Loughlin, D. (2008). Exploring e-service quality: A study of Irish online banking. *Marketing Intelligence & Planning,* 26(7), 759-780.

Lovelock, C. & Wright, L. (1999). *Principles of marketing and management*. New York: Pearson Prentice-Hall.

Malhotra, N.K. (2010). *Marketing research: An applied orientation*. 6th ed. Upper New Jersey: Pearson Education Prentice Hall.

Mueller, R.O. (1996). *Basic principles of structural equation modeling: An introduction to LISREL and EQS*. New York, NY: Springer.

Narteh, B. (2013). Service quality in automated teller machines: An empirical investigation. *Managing Service Quality*, 23(1), 62-89.

Redda, E.H., Surujlal, J. & Leendertz, V. (2015). Internet Banking Service Quality Scale Development in South Africa: A Qualitative Approach. 27th Annual SAIMS Conference 2015, Cape Town, South Africa.

Pallant, J. (2013). A step by step guide to data analysis using IBM SPSS: Survival manual. 5th ed. New York: McGraw-Hill.

Parasuraman, A., Zeithaml V.A., & Berry, L.L. (1985). A conceptual model of service quality and its implications for future research. *Journal of Marketing,* 49, 41-50.

Parasuraman, A., Zeithaml V.A., & Berry, L.L. (1988). SERVQUAL: a multiple-item scale for measuring consumer perceptions of service quality. *Journal of Retailing*, 64, 12-40.

Parasuraman, A., Zeithaml, V. & Malhotra, A. (2005). E-S-QUAL: A multiple-item scale for assessing electronic service quality. *Journal of Service Research,* 7(3), 213-233.

Patterson, P.G. & Spreng, R.A. (1997). Modeling the relationship between perceived value, satisfaction and repurchase intentions in a business-to-business, services context: an empirical examination. *International Journal of Service Industry Management*, 8(5), 414-434.

Ruiz-Molina, M. & Gil-Saura, I. (2008). Perceived value, customer attitude and loyalty in Retailing. *Journal of Retail and Leisure Property,* 7, 305-314.

Santos, F. (2003). E-service quality: A model of virtual service quality dimensions. *Managing Service Quality,* 13(3), 233-246.

Siddiqi, K.M. (2011). The drivers of customer loyalty to retail banks: an empirical study in Bangladesh. *Industrial Engineering Letter*, 1(1), 40-50.

Torres, M., Summers, T. & Belleau, D. (2001). Men's shopping satisfaction and store references. *Journal of Retailing and Consumer Services*, 8, 205-212.

Van Riel, A., Liljander, V. & Jurriens, P. (2001). Exploring consumer evaluations of e-services: a portal site. *International Journal of Service Industry Management*, 12(4), 359-377.

Zeithaml, V.A. (1988). Customer perceptions of price, quality, and value: A measure-end model and synthesis of evidence. *Journal of Marketing*, (5), 2-22.

# TRENDS IN eBUSINESS AND eGOVERNMENT

Technology affects all areas. Business and government processes are changing with the use of the internet, mobile devices, internet of things, blockchain, machine learning, artificial intelligence and many other new technologies. In this book, it is aimed to focus the use of technology, new trends in business life and government covering the studies in all sub-areas of Information Systems, Knowledge Management, eBusiness, eCommerce, eMarketing, mCommerce, eGovernment, ePublic Services, eGovernance etc. The book consists of 7 chapters. Book chapter authors are reputable scientists from different countries of the world. The first chapter is a critical review and a case study in e-Business, with special attention to the digital currencies resource and its possibilities. The second chapter attempts to incorporate the Unified Theory of Acceptance and Use of Technology (UTAUT) model with perceived risk theory (security risk and privacy risk) to explore its impact towards the intention to use m-government services. The third chapter aims to assess the level of gender inclusivity in the municipal e-procurement processes in the City of Johannesburg as a case study. The fourth chapter examines the impediments that derail the intensive uptake of eLearning programmes in a particular higher education institution. The fifth chapter investigated the role of Knowledge Management Systems (KMS) in enhancing the export performance of firms operating within the manufacturing sector in Zimbabwe. In the sixth chapter, a survey was undertaken on 131 small and medium-sized enterprises (SMEs) from Pelagonija region in order to determine the current level of SME digitalization within the region. It is aimed to compare with European Union (EU) average and to make conclusions on the impact of the SME digitalization to region gross domestic product (GDP) growth as well as revenues collection. The last chapter's purpose was to develop a measuring and modelling framework/instrument of Internet banking service quality (IBSQ) for the South African banking sector.

9 786257 729468

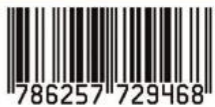

EFEACADEMY